\documentclass[preprint,trackchanges]{aastex63}

\usepackage{amsmath}

\newcommand\aanda{A&A}

\usepackage{color}
\definecolor{msp}{rgb}{0.0,0.5,0.5}
\shorttitle{Gaia-EDR3 dwarf galaxy orbits}
\shortauthors{Li et al.}

\begin{document}

\title{Gaia EDR3 proper motions of Milky Way dwarfs I: 3D Motions and Orbits}

\correspondingauthor{Francois Hammer, Hefan Li}
\email{francois.hammer@obspm.fr, lihefan16@mails.ucas.edu.cn}

\author{Hefan Li}
\affiliation{School of Physical Sciences, University of Chinese Academy of Sciences, Beijing 100049, P. R. China}

\author{Francois Hammer}
\affiliation{GEPI, Observatoire de Paris, Universit\'e PSL, CNRS, Place Jules Janssen 92195, Meudon, France}

\author{Carine Babusiaux}
\affiliation{Universit\'e de Grenoble-Alpes, CNRS, IPAG, F-38000 Grenoble, France }
\affiliation{GEPI, Observatoire de Paris, Universit\'e PSL, CNRS, Place Jules Janssen 92195, Meudon, France}

\author{Marcel S. Pawlowski}
\affiliation{Leibniz-Institut fuer Astrophysik Potsdam (AIP), An der Sternwarte 16, D-14482 Potsdam Germany}

\author{Yanbin Yang}
\affiliation{GEPI, Observatoire de Paris, Universit\'e PSL, CNRS, Place Jules Janssen 92195, Meudon, France}

\author{Frederic Arenou}
\affiliation{GEPI, Observatoire de Paris, Universit\'e PSL, CNRS, Place Jules Janssen 92195, Meudon, France}

\author{Cuihua Du}
\affiliation{School of Physical Sciences, University of Chinese Academy of Sciences, Beijing 100049, P. R. China}

\author{Jianling Wang}
\affiliation{CAS Key Laboratory of Optical Astronomy, National Astronomical Observatories, Beijing 100101, China}

\begin{abstract}
\par Based on Gaia Early Data Release 3 (EDR3), we estimate the proper motions for 46 dwarf galaxies of the Milky Way. The uncertainties in proper motions, determined by combining both statistical and systematic errors, are smaller by a factor 2.5, when compared with Gaia Data Release 2. We have derived orbits in four Milky Way potential models that are consistent with the MW rotation curve, with total mass ranging from $2.8\times10^{11}$ $M_{\odot}$ to $15\times10^{11}$ $M_{\odot}$.  Although the type of orbit (ellipse or hyperbola) are very dependent on the potential model, the pericenter values are firmly determined, largely independent of the adopted MW mass model. By analyzing the orbital phases, we found that the dwarf galaxies are highly concentrated close to their pericenter, rather than to their apocenter as expected from Kepler's law. This may challenge the fact that most dwarf galaxies are Milky Way satellites, or alternatively indicates an unexpected large number of undiscovered dwarf galaxies lying very close to their apocenters.
Between half and two thirds of the satellites have orbital poles that indicate them to orbit along the Vast Polar Structure (VPOS), with the vast majority of these co-orbiting in a common direction also shared by the Magellanic Clouds, which  is indicative of a real structure of dwarf galaxies.
%By comparing dwarf galaxies' orbital poles and sky coordinates, from an half to two-thirds of dwarf galaxies are indeed lying and orbiting into the vast polar structure (VPOS). 

\end{abstract}


\section{introduction}

Gaia DR2 \citep{helmi18Gaia} revolutionized the knowledge of Milky Way (MW) dwarf galaxies by revealing their proper motions (PMs) \citep{fritz18Gaia}. Because they lie in our neighborhood, MW dwarf galaxies are unique galaxies for which we can firmly establish 6D phase diagrams (3D locations and 3D velocities). Gaia EDR3 \citep{2016A&A...595A...1G,2020arXiv201201533G} provides a significant step forward by improving PM precision by a factor $\sim 2$, including for very faint dwarf galaxies.
 
In the meantime between Gaia DR2 and EDR3 epochs, several new dwarf galaxies have been discovered, or have been spectroscopically observed to derive their radial velocities, which is the essential complement to Gaia PMs for establishing their 3D motions (see references in Table~\ref{tab:pm}). Combined with Gaia EDR3 this allows the determination of accurate orbits for several tens of dwarf galaxies much less massive than the classical dwarf spheroidal galaxies\footnote{Following \citet{Tolstoy2009} MW dwarfs include dwarf spheroidals (dSphs) that are the 9 classical dSphs and Canes Venaciti I, and ultra-faint dwarfs (UFDs), which are less luminous, with $L_V$ $<$ $10^5 M_{\odot}$, or $M_V$ $>$ -7.5.}, a yet unprecedented number. In principle, one may now investigate their past and future history.
  
However, besides observed (instantaneous) quantities (distances and velocities), evaluating integrated orbital  quantities (pericenter, apocenter, orbit shape and eccentricity) requires the knowledge of the potential, i.e., namely that of the main Galaxy, the MW. It has been shown \citep{fritz18Gaia,Hammer2020} that the determination of dwarf galaxy orbital quantities is limited by Gaia DR2 PM uncertainties, but perhaps even more by the systematics due to the inaccurate knowledge of the MW mass profile. Reducing PM errors by at least a factor $\sim 2$ with EDR3, leads to a problem for which the limited knowledge of the MW mass-profile dominates the uncertainties in the derived orbital properties. Another possible limitation is the impact of the second massive body in the MW halo, the Large Magellanic Cloud (LMC), which can, if it has a total mass exceeding $10^{11}$ $M_{\odot}$, also affect stellar streams such as the Orphan stream \citep{erkal19total}.

In other words, the MW potential uncertainties hamper our knowledge of dwarf galaxy orbits. The circular velocity curve of the MW has been accurately provided by \citet{eilers19Circular} and \citet{mroz19Rotation}. This result can be associated to other possible probes of the MW potential, for example dwarf galaxy \citep{cautun20milky} or globular cluster motions \citet{wang21}, leading to quite discrepant values for the total MW mass. This is not unexpected because these methods depend on whether or not these additional probes share a similar equilibrium than rotating disk stars in the MW potential. \citet{jiao21Which} showed that MW mass profiles derived from the rotation curve are also affected by the choice of the dark-matter mass profile. In a generalized study, they determine the range of total MW mass that can be consistent with the MW rotation curve fit. They found that the MW mass can be as small as $2.8\times10^{11}$ $M_{\odot}$ (see also \citealt{desalas19estimation} and \citealt{karukes20robust}) or as large as $15\times10^{11}$ $M_{\odot}$. This somewhat constraints the available range of MW potential, but it also emphasizes that it is still required to consider a wide range of total MW masses when performing orbital analyses.

Dwarf galaxies present a large variety of properties in mass, radius and velocity dispersion (see an analysis in \citealt{hammer19Absence}). Few of them (Hercules, Tucana III) show sign of tidal disruption, which for Tucana III could be related to its passage at low pericenter \citep{Li2018}. More enigmatic is the fact that most dwarf galaxies appear to lie and to orbit within a gigantic disk almost perpendicular to the MW disk, the Vast Polar Structure \citep{pawlowski14Vast}. Dwarf galaxies appear also to be excessively close to their pericenters \citep{fritz18Gaia,simon18Gaia}. Their internal properties also show unusual properties. For example, their gravity at half-light radius declines with Galactic distance \citep{hammer19Absence}. DSphs have had their own complex star-formation history, which is revealed when they possess two distinct main stellar populations \citep[Sculptor and Ursa Minor,][]{tolstoy04Two,pace2020}, or more than two \citep[Carina and Fornax,][]{savino15Inclusion, deboer12star}. However, Draco appears to  possess only a single stellar population \citep{aparicio01Star}. %Another possible enigmatic property is the fact that each time a dwarf galaxy is intensively studied, it appears to be populated by two stellar populations with different characteristic radius, age, metal abundance and velocity dispersion \citep[and Andrew Pace, 2020, private communication]{pace2020}.

This paper aims to provide updated and new proper motion measurements for MW dwarf galaxies, and determines their resulting integrated orbital parameters in the widest range of MW masses that remain consistent with the MW rotation curve. The goal is to avoid possible systematics or biases that could affect our understanding of dwarf galaxies. Section 2 describes the dataset, the determination of observed properties deduced from Gaia EDR3 and radial velocity measurements. In Section 3, we have calculated integrated orbit properties for four different MW potentials derived from  the \citet{jiao21Which} analysis. Results are given and discussed in Section 4, including the dwarf galaxy 3D phase-diagram, the VPOS, and the fraction of dwarf galaxies lying near their pericenters.

\section{data}
\subsection{Sample of dwarf galaxies}
The sample has been selected from the literature \citep{fritz18Gaia,Simon2019,mcconnachie20Updated}, with the supplementary request to have at least four spectroscopically measured stars detected by Gaia EDR3 (the full reference list is given in Table~\ref{tab:pm}). To this we have added five dwarf galaxies having three measured stars or less to investigate the behavior of PM accuracy at low signal (name in italic in Table~\ref{tab:pm}).  We have also added Eridanus II, even though its very large distance precludes accurate velocity measurements. Following \citet{fritz18Gaia} we have kept Crater I and Draco II in the dwarf galaxy list, since their precise nature (globular cluster or dwarf galaxy) is still under discussion. The total list includes 46 galaxies.

\subsection{Proper Motions, uncertainties, and comparison with other studies}
\label{sec:pm}
We have first selected dwarf galaxy member stars from literature catalogues based, among others, on their radial velocities (see references in Table~\ref{tab:pm}), for which we have calculated the median.
We then have further selected only stars with Gaia parallaxes and proper motions consistent with the median value of the dwarf galaxy at 5$\sigma$.  Finally, stars with a dubious astrometry have been also removed if the Gaia renormalized unit weight error (ruwe) is larger than 1.4 \citep{lindegren20Gaiaa}.

%\msp{MSP: CAN YOU PROVIDE A BIT MORE EXPLANATION, I.E. WHAT'S THE EXACT PROCESS? IN PARTICULAR, HOW TO YOU SELECT THE FIVE SIGMA AROUND THE MEDIAN PROPER MOTION WHEN IT IS THE PROPOER MOTION YOU WANT TO DERIVE? I ASSUME THIS IS AN ITERATIVE PROCESS? ALSO, CAN YOU EXPLAN WHAT THE RUWE CRITERIA IS, I DONT THINK THE AVERAGE READER WOULD KNOW (I DON'T).}

To compute the dwarf galaxy proper motions and their errors, we have adopted the method described in \citet{vasiliev19Systematic} who used both the statistical covariance matrix (as derived from the formal errors and correlations provided in the Gaia catalogue) and the systematic one derived from the spatial correlations.
The proper motion covariance function has been constructed using the formulae of \citet{vasiliev19Systematic} adapted to EDR3 and using the values of \citet{lindegren20Gaiaa}:
\begin{equation}
 V_\mu(\theta) = 292 \exp(-\theta/12^\circ) +  258 \exp(-\theta/0.25^\circ)\,\mu\mathrm{as^2\,yr}^{-2}
\end{equation} 

\par \citet{mcconnachie20Updated} promptly published PM estimates just after the Gaia EDR3. Appendix~\ref{compPMs} compares their values to those in Table~\ref{tab:pm}, which show an excellent agreement, except for error bars that are systematically smaller in \citet{mcconnachie20Updated} by large factors, almost always larger than 2 but that can reach 11 (for Sculptor), and even 20 for Fornax. We interpret this as a difference in the treatment of systematics (not accounted by \citet{mcconnachie20Updated}), which is still an important limitation for determining PMs of dwarf galaxies, and especially to evaluate their orbital motions. Figure~\ref{fig:compPM} indicates the name of the few dwarf galaxies for which both studies show disagreement, which further points out the uncertain determination of PMs for 5 galaxies with 3 stars or less: Aquarius II, Columba I, Horologium II, Pisces II, and Reticulum III.

\par Appendix~\ref{compPMs} also compares EDR3 results to \citet{fritz18Gaia} DR2 values. EDR3 and DR2 PM values are often consistent within the pretty conservative DR2 and EDR3 error bars (see Figure~\ref{fig:compPMDR2}) . However, for some dwarf galaxies there is some changes of the PM from DR2 to EDR3. Specifically, changes exceeding the DR2  $1\sigma$\ error bars from \citet{fritz18Gaia} are affecting RA PMs of Aquarius II, Crater II, Draco, Grus I, Hercules,  Leo IV, Pisces II, Segue I, Segue II, Tucana III, UMa I, and UMi, and for DEC PMs, Eridanus II, Hydrus, and Triangulum II. Top panels of Figure~\ref{fig:compPMDR2} also show that on average, typical errors (including systematics) have decreased by a factor of about 2.5 from DR2 to EDR3.
Note however that the number of dwarf galaxies with different measurements from DR2 to EDR3 is consistent with expectations: there are 37 satellites in common between the two data sets, and one expects on average 32\% of data points to disagree by $1\sigma$ or more, which corresponds to about 12 dwarf galaxies in this case. This number well consistent with the numbers of measurements differing by $1\sigma$ or more.

\startlongtable
\begin{deluxetable*}{lRCCCCCCc}
\label{tab:pm}
\tabletypesize{\footnotesize}
\tablecaption{Origin data of dwarf galaxies.}
\tablehead{
	\colhead{name} & \colhead{$dm$\tablenotemark{a}} & \colhead{$N_\mathrm{star}$} & \colhead{$\varpi$} &
	\colhead{$\mu_{\alpha^*}$} & \colhead{$\mu_{\delta}$} &
	\colhead{$\rho_{\mu_{\alpha^{*}}}^{\mu_{\delta}}$} & \colhead{$rv$} & \colhead{Ref.} \\
	\colhead{} & \colhead{}& \colhead{}  & \colhead{mas} &
	\colhead{(mas yr$^{-1}$)} & \colhead{(mas yr$^{-1}$)} &
	\colhead{} & \colhead{(km s$^{-1}$)} & \colhead{}
}
\colnumbers
\startdata
AntII &  20.6 \pm 0.11 &  157 & -0.017 \pm 0.018 &  -0.101 \pm 0.02 &   0.113 \pm 0.02 &            0.044 &  294.0 \pm 0.1 &        1 \\
\textit{AquII} & 20.17 \pm 0.07 &    2 & -0.845 \pm 0.564 &  0.647 \pm 0.588 & -0.298 \pm 0.548 &            0.267 &  -74.4 \pm 1.0 &        2 \\
BooI &  19.1 \pm 0.07 &   37 &  -0.04 \pm 0.049 & -0.307 \pm 0.052 & -1.157 \pm 0.043 &           -0.107 &  100.3 \pm 0.1 &      3,4 \\
BooII & 18.12 \pm 0.05 &    4 &  0.215 \pm 0.148 & -2.273 \pm 0.151 & -0.361 \pm 0.115 &           -0.164 & -114.2 \pm 1.4 &        5 \\
CVenI & 21.62 \pm 0.06 &   53 & -0.019 \pm 0.059 & -0.084 \pm 0.052 & -0.127 \pm 0.037 &            0.149 &   26.4 \pm 0.2 &      6,3 \\
CVenII & 21.02 \pm 0.05 &   15 &   0.15 \pm 0.124 & -0.138 \pm 0.111 &  -0.32 \pm 0.082 &            0.333 & -129.5 \pm 0.7 &        7 \\
CarI &  20.13 \pm 0.1 &  882 & -0.004 \pm 0.018 &  0.533 \pm 0.021 &   0.12 \pm 0.021 &           -0.009 &  221.8 \pm 0.1 &      8,9 \\
CarII & 17.79 \pm 0.04 &   18 &   0.05 \pm 0.036 &  1.887 \pm 0.043 &  0.164 \pm 0.043 &            0.024 &  480.3 \pm 0.3 &       10 \\
CarIII & 17.22 \pm 0.05 &    4 &  0.041 \pm 0.054 &  3.082 \pm 0.068 &  1.394 \pm 0.072 &            0.021 &  283.2 \pm 0.6 &       10 \\
\textit{ColI} & 21.31 \pm 0.12 &    3 &  0.065 \pm 0.133 &  0.189 \pm 0.119 & -0.556 \pm 0.134 &           -0.069 &  152.3 \pm 3.7 &       11 \\
CberI & 18.12 \pm 0.08 &   17 &   0.03 \pm 0.056 &  0.374 \pm 0.059 & -1.699 \pm 0.056 &           -0.293 &   97.0 \pm 0.6 &        7 \\
CraI & 20.82 \pm 0.03 &    6 &  0.017 \pm 0.151 &  0.056 \pm 0.127 & -0.122 \pm 0.111 &           -0.292 &  149.3 \pm 0.7 &    12,13 \\
CraII & 20.35 \pm 0.02 &   59 &  0.004 \pm 0.029 & -0.072 \pm 0.032 & -0.123 \pm 0.025 &           -0.024 &   87.9 \pm 0.1 &       14 \\
DraI & 19.57 \pm 0.16 &  536 & -0.021 \pm 0.018 &   0.039 \pm 0.02 &  -0.181 \pm 0.02 &            0.021 & -292.1 \pm 0.0 & 15,16,17 \\
DraII & 16.66 \pm 0.04 &    7 & -0.024 \pm 0.105 &  1.011 \pm 0.115 &  0.956 \pm 0.126 &           -0.082 & -347.1 \pm 1.3 &       18 \\
EriII &  22.82 \pm 0.1 &   13 & -0.256 \pm 0.127 &  0.095 \pm 0.133 & -0.175 \pm 0.164 &           -0.213 &   74.6 \pm 0.3 &       19 \\
FnxI & 20.72 \pm 0.05 & 2463 &  -0.01 \pm 0.016 &  0.384 \pm 0.018 & -0.364 \pm 0.019 &           -0.014 &   55.3 \pm 0.0 &     8,20 \\
GruI &  20.4 \pm 0.21 &    7 & -0.068 \pm 0.078 &  0.071 \pm 0.056 &  -0.27 \pm 0.078 &            0.071 & -139.6 \pm 0.3 &       21 \\
GruII & 18.62 \pm 0.21 &   45 &  -0.032 \pm 0.04 &  0.389 \pm 0.034 & -1.526 \pm 0.036 &            0.159 & -109.9 \pm 0.3 &       22 \\
HerI &   20.6 \pm 0.1 &   20 &    0.11 \pm 0.07 & -0.051 \pm 0.059 & -0.332 \pm 0.051 &            0.431 &   44.7 \pm 0.4 &       23 \\
HorI &   19.7 \pm 0.3 &    5 & -0.058 \pm 0.045 &  0.865 \pm 0.047 & -0.601 \pm 0.048 &            0.029 &  116.5 \pm 0.1 &       24 \\
\textit{HorII} & 19.46 \pm 0.21 &    1 &  0.469 \pm 0.331 &   0.915 \pm 0.33 & -0.913 \pm 0.439 &           -0.113 & 158.3 \pm 11.3 &       11 \\
HyaII & 20.89 \pm 0.11 &    6 &   0.48 \pm 0.246 &  -0.576 \pm 0.28 & -0.101 \pm 0.212 &            0.054 &  303.8 \pm 1.0 &       12 \\
HyiI &  17.2 \pm 0.04 &   32 & -0.004 \pm 0.026 &  3.775 \pm 0.027 & -1.513 \pm 0.027 &            0.004 &   80.3 \pm 0.2 &       25 \\
LeoI & 22.02 \pm 0.13 &  368 & -0.065 \pm 0.032 & -0.066 \pm 0.029 & -0.107 \pm 0.026 &           -0.169 &  283.0 \pm 0.1 &    26,27 \\
LeoII & 21.84 \pm 0.13 &  221 &  0.031 \pm 0.037 & -0.125 \pm 0.039 & -0.122 \pm 0.036 &           -0.158 &   79.6 \pm 0.1 &    28,29 \\
LeoIV & 20.94 \pm 0.07 &    8 &  0.118 \pm 0.178 &  0.007 \pm 0.174 & -0.261 \pm 0.134 &           -0.168 &  131.2 \pm 0.5 &     7,30 \\
LeoV & 21.14 \pm 0.05 &    7 &  0.067 \pm 0.173 &  0.118 \pm 0.212 & -0.387 \pm 0.151 &           -0.137 &  174.0 \pm 0.4 &    31,30 \\
PhxII &  19.63 \pm 0.1 &    5 &  0.042 \pm 0.078 &  0.501 \pm 0.062 & -1.199 \pm 0.076 &           -0.389 &   32.2 \pm 1.5 &       11 \\
\textit{PisII} & 21.31 \pm 0.18 &    3 &  0.135 \pm 0.223 &  0.675 \pm 0.299 & -0.631 \pm 0.212 &           -0.016 & -227.5 \pm 1.1 &       12 \\
RetII &   17.5 \pm 0.1 &   28 &  0.025 \pm 0.029 &  2.391 \pm 0.029 & -1.379 \pm 0.032 &           -0.068 &   63.0 \pm 0.3 &       32 \\
\textit{RetIII} & 19.82 \pm 0.31 &    3 & -0.129 \pm 0.194 &  0.519 \pm 0.222 & -0.173 \pm 0.247 &            0.116 &  274.4 \pm 4.5 &       11 \\
SgrII & 19.23 \pm 0.07 &    7 &  0.015 \pm 0.072 &  -0.71 \pm 0.077 & -0.905 \pm 0.051 &           -0.047 & -175.3 \pm 0.4 &       33 \\
SclI & 19.67 \pm 0.13 & 1405 &  0.011 \pm 0.018 &  0.096 \pm 0.019 & -0.159 \pm 0.019 &           -0.020 &  111.5 \pm 0.0 &     8,34 \\
SegI & 16.81 \pm 0.19 &   22 & -0.023 \pm 0.057 & -2.074 \pm 0.052 & -3.411 \pm 0.043 &           -0.284 &  208.9 \pm 0.7 &       35 \\
SegII & 17.84 \pm 0.18 &   14 & -0.075 \pm 0.057 &  1.425 \pm 0.061 & -0.313 \pm 0.052 &            0.207 &  -41.0 \pm 0.6 &       36 \\
SxtI & 19.89 \pm 0.07 &  511 & -0.007 \pm 0.019 & -0.403 \pm 0.021 &  0.029 \pm 0.021 &           -0.090 &  224.9 \pm 0.1 &     8,37 \\
TriII & 17.27 \pm 0.12 &    7 & -0.008 \pm 0.076 &  0.602 \pm 0.081 &  0.085 \pm 0.093 &            0.229 & -381.3 \pm 0.9 &       38 \\
TucII &  18.82 \pm 0.3 &   15 & -0.011 \pm 0.037 &  0.935 \pm 0.031 & -1.243 \pm 0.036 &           -0.168 & -127.8 \pm 0.2 &       21 \\
TucIII & 16.99 \pm 0.17 &   44 &  0.014 \pm 0.021 & -0.111 \pm 0.023 & -1.629 \pm 0.023 &           -0.098 & -104.3 \pm 0.2 &    39,40 \\
TucIV & 18.41 \pm 0.18 &   39 &  0.038 \pm 0.039 &  0.618 \pm 0.036 & -1.696 \pm 0.038 &           -0.111 &   13.7 \pm 0.3 &       22 \\
TucV &  18.7 \pm 0.36 &    6 & -0.134 \pm 0.094 &  -0.27 \pm 0.083 &  -1.253 \pm 0.11 &           -0.097 &  -36.7 \pm 0.7 &       22 \\
UMaI & 19.94 \pm 0.13 &   10 &   0.111 \pm 0.07 & -0.387 \pm 0.058 & -0.641 \pm 0.068 &            0.004 &  -58.3 \pm 0.4 &        3 \\
UMaII &  17.7 \pm 0.12 &    5 & -0.237 \pm 0.141 &   1.701 \pm 0.12 & -1.845 \pm 0.128 &            0.061 & -115.6 \pm 0.9 &        3 \\
UMiI &  19.4 \pm 0.11 &  782 & -0.019 \pm 0.018 &  -0.114 \pm 0.02 &   0.069 \pm 0.02 &           -0.009 & -245.1 \pm 0.1 &    17,41 \\
WilI & 18.27 \pm 0.49 &    7 &  0.084 \pm 0.123 &  0.295 \pm 0.086 & -1.074 \pm 0.131 &           -0.098 &  -20.7 \pm 0.6 &        3 \\
\enddata
\tablecomments{Columns 1 lists the abbreviated dwarf galaxy name, with names in italics represents dwarf galaxies for which the number of representative stars is small; Column 2-6 gives the distance modulus, number of member stars, parallax, and proper motions in both dimension; Column 7 is the correlation coefficient between $\mu_{\alpha^*}$ and $\mu_{\delta}$; Column 8-9 is the heliocentric radial velocity and the references of member stars.}

\tablerefs{(1) \citet{2019MNRAS.488.2743T}; (2) \citet{2016MNRAS.463..712T}; (3) \citet{2007MNRAS.380..281M}; (4) \citet{2011ApJ...736..146K}; (5) \citet{2009ApJ...690..453K}; (6) \citet{2010MNRAS.402.1357U}; (7) \citet{2007ApJ...670..313S}; (8) \citet{2009AJ....137.3100W}; (9) \citet{2006ApJ...649..201M}; (10) \citet{Li2018}; (11) \citet{2019\aanda...623A.129F}; (12) \citet{2015ApJ...810...56K}; (13) \citet{2016MNRAS.460.3384V}; (14) \citet{2017ApJ...839...20C}; (15) \citet{2002MNRAS.330..792K}; (16) \citet{2015MNRAS.448.2717W}; (17) \citet{1995AJ....110.2131A}; (18) \citet{2016MNRAS.458L..59M}; (19) \citet{2017ApJ...838....8L}; (20) \citet{2006\aanda...459..423B}; (21) \citet{2016ApJ...819...53W}; (22) \citet{2020ApJ...892..137S}; (23) \citet{2009\aanda...506.1147A}; (24) \citet{2015ApJ...811...62K}; (25) \citet{2018MNRAS.479.5343K}; (26) \citet{2008ApJ...675..201M}; (27) \citet{2007ApJ...663..960S}; (28) \citet{2017ApJ...836..202S}; (29) \citet{2007AJ....134..566K}; (30) \citet{2021arXiv210100013J}; (31) \citet{2009ApJ...694L.144W}; (32) \citet{2017MNRAS.466.2006K}; (33) \citet{2015ApJ...808...95S}; (34) \citet{2019\aanda...626A..15H}; (35) \citet{2011ApJ...733...46S}; (36) \citet{2013ApJ...770...16K}; (37) \citet{2011MNRAS.411.1013B}; (38) \citet{2020MNRAS.491..356L}; (39) \citet{2017ApJ...838...83K}; (40) \citet{2017ApJ...838...11S}; (41) \citet{Li2018}; (42) \citet{pace2020}.}
\tablenotetext{a}{The reference of distance modulus is the same as \citet{fritz18Gaia} and updated with reference of \citet{Simon2019}, except the folowing galaxy: Antlia II \citep{torrealba19hidden}.}
%\tablenotetext{b}{The reference of heliocentric radial velocity is the same as \citet{fritz18Gaia}, except the folowing galaxies: Antlia II \citep{torrealba19hidden}; Columba I \citep{fritz19Gaia}; Grus II \citep{simon20Birds}; Horologium II \citep{fritz19Gaia}; Phoenix II \citep{fritz19Gaia}; Reticulum III \citep{fritz19Gaia}; Sagittarius II \citep{longeard20Pristine}; Tucana IV \citep{simon20Birds}; Tucana V \citep{simon20Birds}.}
\end{deluxetable*}

\subsection{Galactocentric coordinates and velocities}
\label{sample}

\par Heliocentric distance ($d$) can be derived from the distance modulus ($dm$):
\begin{equation}
d=10^{dm/5-2}\ \mathrm{kpc}.
\end{equation}
We then transform it and the Galactic coordinates ($l$, $b$) for dwarf galaxies into a Galactocentric Cartesian coordinate system ($x$, $y$, $z$):
\begin{eqnarray}
x & = & R_\odot - d \cos(b) \cos(l) \nonumber \\
y & = & -d \cos(b) \sin(l) \\
z & = & d \sin(b) + z_\odot, \nonumber
\end{eqnarray}
where $R_{\odot} = 8.122$ kpc is the distance from the Sun to Galactic center \citep{gravitycollaboration18Detection} and $z_{\odot} = 25$ pc is the solar offset from the Galactic midplane \citep{juric08Milky}. We use a right-handed Galactocentric frame, the sign of $x$ and $z$ are positive in the directions of the Sun, and the North Galactic Pole (NGP), respectively. The Galactic space-velocity components ($U$, $V$, $W$) can be calculated from proper motions, radial velocities and distances \citep{johnson87Calculating}. $U$, $V$ and $W$ point toward the Galactic center, Galactic rotation, and the North Galactic Pole (NGP), respectively. We assume a solar motion $(U_{\odot}, V_{\odot}, W_{\odot}) = (10., 11., 7.)$ km s$^{-1}$ \citep{bland-hawthorn16Galaxy} and a circular velocity at the location of the Sun $V_\mathrm{LSR} = 229$ km s$^{-1}$ \citep{eilers19Circular}. 

\par We then transform the Cartesian coordinate system ($x$, $y$, $z$, $v_x$, $v_y$, $v_z$) to a spherical coordinate system ($r_\mathrm{gc}$, $\theta$, $\phi$, $v_r$, $v_\theta$, $v_\phi$), where ($v_x$, $v_y$, $v_z$) are related to the ($U$, $V$, $W$) by, $v_x=-U$, $v_y=-V$, and $v_z=W$. We generate 2,000 realizations for each galaxy using a Monte Carlo (MC) method and calculate their 3D position $(r, \theta, \phi)$ and 3D velocity $(v_r, v_\theta, v_\phi)$ in the spherical coordinate system. The result is listed in Table~\ref{tab:sph}.

\par The Galactocentric tangential velocity is given by:
\begin{equation}
	v_\mathrm{tan} = \sqrt{v_\theta^2 + v_\phi^2}.
\end{equation}
The value of $v_\mathrm{tan,obs}$ is a biased estimator of the true tangential velocity $v_\mathrm{tan}$ \citep{vandermarel08M31}, especially when the relative error is close to or greater than 1 \citep{fritz18Gaia}. In order to correct this bias and derive accurate estimates, \citet{vandermarel08M31} use a Bayesian approach to infer tangential velocity. The posterior PDF is:
\begin{equation}
P(v_\mathrm{tan} | v_\mathrm{tan,obs}) \propto P(v_\mathrm{tan,obs} | v_\mathrm{tan}) P(v_\mathrm{tan}).
\end{equation}
We assume uniform priors on $v_\mathrm{tan}$. The likelihood distribution is obtained by sampling the observational data using a MC method. We characterize the posterior via rejection sampling and obtain 2,000 samples for each galaxy. 

\par
\begin{figure*}[htb!]
	\epsscale{0.9}
	\plotone{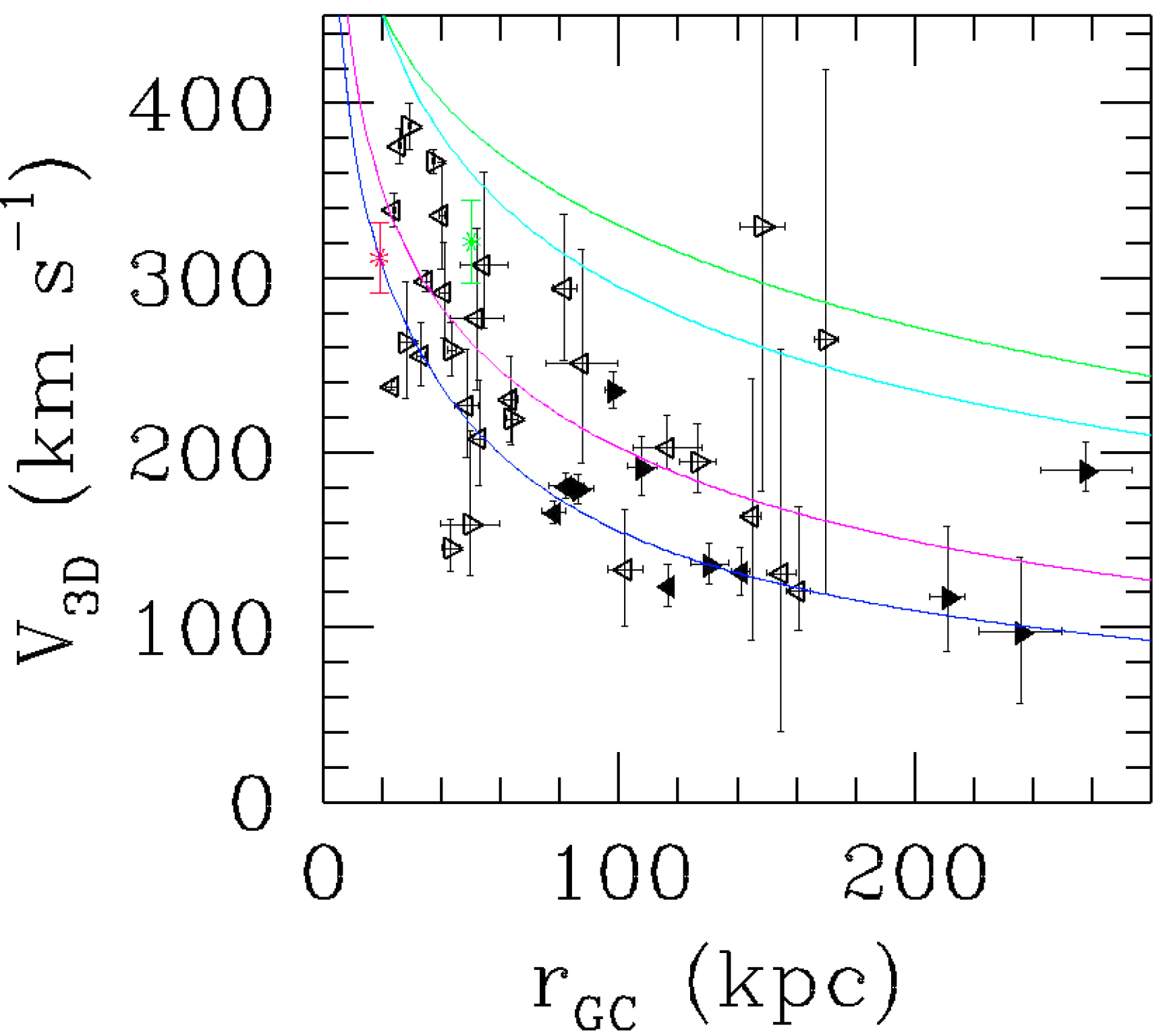}
	\caption{3D phase diagram for dwarf galaxies, which are represented by open and full triangles for visible luminosity smaller or larger than $10^{5}$ $L_{\odot}$, respectively. The  triangle orientation towards $r_\mathrm{gc}=0$ or in the opposite direction indicates whether dwarf galaxies are on an approaching ($v_r < 0$) or on a receding ($v_r > 0$) orbits, respectively. Green, cyan, magenta, and blue lines indicate the escape velocity of Einasto high-mass (PE$_\mathrm{HM}$), PNFW, intermediate mass (PE$_\mathrm{IM}$), and low mass (PE$_\mathrm{LM}$), respectively. Star-points indicates Sagittarius (red) and the LMC (green) positions.  \label{fig:r_v}}
\end{figure*}

 In the Bayesian approach, the effect of proper motions and other observed data is included in the likelihood function. For the 46 dwarf galaxies this does not affect the value of $v_\mathrm{tan,obs}$, but only the error bars for the few objects with very large uncertainties. We adopt the Bayesian calculations in Table~\ref{tab:sph} for error bars of  tangential velocity $v_\mathrm{tan}$ and of total velocity $v_\mathrm{3D}$. 
Figure~\ref{fig:r_v} shows the Galactocentric total velocity $v_{\mathrm{3D}}$ as a function of Galactocentric distance $r_{\mathrm{gc}}$ for all galaxies.

%\par \textbf{We introduce two sets of samples that are generated by MC method and Bayesian method in above. They are used in different parts. The Bayesian sample is used to calculate parameters that contain the term $(v_\theta^2 + v_\phi^2)$, such as tangential velocity $v_\mathrm{tan}$ and total velocity $v_\mathrm{3D}$. While the MC sample is applied in the estimation of other parameters, such as 3D position $(r, \theta, \phi)$ and 3D velocity $(v_r, v_\theta, v_\phi)$.} 

\startlongtable
\begin{deluxetable*}{lCCCCCCCC}
\label{tab:sph}
\tablecaption{Kinematic properties of dwarf galaxies in Galactocentric spherical coordinate system.}
\tablehead{
	\colhead{name} & \colhead{$r_\mathrm{GC}$} & \colhead{$\theta$} & \colhead{$\phi$} & 
	\colhead{$v_r$} & \colhead{$v_\theta$} & \colhead{$v_\phi$} &
	\colhead{$v_\mathrm{tan}$} & \colhead{$v_\mathrm{3D}$}\\
	\colhead{} & \colhead{(kpc)} & \colhead{(deg)} & \colhead{(deg)} &
	\colhead{(km s$^{-1}$)} & \colhead{(km s$^{-1}$)} & \colhead{(km s$^{-1}$)} &
	\colhead{(km s$^{-1}$)} & \colhead{(km s$^{-1}$)}
}
\colnumbers
\startdata
    AntII &   132.8^{+6.7}_{-6.3} &  78.8^{+0.0}_{-0.0} &  81.3^{+0.2}_{-0.2} &    66.4^{+0.8}_{-0.8} &    -61.2^{+12.7}_{-12.5} &   -102.6^{+12.8}_{-13.3} &   119.7^{+12.9}_{-13.1} &   136.9^{+11.7}_{-11.7} \\
    AquII &   105.4^{+3.2}_{-3.3} & 144.9^{+0.1}_{-0.0} & 241.4^{+0.2}_{-0.2} &  59.3^{+24.0}_{-23.5} &  182.4^{+241.9}_{-257.2} &  191.3^{+323.5}_{-326.1} & 300.3^{+288.5}_{-205.7} & 305.2^{+288.6}_{-201.0} \\
    BooI &    63.7^{+2.0}_{-1.9} &  13.5^{+0.2}_{-0.2} & 177.0^{+0.0}_{-0.0} &    87.4^{+1.9}_{-1.9} &    178.4^{+16.1}_{-16.4} &    -92.6^{+17.0}_{-17.1} &   201.7^{+16.7}_{-17.6} &   219.7^{+15.0}_{-15.2} \\
    BooII &    39.8^{+1.0}_{-1.0} &  10.3^{+0.3}_{-0.3} & 166.6^{+0.4}_{-0.4} &   -54.2^{+5.7}_{-5.8} &   -283.9^{+27.1}_{-29.1} &   -169.4^{+26.1}_{-26.4} &   331.7^{+31.4}_{-32.0} &   336.1^{+30.5}_{-30.4} \\
    CVenI &   210.7^{+6.3}_{-6.1} &   9.8^{+0.0}_{-0.0} & 266.9^{+0.4}_{-0.4} &    73.8^{+1.6}_{-1.6} &     85.7^{+46.6}_{-45.5} &    -18.2^{+40.5}_{-40.9} &    91.2^{+49.2}_{-47.5} &   117.6^{+40.7}_{-31.2} \\
    CVenII &   160.7^{+4.4}_{-4.0} &   8.9^{+0.0}_{-0.1} & 311.1^{+0.4}_{-0.4} &   -97.8^{+3.4}_{-3.6} &    -51.4^{+57.9}_{-56.1} &    -26.1^{+88.7}_{-82.6} &    71.3^{+64.5}_{-46.7} &   121.0^{+48.1}_{-22.3} \\
    CarI &   107.6^{+4.9}_{-4.9} & 111.9^{+0.0}_{-0.0} &  75.5^{+0.2}_{-0.2} &     0.8^{+0.8}_{-0.7} &   -191.6^{+17.5}_{-16.4} &      -8.6^{+9.8}_{-10.2} &   192.2^{+17.4}_{-15.9} &   192.2^{+17.4}_{-15.9} \\
    CarII &    37.1^{+0.6}_{-0.6} & 106.7^{+0.0}_{-0.0} &  76.8^{+0.2}_{-0.2} &   217.6^{+1.6}_{-1.7} &     -231.4^{+9.7}_{-9.5} &      183.5^{+7.4}_{-6.7} &     295.4^{+8.8}_{-8.6} &     366.8^{+6.8}_{-6.5} \\
    CarIII &    29.0^{+0.6}_{-0.6} & 106.1^{+0.0}_{-0.0} &  73.0^{+0.3}_{-0.3} &    46.6^{+2.6}_{-2.7} &   -379.9^{+13.8}_{-13.4} &       57.1^{+9.1}_{-8.4} &   384.0^{+13.5}_{-12.8} &   386.9^{+13.3}_{-12.7} \\
    ColI &  187.1^{+10.5}_{-9.3} & 118.1^{+0.0}_{-0.0} &  49.3^{+0.1}_{-0.1} &   -34.4^{+5.6}_{-5.9} &    84.0^{+97.8}_{-105.5} &  360.6^{+124.1}_{-122.4} & 372.9^{+122.7}_{-113.3} & 374.4^{+122.8}_{-113.4} \\
    CberI &    43.3^{+1.5}_{-1.6} &  14.9^{+0.4}_{-0.3} &  21.8^{+0.6}_{-0.6} &    32.5^{+2.6}_{-2.6} &   -230.6^{+12.3}_{-12.8} &    111.9^{+18.7}_{-17.0} &   256.9^{+16.0}_{-15.2} &   258.9^{+15.7}_{-14.7} \\
    CraI &   145.9^{+1.9}_{-2.1} &  40.5^{+0.0}_{-0.0} &  88.8^{+0.1}_{-0.1} &    -5.8^{+5.4}_{-5.2} &   -134.5^{+69.7}_{-66.2} &     88.0^{+92.0}_{-95.5} &   163.8^{+79.1}_{-70.4} &   163.9^{+79.1}_{-70.5} \\
    CraII &   116.5^{+1.1}_{-1.1} &  47.5^{+0.0}_{-0.0} &  97.6^{+0.0}_{-0.0} &   -81.7^{+1.3}_{-1.1} &    -85.7^{+15.0}_{-14.6} &     37.5^{+15.7}_{-18.4} &    93.6^{+15.8}_{-16.5} &   124.2^{+12.6}_{-12.1} \\
    DraI &    81.8^{+6.1}_{-5.7} &  55.3^{+0.0}_{-0.0} & 273.3^{+0.5}_{-0.5} &   -97.4^{+0.8}_{-0.8} &      136.9^{+7.9}_{-7.4} &      -66.2^{+9.5}_{-9.7} &     153.1^{+8.3}_{-8.6} &     181.5^{+7.2}_{-7.5} \\
    DraII &    23.8^{+0.4}_{-0.4} &  52.0^{+0.1}_{-0.1} & 303.7^{+0.4}_{-0.4} &  -153.9^{+4.5}_{-4.7} &    299.5^{+11.7}_{-11.7} &    -43.5^{+11.7}_{-12.2} &   302.0^{+11.7}_{-11.3} &    338.9^{+10.2}_{-9.6} \\
    EriII & 367.9^{+17.4}_{-16.2} & 141.3^{+0.0}_{-0.0} &  67.9^{+0.1}_{-0.1} &   -76.7^{+5.7}_{-5.4} &  -21.2^{+225.2}_{-249.0} &  193.2^{+280.3}_{-264.6} & 267.8^{+269.7}_{-184.1} & 278.8^{+265.2}_{-167.7} \\
    FnxI &   141.0^{+3.1}_{-3.1} & 153.9^{+0.0}_{-0.0} &  50.9^{+0.1}_{-0.1} &   -36.8^{+0.7}_{-0.7} &   -100.0^{+14.6}_{-12.9} &     78.5^{+12.8}_{-13.2} &   126.9^{+14.7}_{-14.2} &   132.2^{+14.1}_{-13.7} \\
    GruI & 116.0^{+11.9}_{-10.7} & 151.4^{+0.3}_{-0.3} & 155.6^{+0.3}_{-0.3} &  -187.9^{+2.1}_{-2.1} &     21.2^{+37.0}_{-42.1} &     73.0^{+41.0}_{-41.6} &    76.9^{+41.6}_{-40.2} &   203.0^{+19.1}_{-11.7} \\
    GruII &    48.6^{+5.4}_{-4.6} & 149.4^{+0.8}_{-0.7} & 168.2^{+0.3}_{-0.4} &  -125.9^{+1.2}_{-1.2} &       17.5^{+9.4}_{-9.3} &   -162.9^{+36.8}_{-40.0} &   166.1^{+40.7}_{-35.1} &   208.3^{+33.8}_{-26.5} \\
    HerI &   126.5^{+6.4}_{-6.2} &  51.2^{+0.1}_{-0.1} & 211.0^{+0.1}_{-0.1} &   143.8^{+1.2}_{-1.3} &    130.9^{+30.4}_{-31.6} &     -4.2^{+38.7}_{-38.9} &   131.7^{+31.1}_{-28.4} &   195.0^{+21.8}_{-17.6} \\
    HorI &  86.6^{+13.6}_{-10.9} & 144.5^{+0.1}_{-0.1} &  82.1^{+1.3}_{-1.3} &   -27.5^{+1.8}_{-1.9} &   -238.2^{+57.2}_{-69.2} &     24.2^{+21.4}_{-18.2} &   249.9^{+65.7}_{-57.7} &   251.3^{+65.5}_{-57.0} \\
    HorII &    79.2^{+7.6}_{-7.0} & 143.1^{+0.1}_{-0.1} &  72.7^{+0.8}_{-0.9} &   2.1^{+19.0}_{-19.1} & -236.6^{+134.1}_{-148.5} &  163.9^{+157.1}_{-157.9} & 310.7^{+167.9}_{-153.5} & 310.8^{+168.1}_{-152.9} \\
    HyaII &   147.8^{+7.8}_{-7.1} &  58.9^{+0.0}_{-0.0} & 112.3^{+0.2}_{-0.2} &   138.7^{+9.6}_{-9.6} &   11.6^{+164.4}_{-154.0} & -286.1^{+195.3}_{-194.2} & 299.2^{+190.6}_{-178.0} & 329.7^{+181.1}_{-151.7} \\
    HyiI &    25.7^{+0.5}_{-0.5} & 129.9^{+0.0}_{-0.0} &  96.0^{+0.4}_{-0.4} &   -47.6^{+1.1}_{-1.0} &     -327.9^{+8.6}_{-8.8} &     -176.1^{+5.8}_{-7.1} &     372.8^{+9.8}_{-9.9} &     375.8^{+9.7}_{-9.7} \\
    LeoI & 257.9^{+17.0}_{-14.6} &  41.8^{+0.1}_{-0.1} &  44.0^{+0.1}_{-0.1} &   172.4^{+1.0}_{-1.0} &    -11.8^{+32.9}_{-32.2} &    -79.2^{+36.2}_{-31.8} &    80.5^{+31.8}_{-32.7} &   190.2^{+15.9}_{-11.9} \\
    LeoII & 235.6^{+13.9}_{-14.1} &  24.2^{+0.1}_{-0.1} &  37.1^{+0.2}_{-0.2} &    25.7^{+1.4}_{-1.5} &     23.7^{+43.1}_{-39.8} &    -91.1^{+44.0}_{-42.2} &    94.5^{+43.2}_{-43.1} &    97.9^{+42.5}_{-41.2} \\
    LeoIV &   154.4^{+5.1}_{-4.8} &  33.8^{+0.0}_{-0.0} &  80.0^{+0.2}_{-0.2} &    -0.6^{+6.8}_{-6.5} &  -55.0^{+102.9}_{-102.5} &   94.8^{+115.7}_{-131.0} &  131.2^{+128.6}_{-90.4} &  131.2^{+128.6}_{-90.3} \\
    LeoV &   169.8^{+3.9}_{-3.8} &  31.9^{+0.0}_{-0.0} &  76.7^{+0.1}_{-0.1} &    44.1^{+7.6}_{-8.0} &  -27.9^{+131.7}_{-127.3} &  233.2^{+158.7}_{-154.5} & 261.5^{+156.7}_{-154.0} & 265.1^{+154.6}_{-145.9} \\
    PhxII &    81.4^{+4.3}_{-3.7} & 154.3^{+0.2}_{-0.2} & 136.9^{+0.4}_{-0.4} &   -37.6^{+2.5}_{-2.5} &   -202.2^{+30.3}_{-29.3} &   -210.0^{+36.0}_{-36.8} &   291.7^{+42.8}_{-41.0} &   294.2^{+42.5}_{-40.7} \\
    PisII & 181.9^{+16.5}_{-15.0} & 137.4^{+0.0}_{-0.0} & 262.9^{+0.3}_{-0.3} & -60.0^{+10.6}_{-10.6} &  606.5^{+226.6}_{-210.0} &  188.9^{+241.0}_{-224.9} & 641.0^{+257.6}_{-234.3} & 644.0^{+256.2}_{-231.8} \\
    RetII &    32.9^{+1.4}_{-1.4} & 137.0^{+0.2}_{-0.2} &  65.1^{+0.9}_{-0.9} &   -98.7^{+1.1}_{-1.2} &   -224.2^{+18.4}_{-19.7} &       68.2^{+7.1}_{-6.8} &   235.9^{+20.4}_{-19.0} &   255.7^{+18.9}_{-17.4} \\
    RetIII &  91.6^{+13.2}_{-11.7} & 135.7^{+0.0}_{-0.0} &  86.6^{+0.9}_{-1.1} & 102.7^{+10.7}_{-10.9} &   -69.6^{+99.4}_{-103.9} &    2.4^{+113.7}_{-105.7} &   99.0^{+104.4}_{-69.4} &   143.3^{+84.2}_{-35.5} \\
    SgrII &    63.1^{+2.3}_{-2.3} & 115.9^{+0.1}_{-0.1} & 200.6^{+0.1}_{-0.1} &  -113.4^{+1.7}_{-1.6} &   -143.5^{+24.0}_{-25.6} &   -138.1^{+22.1}_{-21.7} &   200.8^{+27.2}_{-27.2} &   230.7^{+24.7}_{-24.1} \\
    SclI &    86.0^{+5.2}_{-4.7} & 172.7^{+0.1}_{-0.1} &  62.7^{+2.2}_{-2.2} &    75.9^{+0.7}_{-0.7} &      154.2^{+8.2}_{-8.4} &    -54.3^{+11.4}_{-10.9} &     162.7^{+9.1}_{-9.2} &     179.6^{+8.3}_{-8.2} \\
    SegI &    27.9^{+2.0}_{-1.8} &  50.4^{+0.8}_{-0.8} &  26.3^{+0.8}_{-0.8} &   130.4^{+1.6}_{-1.7} &    182.9^{+30.5}_{-28.6} &    133.2^{+24.7}_{-24.0} &   229.0^{+38.6}_{-37.6} &   263.5^{+34.4}_{-31.8} \\
    SegII &    42.8^{+3.0}_{-2.9} & 122.1^{+0.4}_{-0.4} & 336.0^{+0.4}_{-0.4} &    60.4^{+2.1}_{-2.1} &   -128.1^{+11.8}_{-12.0} &     23.8^{+24.5}_{-21.1} &   132.2^{+17.8}_{-13.1} &   145.5^{+17.0}_{-12.7} \\
    SxtI &    98.1^{+3.0}_{-2.9} &  49.3^{+0.1}_{-0.0} &  57.9^{+0.2}_{-0.2} &    83.3^{+0.8}_{-0.8} &     -8.8^{+10.0}_{-10.4} &    -219.7^{+9.9}_{-11.0} &   220.7^{+11.3}_{-10.4} &   235.9^{+10.8}_{-10.0} \\
    TriII &    34.7^{+1.6}_{-1.5} & 109.3^{+0.2}_{-0.2} & 329.9^{+0.4}_{-0.4} &  -259.4^{+2.2}_{-2.3} &   -117.1^{+12.2}_{-12.8} &    -89.0^{+10.3}_{-10.0} &   147.4^{+12.1}_{-12.0} &     298.5^{+5.9}_{-5.9} \\
    TucII &    54.0^{+8.8}_{-7.4} & 147.9^{+0.8}_{-0.7} & 139.4^{+1.4}_{-1.6} &  -182.9^{+1.1}_{-1.0} &     31.9^{+13.8}_{-16.1} &   -235.6^{+51.3}_{-61.0} &   247.6^{+63.7}_{-47.0} &   307.5^{+53.2}_{-35.6} \\
    TucIII &    23.0^{+2.0}_{-1.7} & 154.4^{+0.2}_{-0.3} & 100.2^{+3.6}_{-3.7} &  -229.2^{+1.0}_{-1.0} &     25.7^{+17.8}_{-21.1} &      57.2^{+9.9}_{-11.5} &      64.3^{+7.7}_{-3.8} &     238.2^{+2.8}_{-1.8} \\
    TucIV &    45.6^{+4.1}_{-4.0} & 150.2^{+0.4}_{-0.3} & 118.2^{+1.4}_{-1.7} &   -93.5^{+1.4}_{-1.5} &   -194.3^{+31.4}_{-32.5} &    -58.4^{+19.4}_{-21.4} &   207.2^{+35.3}_{-33.7} &   227.3^{+32.4}_{-30.0} \\
    TucV &    52.2^{+9.0}_{-7.9} & 146.5^{+0.7}_{-0.6} & 125.1^{+1.8}_{-2.3} &  -166.8^{+3.3}_{-3.3} &   -175.9^{+60.4}_{-67.5} &    105.6^{+22.6}_{-25.7} &   221.0^{+62.5}_{-53.3} &   276.9^{+51.7}_{-40.5} \\
    UMaI &   102.0^{+5.9}_{-5.5} &  39.1^{+0.2}_{-0.2} & 342.0^{+0.1}_{-0.1} &    -2.2^{+1.8}_{-1.9} &     61.9^{+28.3}_{-26.7} &    114.7^{+37.9}_{-36.0} &   133.3^{+34.9}_{-32.1} &   133.3^{+34.9}_{-32.1} \\
    UMaII &    40.8^{+2.0}_{-1.9} &  58.9^{+0.3}_{-0.3} & 338.6^{+0.3}_{-0.3} &   -59.7^{+3.1}_{-3.2} &   -280.5^{+26.5}_{-28.2} &     30.7^{+26.3}_{-25.4} &   285.9^{+29.3}_{-26.7} &   292.1^{+28.9}_{-26.3} \\
    UMiI &    77.8^{+4.2}_{-3.9} &  46.5^{+0.1}_{-0.1} & 293.0^{+0.4}_{-0.4} &   -76.9^{+0.8}_{-0.8} &      143.9^{+7.3}_{-6.7} &      -31.9^{+8.1}_{-8.0} &     147.6^{+7.1}_{-7.4} &     166.4^{+6.4}_{-6.5} \\
    WilI &   49.3^{+11.6}_{-8.8} &  40.7^{+1.7}_{-1.4} & 343.9^{+0.9}_{-0.9} &     8.1^{+2.9}_{-2.8} &   -124.7^{+36.9}_{-49.7} &    -67.8^{+49.3}_{-39.3} &   159.2^{+53.7}_{-29.3} &   159.4^{+53.7}_{-29.2} \\
\enddata
\tablecomments{Column 1 lists the abbreviated dwarf galaxy name; Column 2-4 is the Galactocentric distance, angle with respect to the North Galactic Pole and azimuthal angle; Column 5-7 gives the velocities in three dimensions; Column 8 provides the Galactocentric tangential velocity and Column 9 lists the total velocity in the Galactic rest frame.}
\end{deluxetable*}

\section{Integrated orbital parameters}

\subsection{Four flavors of the Milky Way potential}

\par Based on APOGEE, WISE, 2MASS and Gaia data, \citet{eilers19Circular} derived the rotation curve of the Milky Way at Galactocentric distances between $5 \leqslant R \leqslant 25$ kpc. Combined with the baryonic components of \citet[their Model I]{pouliasis17Milky}, they estimate the parameters of a NFW \citep{navarro97Universal} dark matter halo by fitting the rotation curve. Model I of \citet{pouliasis17Milky} includes three baryonic components: a Plummer \citep{binney87Galactic} bulge, and two Miyamoto-Nagai \citep{miyamoto75Threedimensional} for the thin and thick disk. We choose this model ($PNFW$, MW total mass: $8.1\times10^{11}$ $M_{\odot}$) that is similar to that of \citet[see a comparison of the two mass profiles in Figure 1 of \citet{Hammer2020}]{bovy15galpy}.

\par \citet{jiao21Which} determined the range of Milky Way mass that can reproduce the rotation curve of \citet{eilers19Circular}. They found that an Einasto profile \citep{retana-montenegro12Analytical} for the dark matter is more appropriate to define this range, which includes MW total masses from $2.8\times10^{11}$ $M_{\odot}$ to $15\times10^{11}$ $M_{\odot}$. We adopt these two models (PE$_\mathrm{LM}$ and PE$_\mathrm{HM}$, also associated to Model I of \citealt{pouliasis17Milky}). PE$_\mathrm{LM}$ provides the best fit of the MW rotation curve, while PE$_\mathrm{HM}$ is among the most massive MW mass available for such a fitting. Both $PNFW$ and PE$_\mathrm{HM}$ are comparable to the 'low' and 'high' MW mass models used by \citet{fritz18Gaia}, respectively.   For a better sampling of possible MW masses that fit the rotation curve, we also consider an intermediate mass model (PE$_\mathrm{IM}$) from \citet{wang21}, obtained after combining constraints from the MW rotation curve and globular cluster orbital motions. Properties of the dark matter halo of the four models are listed in Table \ref{tab:halo}, all of them being associated to Model I of \citet{pouliasis17Milky} for baryons, with a total baryonic mass of $0.89\times10^{11}$ $M_{\odot}$. 

\begin{deluxetable}{LCCCC}
\label{tab:halo}
\tablecaption{Properties of the four Milky Way dark-matter mass models.}
\tablehead{
	\colhead{Parameters} & \colhead{PE$_\mathrm{HM}$} & \colhead{PNFW} & \colhead{PE$_\mathrm{IM}$} & \colhead{PE$_\mathrm{LM}$}
}
\decimals
\startdata
M_\mathrm{DM}(10^{11}M_\odot) & 14.1 & 7.2 &  4.2 & 1.9\\
M_\mathrm{tot}(10^{11}M_\odot) & 15 & 8.1 &  5.1 & 2.8\\
r_\mathrm{200}(\mathrm{kpc}) & 236 & 189 & 164 & 135\\
\enddata
%\tablecomments{All dark-matter models are associated to Model I of \citet{pouliasis17Milky} for baryons, with a total baryonic mass of $0.89\times10^{11}M_{\odot}$}.}
\end{deluxetable}

\subsection{Orbit integration}
\label{sec:orbit_int}
We have used \texttt{galpy} \citep{bovy15galpy} to investigate the orbital properties of the dwarf galaxies by adopting a Milky Way potential model. We use the sample derived from the Bayesian method, because the orbital properties are related to total energy, which is also affected by the bias in $v_\mathrm{tan}$. For example, for the PNFW model, we have integrated the orbit from $-10$ to 10 Gyr for each dwarf galaxy.
Thus, we can derive the pericenter, $r_\mathrm{peri}$, the closest approach of an orbit to the the Milky Way center (i.e., the Galaxy center, GC) for each orbit.  
However, due to long orbital periods or hyperbolic orbits, it is not always possible to derive the apocenter, $r_\mathrm{apo}$, the farthest extent of an orbit from the GC.
In case the chosen integration time is too short to fully cover one orbit, we at least may ensure that each dwarf galaxy can reach 300 kpc from the Milky Way center so that the Milky Way can be considered as a point source. 
\par By assuming conservation of energy and angular momentum, $r_\mathrm{apo}$ can be derived from the following equation\footnote{$r_\mathrm{apo}$ is the larger one of the two roots.}: 
\begin{eqnarray}
\label{equ:apo}
E & = & \Phi(\vec{r}_\mathrm{apo}) + \frac{v_\mathrm{apo}^2}{2} \nonumber \\
& = & -\frac{GM}{r_\mathrm{apo}} + \frac{L_\mathrm{apo}^2}{2 r_\mathrm{apo}^2} \nonumber \\
& = & -\frac{GM}{r_\mathrm{apo}} + \frac{L_\mathrm{10Gyr}^2}{2 r_\mathrm{apo}^2},
\end{eqnarray}
where $E$ is the total energy, $M$ is the total mass of the MW, and $L_\mathrm{10Gyr}$ is the dwarf galaxy's angular momentum at 10 Gyr.

\par  We define the eccentricity of an elliptical orbit by:
\begin{equation}
e = \frac{r_\mathrm{apo}-r_\mathrm{peri}}{r_\mathrm{apo}+r_\mathrm{peri}}.
\end{equation}

Hyperbolic orbits in extended mass profile require another definition of the eccentricity. This is why hyperbolic orbits are characterized in Tables~\ref{tab:dyn1} and ~\ref{tab:dyn2} by e$>$1 for purely hyperbolic orbits (hence without values for apocenter) and e $\ge$ 1 for orbits being hyperbolic on average but having few MC solutions consistent with elliptical orbits (and then with a quoted value for apocenter).\\

We notice that pericenters are well determined and do not depend much on the adopted MW mass profile (compare Tables~\ref{tab:dyn1} and ~\ref{tab:dyn2}), while the opposite is true for eccentricity and apocenter values. Such a property will be further investigated in a future paper \citep{hammer21}. We also notice that the more precise Gaia-EDR3 values have led to only two dwarf galaxies (Triangulum II and Tucana III) having pericenters lower than 20 kpc. This contrasts with \citet{fritz18Gaia} who were finding between seven to eight such low-pericenter dwarf galaxies, and this may affect some analyses based on the tidal disruption scenario caused by, e.g., the MW disk.

\startlongtable
\begin{deluxetable*}{lCCCCCCCCCC}
\label{tab:dyn1}
\tabletypesize{\footnotesize}%\small (11pt), \footnotesize (10pt), or \scriptsize (8pt)
\tablecaption{Orbital properties of dwarf galaxies for Model PE$_\mathrm{HM}$ and Model PNFW.}
\tablehead{
	\colhead{} & \multicolumn{5}{c}{Model PE$_\mathrm{HM}$} & \multicolumn{5}{c}{Model $PNFW$}\\
	\colhead{name} & \colhead{$r_\mathrm{peri}$} & \colhead{$r_\mathrm{apo}$\tablenotemark{a,c}} & \colhead{$e$} & \colhead{$P_\mathrm{unb}$} & \colhead{$P$\tablenotemark{b,c}} &
	\colhead{$r_\mathrm{peri}$} & \colhead{$r_\mathrm{apo}$\tablenotemark{a,c}} & \colhead{$e$} & \colhead{$P_\mathrm{unb}$} & \colhead{$P$\tablenotemark{b,c}}\\
	\colhead{} & \colhead{(kpc)} & \colhead{(kpc)} & \colhead{} & \colhead{} & \colhead{} &
	\colhead{(kpc)} & \colhead{(kpc)} & \colhead{} & \colhead{} & \colhead{} 
}
\colnumbers
\startdata
AntII &   56^{+12}_{-10} &      149^{+8}_{-7} & 0.45^{+0.06}_{-0.06} &   0.0\% & 0.63^{+0.03}_{-0.04} &   69^{+15}_{-13} &     168^{+19}_{-12} & 0.41^{+0.06}_{-0.04} &   0.0\% & 0.48^{+0.06}_{-0.08} \\
AquII &   101^{+4}_{-71} &    116^{+190}_{-8} & 0.95^{+6.15}_{-0.61} &  47.3\% & 0.07^{+0.72}_{-0.02} &   102^{+4}_{-66} &     116^{+163}_{-7} &          \geqslant 1 &  58.9\% & 0.07^{+0.65}_{-0.02} \\
BooI &     49^{+5}_{-6} &     97^{+15}_{-10} & 0.33^{+0.02}_{-0.01} &   0.0\% & 0.28^{+0.08}_{-0.07} &     52^{+4}_{-5} &     135^{+39}_{-25} & 0.44^{+0.07}_{-0.04} &   0.0\% & 0.15^{+0.08}_{-0.06} \\
BooII &     39^{+1}_{-1} &   181^{+104}_{-56} & 0.65^{+0.11}_{-0.11} &  0.45\% & 0.02^{+0.02}_{-0.01} &     39^{+1}_{-1} &   387^{+738}_{-198} &  0.92^{+0.84}_{-0.2} & 36.95\% &   0.02^{+0.0}_{-0.0} \\
CVenI &   62^{+59}_{-39} &     242^{+26}_{-8} &   0.6^{+0.23}_{-0.2} &  0.25\% &  0.6^{+0.04}_{-0.15} &   81^{+73}_{-55} &    290^{+220}_{-31} & 0.66^{+0.25}_{-0.12} &   6.5\% & 0.48^{+0.03}_{-0.06} \\
CVenII &   31^{+48}_{-23} &     193^{+27}_{-7} & 0.73^{+0.19}_{-0.23} &   0.3\% & 0.55^{+0.03}_{-0.13} &   37^{+59}_{-28} &    222^{+103}_{-16} & 0.77^{+0.18}_{-0.15} &  4.55\% &  0.4^{+0.06}_{-0.08} \\
CarI &   108^{+5}_{-17} &    117^{+36}_{-13} & 0.08^{+0.09}_{-0.05} &   0.0\% & 0.02^{+0.96}_{-0.02} &    108^{+5}_{-5} &    201^{+154}_{-61} &  0.3^{+0.22}_{-0.15} &   0.7\% &   0.0^{+0.01}_{-0.0} \\
CarII &     27^{+1}_{-1} &    252^{+34}_{-26} &  0.8^{+0.02}_{-0.01} &   0.0\% & 0.04^{+0.01}_{-0.01} &     28^{+1}_{-1} &                   - &                   >1 & 94.05\% &                    - \\
CarIII &     29^{+1}_{-0} &    227^{+64}_{-42} & 0.78^{+0.04}_{-0.04} &   0.0\% &   0.01^{+0.0}_{-0.0} &     29^{+1}_{-0} & 1196^{+2881}_{-651} &          \geqslant 1 & 65.65\% &   0.01^{+0.0}_{-0.0} \\
ColI &    186^{+8}_{-8} & 404^{+1734}_{-208} &          \geqslant 1 &  83.6\% & 0.08^{+0.03}_{-0.01} &    186^{+8}_{-8} &                   - &                   >1 &  94.1\% &                    - \\
CberI &     42^{+1}_{-1} &     83^{+18}_{-13} & 0.32^{+0.07}_{-0.06} &   0.0\% & 0.06^{+0.04}_{-0.02} &     43^{+1}_{-1} &     119^{+41}_{-26} &  0.47^{+0.1}_{-0.09} &   0.0\% & 0.03^{+0.02}_{-0.01} \\
CraI &  119^{+27}_{-75} &    147^{+144}_{-2} & 0.35^{+0.41}_{-0.24} &   5.9\% & 0.91^{+0.08}_{-0.89} &   144^{+2}_{-89} &     149^{+338}_{-3} & 0.53^{+1.04}_{-0.38} & 24.35\% & 0.04^{+0.94}_{-0.02} \\
CraII &     33^{+8}_{-8} &      133^{+2}_{-2} &  0.6^{+0.08}_{-0.07} &   0.0\% & 0.61^{+0.02}_{-0.02} &   38^{+11}_{-10} &       145^{+7}_{-4} & 0.58^{+0.08}_{-0.07} &   0.0\% & 0.51^{+0.03}_{-0.04} \\
DraI &     45^{+6}_{-5} &     108^{+10}_{-7} & 0.41^{+0.02}_{-0.02} &   0.0\% & 0.45^{+0.03}_{-0.03} &     50^{+6}_{-5} &     133^{+19}_{-13} & 0.46^{+0.01}_{-0.01} &   0.0\% &  0.3^{+0.04}_{-0.05} \\
DraII &     20^{+1}_{-1} &     98^{+12}_{-10} & 0.66^{+0.03}_{-0.02} &   0.0\% & 0.06^{+0.01}_{-0.01} &     20^{+1}_{-0} &     146^{+32}_{-22} & 0.76^{+0.04}_{-0.03} &   0.0\% & 0.03^{+0.01}_{-0.01} \\
EriII & 345^{+23}_{-237} &   485^{+822}_{-50} &          \geqslant 1 & 67.85\% &    1.0^{+0.0}_{-0.0} & 348^{+21}_{-187} &   617^{+745}_{-102} &          \geqslant 1 &  78.3\% &    1.0^{+0.0}_{-0.0} \\
FnxI &   69^{+16}_{-13} &      147^{+4}_{-3} & 0.36^{+0.08}_{-0.08} &   0.0\% & 0.77^{+0.03}_{-0.04} &   92^{+21}_{-19} &      157^{+16}_{-7} & 0.26^{+0.09}_{-0.05} &   0.0\% & 0.61^{+0.09}_{-0.16} \\
GruI &   22^{+15}_{-13} &    224^{+28}_{-20} &  0.82^{+0.1}_{-0.08} &   0.0\% & 0.25^{+0.02}_{-0.04} &   24^{+17}_{-14} &    430^{+281}_{-91} & 0.91^{+0.05}_{-0.03} &  3.35\% & 0.23^{+0.04}_{-0.05} \\
GruII &    27^{+11}_{-9} &     74^{+28}_{-13} & 0.48^{+0.07}_{-0.03} &  0.05\% & 0.35^{+0.08}_{-0.13} &   28^{+12}_{-10} &      87^{+57}_{-21} & 0.54^{+0.07}_{-0.02} &  0.75\% & 0.26^{+0.11}_{-0.14} \\
HerI &   52^{+17}_{-15} &    212^{+37}_{-20} & 0.61^{+0.07}_{-0.04} &   0.0\% & 0.31^{+0.05}_{-0.07} &   59^{+18}_{-17} &   390^{+378}_{-107} & 0.76^{+0.11}_{-0.03} &  6.95\% & 0.25^{+0.03}_{-0.05} \\
HorI &   87^{+13}_{-17} &   156^{+258}_{-72} &  0.37^{+0.5}_{-0.22} &  12.2\% & 0.05^{+0.37}_{-0.02} &   87^{+13}_{-13} &   194^{+517}_{-104} & 0.79^{+1.61}_{-0.53} &  41.3\% &   0.03^{+0.1}_{-0.0} \\
HorII &    78^{+7}_{-26} &   127^{+305}_{-51} &  0.74^{+2.99}_{-0.5} & 39.45\% & 0.01^{+0.88}_{-0.01} &    79^{+6}_{-13} &    107^{+384}_{-31} &          \geqslant 1 &  57.3\% & 0.01^{+0.74}_{-0.01} \\
HyaII &  128^{+14}_{-71} &   254^{+429}_{-58} &          \geqslant 1 & 59.35\% & 0.24^{+0.12}_{-0.07} &  130^{+12}_{-63} &    291^{+627}_{-63} &          \geqslant 1 &  75.4\% & 0.24^{+0.08}_{-0.07} \\
HyiI &     25^{+0}_{-0} &    167^{+29}_{-23} & 0.74^{+0.03}_{-0.03} &   0.0\% &   0.01^{+0.0}_{-0.0} &     25^{+0}_{-0} &   518^{+678}_{-201} & 0.91^{+0.06}_{-0.06} &   9.6\% &   0.01^{+0.0}_{-0.0} \\
LeoI &   56^{+30}_{-27} &  888^{+578}_{-185} &  0.9^{+0.04}_{-0.02} &  3.75\% & 0.81^{+0.05}_{-0.05} &   67^{+35}_{-33} &                   - &                   >1 & 100.0\% &                    - \\
LeoII &   79^{+72}_{-47} &    242^{+14}_{-11} & 0.51^{+0.26}_{-0.25} &  0.05\% & 0.85^{+0.03}_{-0.11} & 114^{+104}_{-75} &    254^{+146}_{-16} & 0.47^{+0.33}_{-0.22} &   4.0\% &  0.72^{+0.1}_{-0.34} \\
LeoIV &   84^{+72}_{-69} &    157^{+116}_{-5} & 0.59^{+0.37}_{-0.41} & 11.75\% & 0.97^{+0.02}_{-0.95} & 123^{+33}_{-107} &     156^{+154}_{-5} & 0.74^{+1.32}_{-0.49} &  27.2\% & 0.94^{+0.05}_{-0.92} \\
LeoV &  164^{+6}_{-101} &   197^{+460}_{-18} & 0.97^{+3.16}_{-0.64} & 48.15\% & 0.13^{+0.58}_{-0.07} &   165^{+5}_{-81} &    203^{+462}_{-18} &          \geqslant 1 &  65.1\% & 0.13^{+0.44}_{-0.07} \\
PhxII &     80^{+4}_{-4} &  276^{+379}_{-117} & 0.58^{+0.33}_{-0.22} &  10.0\% &  0.03^{+0.03}_{-0.0} &     80^{+4}_{-4} &  491^{+1283}_{-279} &          \geqslant 1 &  66.9\% &   0.04^{+0.0}_{-0.0} \\
PisII &  181^{+12}_{-12} &                  - &                   >1 & 95.85\% &                    - &  181^{+12}_{-12} &                   - &                   >1 &  97.9\% &                    - \\
RetII &     28^{+2}_{-3} &     63^{+15}_{-10} & 0.39^{+0.05}_{-0.03} &   0.0\% & 0.18^{+0.07}_{-0.05} &     28^{+2}_{-3} &      77^{+28}_{-17} & 0.47^{+0.08}_{-0.06} &   0.0\% & 0.13^{+0.06}_{-0.05} \\
RetIII &   27^{+46}_{-22} &    114^{+50}_{-15} & 0.67^{+0.25}_{-0.22} &  3.05\% & 0.53^{+0.05}_{-0.32} &   30^{+47}_{-25} &     124^{+74}_{-18} & 0.74^{+0.23}_{-0.21} &  11.2\% & 0.43^{+0.07}_{-0.28} \\
SgrII &     45^{+6}_{-7} &    109^{+28}_{-17} & 0.42^{+0.04}_{-0.01} &   0.0\% &  0.25^{+0.1}_{-0.09} &     47^{+5}_{-7} &     158^{+91}_{-41} & 0.54^{+0.12}_{-0.05} &   0.4\% &  0.13^{+0.1}_{-0.06} \\
SclI &     54^{+3}_{-3} &      107^{+4}_{-4} & 0.32^{+0.03}_{-0.02} &   0.0\% & 0.48^{+0.04}_{-0.04} &     62^{+3}_{-3} &       133^{+9}_{-7} & 0.36^{+0.01}_{-0.01} &   0.0\% &  0.3^{+0.05}_{-0.05} \\
SegI &     21^{+4}_{-5} &     58^{+26}_{-15} & 0.47^{+0.07}_{-0.02} &  0.05\% & 0.19^{+0.11}_{-0.08} &     21^{+4}_{-5} &      68^{+46}_{-21} & 0.53^{+0.12}_{-0.04} &   0.5\% & 0.14^{+0.11}_{-0.08} \\
SegII &     19^{+6}_{-4} &       47^{+4}_{-3} & 0.42^{+0.06}_{-0.08} &   0.0\% & 0.69^{+0.03}_{-0.06} &     20^{+6}_{-4} &        48^{+5}_{-4} &  0.4^{+0.06}_{-0.08} &   0.0\% & 0.65^{+0.04}_{-0.08} \\
SxtI &     84^{+4}_{-4} &    188^{+28}_{-21} & 0.39^{+0.04}_{-0.03} &   0.0\% & 0.17^{+0.04}_{-0.04} &     87^{+3}_{-3} &   592^{+713}_{-217} & 0.75^{+0.14}_{-0.11} &   4.3\% & 0.12^{+0.01}_{-0.01} \\
TriII &     12^{+1}_{-1} &      110^{+7}_{-6} &  0.8^{+0.01}_{-0.01} &   0.0\% & 0.13^{+0.01}_{-0.01} &     12^{+1}_{-1} &     161^{+18}_{-14} &   0.86^{+0.0}_{-0.0} &   0.0\% & 0.07^{+0.01}_{-0.01} \\
TucII &   39^{+13}_{-11} &   191^{+243}_{-73} & 0.68^{+0.24}_{-0.06} &  11.7\% &  0.1^{+0.06}_{-0.03} &   40^{+13}_{-11} &   287^{+630}_{-138} & 0.96^{+1.48}_{-0.22} & 45.75\% & 0.09^{+0.01}_{-0.03} \\
TucIII &      3^{+1}_{-0} &       47^{+6}_{-4} & 0.87^{+0.01}_{-0.02} &   0.0\% &  0.24^{+0.0}_{-0.01} &      3^{+1}_{-0} &        51^{+8}_{-4} & 0.88^{+0.01}_{-0.01} &   0.0\% & 0.21^{+0.01}_{-0.01} \\
TucIV &     36^{+8}_{-9} &     72^{+31}_{-16} & 0.36^{+0.07}_{-0.02} &   0.0\% & 0.27^{+0.15}_{-0.13} &     37^{+7}_{-9} &      90^{+66}_{-28} & 0.43^{+0.14}_{-0.05} &   0.8\% & 0.18^{+0.16}_{-0.11} \\
TucV &   37^{+14}_{-14} &   138^{+149}_{-52} & 0.61^{+0.14}_{-0.03} &   5.6\% & 0.14^{+0.11}_{-0.06} &   38^{+14}_{-14} &    182^{+385}_{-81} & 0.77^{+1.04}_{-0.11} &  26.9\% &  0.1^{+0.07}_{-0.03} \\
UMaI &   53^{+33}_{-20} &      102^{+5}_{-5} & 0.32^{+0.19}_{-0.19} &   0.0\% & 0.98^{+0.01}_{-0.01} &   68^{+37}_{-30} &      103^{+24}_{-5} & 0.25^{+0.22}_{-0.17} &  0.15\% & 0.97^{+0.01}_{-0.96} \\
UMaII &     39^{+2}_{-2} &    110^{+52}_{-29} & 0.48^{+0.12}_{-0.11} &  0.05\% & 0.06^{+0.04}_{-0.03} &     39^{+2}_{-2} &    178^{+206}_{-68} &  0.65^{+0.2}_{-0.15} &  5.35\% & 0.03^{+0.03}_{-0.01} \\
UMiI &     42^{+3}_{-3} &       92^{+4}_{-3} & 0.37^{+0.03}_{-0.02} &   0.0\% & 0.55^{+0.02}_{-0.03} &     48^{+4}_{-4} &       105^{+6}_{-5} & 0.37^{+0.02}_{-0.01} &   0.0\% & 0.41^{+0.04}_{-0.04} \\
WilI &   35^{+31}_{-14} &      54^{+30}_{-9} & 0.25^{+0.14}_{-0.16} &  0.55\% & 0.94^{+0.02}_{-0.87} &   40^{+27}_{-17} &       53^{+55}_{-9} & 0.25^{+0.17}_{-0.16} &  2.95\% & 0.91^{+0.04}_{-0.89} \\
\enddata
\tablecomments{Column 1 lists the abbreviated dwarf galaxy name; Column 2 and 7 gives the pericenter of the orbit for the two potential models; Column 3 and 8 gives the apocenter of the orbit for the two potential models; Column 4 and 9 is the eccentricity of the orbit for the two potential models; in Column 5 and 9, we provide the probability of the galaxy being unbound for the two potential models. Column 5 and 11 gives the orbital phase.}
\tablenotetext{a}{The apocenter is only for samples with elliptical orbit.}
\tablenotetext{b}{The orbital phase probability defined in Eq.~\ref{eq:orbitphase} to characterize the chance for a dwarf galaxy to be that close to its pericenter.}
\tablenotetext{c}{When the value of total energy minus 1$\sigma$ is larger than 0 km$^2$ s$^{-2}$ ($P_\mathrm{unb} > 84.13\%$), we only quote ``$-$'' in the table.}
\end{deluxetable*}

\startlongtable
\begin{deluxetable*}{lCCCCCCCCCC}
\label{tab:dyn2}
\tabletypesize{\footnotesize}%\small (11pt), \footnotesize (10pt), or \scriptsize (8pt)
\tablecaption{Orbital properties of dwarf galaxies for Model PE$_\mathrm{IM}$ and Model PE$_\mathrm{LM}$.}
\tablehead{
	\colhead{} & \multicolumn{5}{c}{Model PE$_\mathrm{IM}$} & \multicolumn{5}{c}{Model PE$_\mathrm{LM}$}\\
\colhead{name} & \colhead{$r_\mathrm{peri}$} & \colhead{$r_\mathrm{apo}$\tablenotemark{a,c}} & \colhead{$e$} & \colhead{$P_\mathrm{unb}$} & \colhead{$P$\tablenotemark{b,c}} &
\colhead{$r_\mathrm{peri}$} & \colhead{$r_\mathrm{apo}$\tablenotemark{a,c}} & \colhead{$e$} & \colhead{$P_\mathrm{unb}$} & \colhead{$P$\tablenotemark{b,c}}\\
\colhead{} & \colhead{(kpc)} & \colhead{(kpc)} & \colhead{} & \colhead{} & \colhead{} &
\colhead{(kpc)} & \colhead{(kpc)} & \colhead{} & \colhead{} & \colhead{} 
}
\colnumbers
\startdata
AntII &   83^{+16}_{-17} &    237^{+117}_{-47} & 0.49^{+0.07}_{-0.02} &  0.25\% &  0.25^{+0.1}_{-0.05} &  102^{+11}_{-13} & 1078^{+2506}_{-584} &          \geqslant 1 &  56.8\% & 0.26^{+0.02}_{-0.02} \\
AquII &   102^{+4}_{-62} &     117^{+185}_{-7} &          \geqslant 1 &  66.9\% & 0.07^{+0.57}_{-0.02} &   102^{+3}_{-43} &    124^{+167}_{-10} &          \geqslant 1 & 75.35\% & 0.07^{+0.31}_{-0.02} \\
BooI &     53^{+4}_{-5} &    244^{+280}_{-88} & 0.65^{+0.18}_{-0.12} &  3.25\% & 0.07^{+0.05}_{-0.01} &     55^{+3}_{-4} &                   - &                   >1 &  95.5\% &                    - \\
BooII &     39^{+1}_{-1} &                   - &                   >1 & 86.35\% &                    - &     39^{+1}_{-1} &                   - &                   >1 & 99.95\% &                    - \\
CVenI &  104^{+67}_{-75} &    392^{+678}_{-87} & 0.83^{+0.55}_{-0.14} & 25.95\% & 0.51^{+0.05}_{-0.11} &  136^{+44}_{-96} &  679^{+1315}_{-232} &          \geqslant 1 &  61.3\% &  0.55^{+0.05}_{-0.1} \\
CVenII &   42^{+66}_{-32} &    288^{+269}_{-40} & 0.87^{+0.25}_{-0.11} & 17.75\% & 0.31^{+0.07}_{-0.04} &   56^{+63}_{-46} &   546^{+945}_{-119} &  1.0^{+1.37}_{-0.06} &  48.8\% & 0.39^{+0.04}_{-0.06} \\
CarI &    108^{+5}_{-5} &  589^{+1348}_{-324} & 0.87^{+0.49}_{-0.35} & 38.05\% &    0.0^{+0.0}_{-0.0} &    108^{+5}_{-5} &                   - &                   >1 & 99.15\% &                    - \\
CarII &     28^{+1}_{-1} &                   - &                   >1 & 100.0\% &                    - &     28^{+1}_{-1} &                   - &                   >1 & 100.0\% &                    - \\
CarIII &     29^{+1}_{-0} &                   - &                   >1 & 100.0\% &                    - &     29^{+1}_{-0} &                   - &                   >1 & 100.0\% &                    - \\
ColI &    186^{+8}_{-7} &                   - &                   >1 & 97.65\% &                    - &    187^{+8}_{-7} &                   - &                   >1 &  99.4\% &                    - \\
CberI &     43^{+1}_{-1} &    201^{+227}_{-72} & 0.66^{+0.18}_{-0.14} &   3.6\% &  0.01^{+0.02}_{-0.0} &     43^{+1}_{-1} &                   - &                   >1 &  94.9\% &                    - \\
CraI &   145^{+2}_{-74} &     150^{+557}_{-5} &  0.9^{+2.12}_{-0.66} & 45.65\% & 0.03^{+0.94}_{-0.01} &    145^{+2}_{-3} &    167^{+751}_{-22} &          \geqslant 1 &  68.0\% & 0.03^{+0.12}_{-0.01} \\
CraII &   43^{+13}_{-12} &     170^{+24}_{-12} &  0.6^{+0.07}_{-0.04} &   0.0\% & 0.36^{+0.05}_{-0.08} &   58^{+15}_{-17} &   396^{+448}_{-137} & 0.76^{+0.13}_{-0.03} &  7.55\% & 0.23^{+0.02}_{-0.05} \\
DraI &     54^{+7}_{-6} &     198^{+79}_{-38} & 0.57^{+0.07}_{-0.03} &   0.0\% & 0.15^{+0.05}_{-0.04} &     61^{+6}_{-5} &                   - &                   >1 &  86.7\% &                    - \\
DraII &     20^{+1}_{-0} &   304^{+308}_{-102} & 0.88^{+0.07}_{-0.05} &  3.95\% &   0.02^{+0.0}_{-0.0} &     20^{+0}_{-0} &                   - &                   >1 & 100.0\% &                    - \\
EriII & 349^{+20}_{-138} &                   - &                   >1 & 84.95\% &                    - & 349^{+20}_{-104} &                   - &                   >1 & 93.55\% &                    - \\
FnxI &  116^{+13}_{-22} &    218^{+148}_{-48} & 0.33^{+0.16}_{-0.04} &  0.35\% & 0.24^{+0.23}_{-0.09} &    130^{+5}_{-7} &  904^{+2232}_{-500} &          \geqslant 1 & 55.95\% & 0.18^{+0.02}_{-0.02} \\
GruI &   25^{+18}_{-15} &                   - &                   >1 & 88.15\% &                    - &   28^{+21}_{-18} &                   - &                   >1 & 100.0\% &                    - \\
GruII &   29^{+12}_{-10} &    102^{+118}_{-31} & 0.59^{+0.17}_{-0.03} &  6.05\% & 0.19^{+0.13}_{-0.11} &   31^{+11}_{-11} &    165^{+352}_{-76} & 0.88^{+0.74}_{-0.19} &  40.1\% & 0.09^{+0.06}_{-0.03} \\
HerI &   64^{+17}_{-19} & 1336^{+3149}_{-677} &          \geqslant 1 &  71.2\% &  0.3^{+0.02}_{-0.02} &   73^{+15}_{-19} &                   - &                   >1 & 100.0\% &                    - \\
HorI &   88^{+13}_{-12} &   212^{+588}_{-126} &          \geqslant 1 &  67.7\% &  0.04^{+0.01}_{-0.0} &   88^{+13}_{-11} &                   - &                   >1 & 90.15\% &                    - \\
HorII &    79^{+6}_{-10} &     87^{+238}_{-12} &          \geqslant 1 &  70.3\% & 0.01^{+0.21}_{-0.01} &     79^{+6}_{-8} &      84^{+304}_{-9} &          \geqslant 1 & 80.85\% & 0.01^{+0.02}_{-0.01} \\
HyaII &  131^{+12}_{-56} &                   - &                   >1 & 85.45\% &                    - &  132^{+11}_{-45} &                   - &                   >1 & 96.85\% &                    - \\
HyiI &     25^{+0}_{-0} &                   - &                   >1 & 98.85\% &                    - &     25^{+0}_{-0} &                   - &                   >1 & 100.0\% &                    - \\
LeoI &   77^{+37}_{-39} &                   - &                   >1 & 100.0\% &                    - &   89^{+36}_{-43} &                   - &                   >1 & 100.0\% &                    - \\
LeoII & 182^{+46}_{-134} &    274^{+498}_{-30} & 0.61^{+0.59}_{-0.29} &  19.0\% &  0.51^{+0.24}_{-0.2} & 215^{+18}_{-134} &    326^{+870}_{-69} &  0.95^{+1.9}_{-0.51} & 47.05\% &  0.42^{+0.2}_{-0.16} \\
LeoIV &  150^{+7}_{-132} &     156^{+110}_{-5} &  0.9^{+2.92}_{-0.57} &  40.1\% & 0.03^{+0.95}_{-0.01} &  152^{+5}_{-131} &     157^{+219}_{-5} &          \geqslant 1 & 52.35\% & 0.02^{+0.94}_{-0.01} \\
LeoV &   166^{+5}_{-56} &    218^{+429}_{-23} &          \geqslant 1 & 75.15\% &  0.12^{+0.2}_{-0.06} &   166^{+4}_{-28} &                   - &                   >1 &  84.4\% &                    - \\
PhxII &     80^{+4}_{-4} &                   - &                   >1 & 94.35\% &                    - &     81^{+4}_{-4} &                   - &                   >1 & 99.95\% &                    - \\
PisII &  181^{+12}_{-12} &                   - &                   >1 & 98.75\% &                    - &  181^{+12}_{-12} &                   - &                   >1 &  99.3\% &                    - \\
RetII &     28^{+2}_{-3} &      93^{+57}_{-26} & 0.54^{+0.13}_{-0.09} &  0.35\% & 0.09^{+0.06}_{-0.05} &     29^{+2}_{-2} &   235^{+488}_{-122} & 0.92^{+0.43}_{-0.23} & 39.95\% & 0.03^{+0.01}_{-0.01} \\
RetIII &   33^{+47}_{-27} &    136^{+109}_{-23} & 0.84^{+0.59}_{-0.19} & 20.85\% &  0.3^{+0.12}_{-0.16} &   40^{+43}_{-34} &    183^{+269}_{-42} &  0.96^{+2.0}_{-0.14} & 39.65\% & 0.19^{+0.06}_{-0.05} \\
SgrII &     49^{+5}_{-6} &   287^{+560}_{-131} & 0.77^{+0.42}_{-0.17} &  20.3\% & 0.09^{+0.04}_{-0.02} &     51^{+4}_{-5} &                   - &                   >1 & 92.95\% &                    - \\
SclI &     66^{+3}_{-3} &     209^{+44}_{-29} & 0.52^{+0.05}_{-0.04} &   0.0\% & 0.13^{+0.04}_{-0.04} &     73^{+3}_{-3} &                   - &                   >1 & 98.25\% &                    - \\
SegI &     21^{+4}_{-5} &      76^{+82}_{-27} & 0.57^{+0.19}_{-0.07} &   3.8\% & 0.11^{+0.12}_{-0.08} &     22^{+4}_{-5} &    106^{+273}_{-50} & 0.82^{+0.64}_{-0.23} &  34.1\% &  0.04^{+0.1}_{-0.01} \\
SegII &     21^{+7}_{-4} &        49^{+6}_{-4} &  0.4^{+0.06}_{-0.07} &   0.0\% &  0.62^{+0.05}_{-0.1} &     23^{+8}_{-5} &       53^{+14}_{-6} &  0.4^{+0.06}_{-0.03} &   0.0\% &  0.5^{+0.09}_{-0.18} \\
SxtI &     88^{+3}_{-3} &                   - &                   >1 & 99.75\% &                    - &     90^{+3}_{-3} &                   - &                   >1 & 100.0\% &                    - \\
TriII &     12^{+1}_{-1} &    363^{+189}_{-87} & 0.93^{+0.02}_{-0.01} &  0.55\% & 0.04^{+0.01}_{-0.01} &     12^{+1}_{-1} &                   - &                   >1 & 100.0\% &                    - \\
TucII &   41^{+13}_{-11} &  350^{+1109}_{-169} &          \geqslant 1 &  79.8\% &  0.1^{+0.01}_{-0.02} &   42^{+13}_{-11} &                   - &                   >1 & 99.45\% &                    - \\
TucIII &      3^{+1}_{-0} &        53^{+9}_{-5} & 0.89^{+0.01}_{-0.01} &   0.0\% & 0.19^{+0.01}_{-0.02} &      3^{+1}_{-0} &       68^{+22}_{-9} &   0.91^{+0.0}_{-0.0} &   0.0\% & 0.12^{+0.02}_{-0.03} \\
TucIV &     37^{+7}_{-9} &    109^{+157}_{-42} &  0.52^{+0.29}_{-0.1} &  7.95\% & 0.11^{+0.17}_{-0.07} &     39^{+6}_{-8} &    176^{+395}_{-96} &          \geqslant 1 & 52.35\% & 0.06^{+0.04}_{-0.01} \\
TucV &   38^{+14}_{-14} &   207^{+472}_{-101} &          \geqslant 1 & 55.75\% &  0.1^{+0.01}_{-0.03} &   40^{+13}_{-14} &                   - &                   >1 &  89.4\% &                    - \\
UMaI &   90^{+17}_{-46} &     104^{+104}_{-6} & 0.27^{+0.29}_{-0.19} &  4.25\% & 0.92^{+0.05}_{-0.92} &   102^{+5}_{-28} &    161^{+507}_{-63} & 0.56^{+0.96}_{-0.41} & 30.05\% &  0.01^{+0.9}_{-0.01} \\
UMaII &     40^{+2}_{-2} &   301^{+673}_{-158} & 0.93^{+0.72}_{-0.27} & 41.55\% &   0.02^{+0.0}_{-0.0} &     40^{+2}_{-2} &                   - &                   >1 & 96.95\% &                    - \\
UMiI &     52^{+4}_{-4} &     129^{+14}_{-11} & 0.43^{+0.01}_{-0.01} &   0.0\% & 0.27^{+0.05}_{-0.05} &     60^{+3}_{-3} &  702^{+1059}_{-282} & 0.86^{+0.11}_{-0.09} & 10.35\% & 0.12^{+0.01}_{-0.01} \\
WilI &   44^{+23}_{-21} &       53^{+70}_{-9} &  0.27^{+0.3}_{-0.16} &   8.2\% & 0.86^{+0.08}_{-0.85} &   52^{+15}_{-25} &     53^{+132}_{-10} & 0.33^{+1.27}_{-0.22} &  23.9\% & 0.07^{+0.84}_{-0.06} \\
\enddata
\tablecomments{Similar to Table~\ref{tab:dyn1}, but using two lighter MW potential models.}
\tablenotetext{a}{The apocenter is only for samples with elliptical orbit.}
\tablenotetext{b}{The orbital phase probability defined in Eq.~\ref{eq:orbitphase} to characterize the chance for a dwarf galaxy to be that close to its pericenter.}
\tablenotetext{c}{When the value of total energy minus 1$\sigma$ is larger than 0 km$^2$ s$^{-2}$ ($P_\mathrm{unb} > 84.13\%$), we only quote ``$-$'' in the table.}
\end{deluxetable*}

\section{Results and discussion}
\subsection{Phase diagram and the bound nature of dwarf galaxies}
Figure~\ref{fig:r_v} shows the escape velocity curves of the four MW potential models superposed to the dwarf galaxy phase diagram. It demonstrates that the dwarf galaxies can be either almost all bound to the MW (models Model PE$_\mathrm{HM}$ \& PNFW), or half of them (model PE$_\mathrm{IM}$) or even most of them (model PE$_\mathrm{LM}$) could be unbound. This underlines how the nature of dwarf galaxies depends on our knowledge of the MW potential \citep{hammer21}. This calls for caution when interpreting the dwarf galaxy orbital properties. In discussing the dwarf galaxy orbits, one thus needs to always account for their dependency to the adopted MW mass profile, and also consider the possible impact of the LMC.

The position of the LMC is also given in Figure~\ref{fig:r_v}. \citet{kallivayalil13Thirdepoch} discussed whether the LMC is bound or not to the MW, and concluded that for most MW mass models, it is likely at its first passage. In fact, their result depends on the LMC mass, for which they assumed a sufficiently high value ($>10^{11}$ $M_{\odot}$) to keep the SMC bound to the LMC for more than 2 Gyr. The goal was to perform the modeling of the Magellanic Stream \citep{besla12role}, assumed to be a tidal tail induced 1 Gyr ago, during an interaction between the Clouds before they entered the MW halo.

However, the Magellanic stream properties are apparently better reproduced by a ``ram-pressure + collision'' model \citep{hammer15Magellanic}, which only requires the unquestionable collision between the two Clouds 250 to 300 Myr ago. This type of model recovers \citep{wang19complete} the HI and stellar properties of the Magellanic Bridge, and additionally the unusual 30 kpc elongated shape of the SMC on the line of sight, discovered after examining the young variable star distribution \citep{ripepi17VMC}. \citet{wang19complete} argued that it is the recent collision between the Clouds that dominates their past orbital history, and that a light LMC ($\leqslant 2\times10^{10}$ $M_{\odot}$) is required to let large amounts of ionized gas ($>10^{9}$ $M_{\odot}$, see \citealt{Fox2014}) be stripped from both LMC and SMC due to the ram-pressure exerted by the MW halo gas. 

In Figure~\ref{fig:r_v} there are several dwarf galaxies that share a kinetic energy similar is to that of the LMC, namely Bootes II, Carina II, Carina III, Hydrus \& Tucana II. Carina II, Carina III \& Hydrus may have their orbits significantly affected by the LMC, or they could be bound to it \citep{patel20Orbital}, an issue that also depends on the total LMC mass. Therefore, and despite the accuracy of the Gaia EDR3 3D velocities (see error bars in Figure~\ref{fig:r_v}), the uncertainties in the mass and potential of the MW are too large for a robust conclusion on the fraction of dwarf galaxies that are bound to the MW.

In contrast to this, we confirm that the pericenter determination is very robust, and almost independent of the mass model \citep{simon18Gaia}. Half of the dwarf galaxies have the same pericenters for all the different MW mass models, within less than 10\% and well within their error bars. Only 3 dwarf galaxies (Antlia II, Crater II, and Fornax) show a pericenter almost 2 times smaller when using the high MW mass model (PE$_\mathrm{HM}$) compared to result from the low MW mass model (PE$_\mathrm{LM}$).

\subsection{Gaia EDR3 confirms the prominence of the VPOS}
\label{sec:VPOS}
\begin{figure*}[htb!]
	\epsscale{1.1}
	\plotone{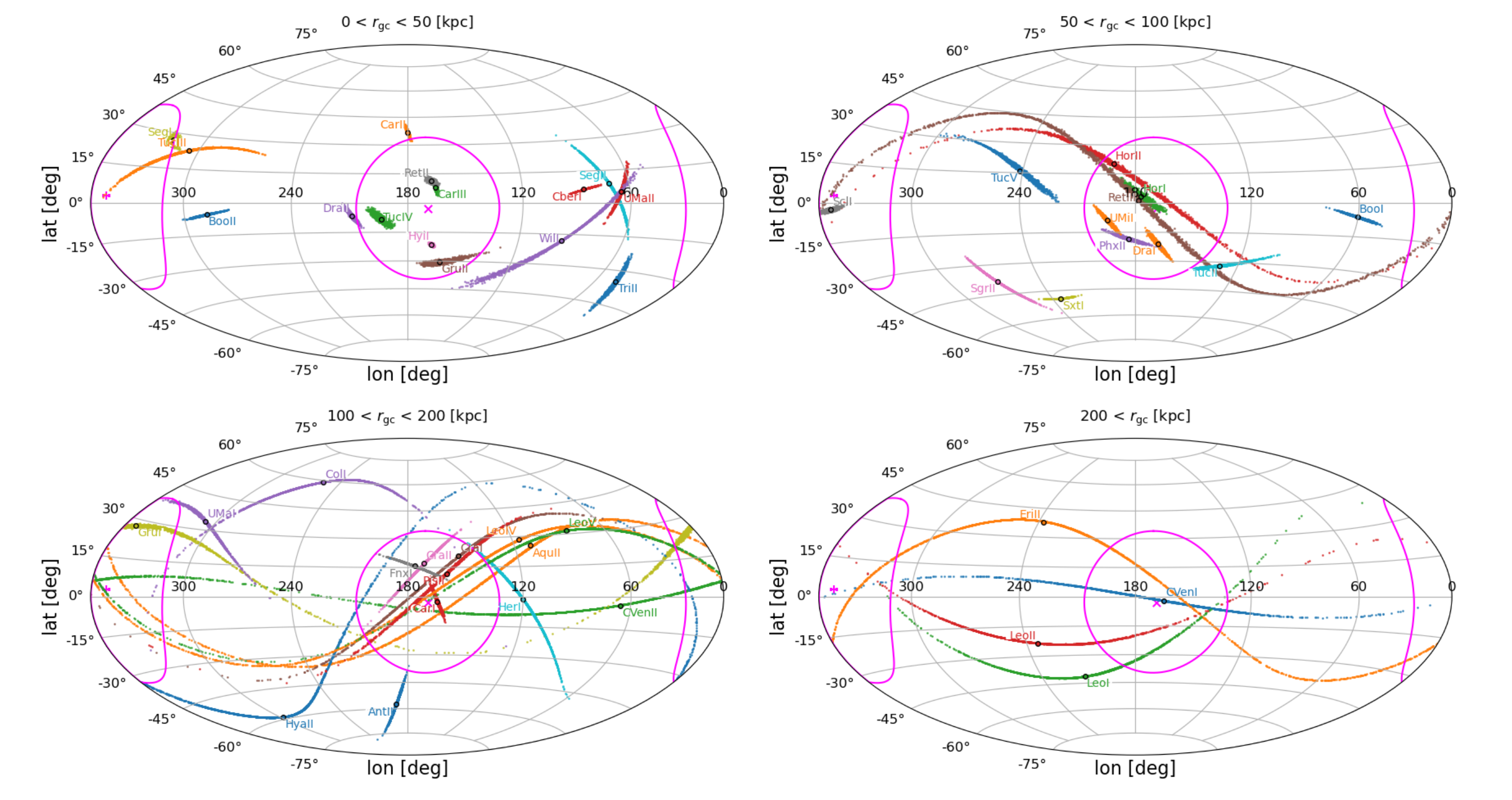}
	\caption{Angular momentum position of dwarf galaxy orbits in an Hammer-Aitoff sky-projection (Galactocentric coordinates), using 4 distance intervals from the Galactic center (from left to right, top to bottom, 0$<$ $r_\mathrm{gc}$ $<$ 50 kpc, 50$<$ $r_\mathrm{gc}$ $<$ 100 kpc, 100$<$ $r_\mathrm{gc}$ $<$ 200 kpc, and 200$<$ $r_\mathrm{gc}$. The magenta circle defines the VPOS location as shown in \citet{fritz18Gaia}, with magenta lines define the opposite direction showing for example that Sculptor lie in the VPOS but it orbits in the opposite direction.  Small points around each dwarf galaxy plot the orbital poles from 2000 Monte Carlo simulations, for which the dot represent the median. \label{fig:vpos}}
\end{figure*}

Figure~\ref{fig:vpos} presents the orbital poles of all 46 dwarf galaxies, projected in an Hammer-Aitoff diagram. The angular momenta are calculated as the cross-product of the Galactocentric position vector $\vec{r}_\mathrm{gc}$, and the 3D velocity in the MW frame, $\vec{v}_\mathrm{3D}$, for each of the 2000 Monte-Carlo realizations. Figure~\ref{fig:vpos} shows that the Gaia EDR3 accuracy is not sufficient to efficiently constraint the poles of dwarf galaxies beyond $r_\mathrm{gc} >$ 200 kpc.
%, except perhaps for Canes Venaciti I, for which more than half of its predicted orbital pole is included into the VPOS (see bottom-right panel of Figure~\ref{fig:vpos}).

Whether a dwarf galaxy is part of the VPOS depends on its 3D position (it has to lie close to the VPOS plane), and orbital pole (which should be close to the VPOS normal vector). Since the latter contains information on both the positon and velocity, we here focus on it to judge possible VPOS membership. Figure ~\ref{fig:vpos} thus already allows a simple visual confirmation of a preferred alignment of orbital poles with the VPOS. It indicates in pink the areas corresponding to the VPOS (from Figure~3 of \citealt{fritz18Gaia}), containing 10\% of the area on the sphere around the adopted VPOS normal vector pointing towards Galactic coordinates $(l, b) = (169.3^\circ, -2.8^\circ)$.

For a more quantitative analysis, we follow the method outlined in \citet{fritz18Gaia}. Table \ref{tab:VPOS2} lists for all 46 dwarf galaxies in our sample, the best-possible alignment $\theta^{\mathrm{predicted}}_{\mathrm{VPOS-3}}$\ of their orbital pole with the VPOS normal vector that is defined by their spatial position \citep[following the geometric method presented in ][]{Pawlowski2013,pawlowski15new}, and the actually observed angular separation $\theta^{\mathrm{measured}}_{\mathrm{VPOS-3}}$\ between the median orbital pole and the VPOS normal. The latter is positive if the object co-orbits in the same sense as the majority of VPOS members (including the LMC and SMC), and negative if the object counter-orbits. We furthermore count the fraction $f_\mathrm{inVPOS}$\ of Monte-Carlo realizations which result in orbital poles that are aligned with the VPOS normal vector to within an angle $\theta_\mathrm{inVPOS}$. For consistency and direct comparability we adopt the same values as in \citet{fritz18Gaia}: $\theta_\mathrm{inVPOS} = 36.9^\circ$\ corresponding to 10\% of the area of the sky, and a VPOS normal vector pointing to Galactic coordinates $(l, b) = (169.3^\circ, -2.8^\circ)$. To further assess the degree to which a misalignment might simply be due to proper motion uncertainties, we also generate 2000 mock orbital pole realizations by placing the object's intrinsic orbital pole in its predicted (best-aligned with the VPOS) position, and then vary its direction by drawing from the orbital pole uncertainty of the observed data. This way we determine the probability $p_\mathrm{outsideVPOS}$\ of these mock-realizations to be found outside of $\theta_\mathrm{inVPOS}$\ despite their intrinsic close alignment (i.e. the false negative rate), and $p_\mathrm{>obs}$, the chance to observe the orbital pole as far from the VPOS normal as in the real data. Note that the latter is only a lower limit, as the best-possible alignment is assumed.

\startlongtable
\begin{deluxetable*}{lCCCCCC}
\label{tab:VPOS2}
\tablecaption{Alignment with the VPOS (see Sect.~\ref{sec:VPOS}) and membership to volume-complete samples (see Sect.~\ref{sec:pericenters}).}
\tablehead{
	\colhead{Name} & \colhead{$\theta^{\mathrm{predicted}}_{\mathrm{VPOS-3}}$} & \colhead{$\theta^{\mathrm{measured}}_{\mathrm{VPOS-3}}$} & \colhead{$f_\mathrm{inVPOS}$} & \colhead{$p_\mathrm{outsideVPOS}$} & \colhead{$p_\mathrm{>obs}$} & \colhead{volume-complete member}
\\
	\colhead{} & \colhead{$(^\circ)$} & \colhead{$(^\circ)$} & \colhead{} & \colhead{} & \colhead{} & \colhead{}
	}
\decimalcolnumbers
\startdata
AntII        &  2.5 &  56.9 & 0.001 & 0.000 & 0.000 &          Near-S  \\
AquII        &  7.8 &  63.1 & 0.297 & 0.455 & 0.205 &                  \\
BooI         & 16.1 & -71.4 & 0.000 & 0.000 & 0.000 &  Near-S, Dist-S  \\
BooII        & 12.8 & -62.9 & 0.000 & 0.000 & 0.000 &          Near-S  \\
CVenI        &  1.5 &   3.9 & 0.800 & 0.194 & 0.887 &          Dist-S  \\
CVenII       &  4.2 & -76.8 & 0.390 & 0.603 & 0.135 &          Dist-S  \\
CarI         &  4.7 &   4.7 & 1.000 & 0.000 & 0.790 &  Near-S, Dist-S  \\
CarII        &  3.5 &  40.5 & 0.005 & 0.000 & 0.000 &  Near-S, Dist-S  \\
CarIII       &  7.1 &  11.7 & 1.000 & 0.000 & 0.000 &          Near-S  \\
ColI         & 27.6 & -89.6 & 0.005 & 0.128 & 0.000 &                  \\
CberI        &  9.6 &  82.9 & 0.000 & 0.000 & 0.000 &  Near-S, Dist-S  \\
CraI         &  9.6 &  28.9 & 0.555 & 0.253 & 0.366 &          Dist-S  \\
CraII        & 15.2 &  20.0 & 0.991 & 0.002 & 0.187 &  Near-S, Dist-S  \\
DraI         & 10.4 &  18.2 & 1.000 & 0.000 & 0.000 &  Near-S, Dist-S  \\
DraII        & 29.9 &  39.3 & 0.036 & 0.000 & 0.000 &                  \\
EriII        &  9.3 &  76.1 & 0.305 & 0.458 & 0.106 &                  \\
FnxI         & 14.6 &  19.8 & 1.000 & 0.000 & 0.017 &          Dist-S  \\
GruI         & 24.9 & -25.0 & 0.675 & 0.323 & 0.933 &          Near-S  \\
GruII        & 27.6 &  28.7 & 0.998 & 0.001 & 0.085 &          Near-S  \\
HerI         & 37.7 &  49.5 & 0.000 & 1.000 & 0.050 &  Near-S, Dist-S  \\
HorI         &  1.0 &  10.0 & 1.000 & 0.000 & 0.050 &  Near-S, Dist-S  \\
HorII        &  6.2 &  31.8 & 0.551 & 0.201 & 0.267 &                  \\
HyaII        & 28.9 & -73.9 & 0.104 & 0.443 & 0.066 &          Dist-S  \\
HyiI         & 10.5 &  18.7 & 1.000 & 0.000 & 0.000 &  Near-S, Dist-S  \\
LeoI         & 20.4 &  54.4 & 0.196 & 0.219 & 0.065 &            \\
LeoII        & 13.4 &  65.3 & 0.129 & 0.189 & 0.035 &          Dist-S  \\
LeoIV        &  2.7 &  59.9 & 0.300 & 0.398 & 0.189 &          Dist-S  \\
LeoV         &  1.1 &  87.1 & 0.143 & 0.269 & 0.009 &          Dist-S  \\
PhxII        & 19.5 &  20.9 & 1.000 & 0.000 & 0.060 &          Near-S  \\
PisII        &  4.6 &  17.2 & 0.853 & 0.081 & 0.383 &          Dist-S  \\
RetII        & 11.9 &  14.3 & 1.000 & 0.000 & 0.000 &  Near-S, Dist-S  \\
RetIII       &  3.0 &  10.2 & 0.443 & 0.549 & 0.866 &          Near-S  \\
SclI         & 47.5 & -85.1 & 0.000 & 1.000 & 0.000 &  Near-S, Dist-S  \\
SegI         &  5.0 &  -5.8 & 1.000 & 0.000 & 0.251 &                  \\
SegII        & 36.1 & -36.2 & 0.710 & 0.001 & 0.077 &                  \\
SxtI         & 58.7 & -81.9 & 0.000 & 1.000 & 0.000 &  Near-S, Dist-S  \\
SgrII        & 14.7 &  71.8 & 0.000 & 0.000 & 0.000 &  Near-S, Dist-S  \\
TriII        & 64.5 & -64.9 & 0.000 & 1.000 & 0.053 &                  \\
TucII        & 25.3 &  46.6 & 0.000 & 0.001 & 0.000 &          Near-S  \\
TucIII       &  6.1 & -46.0 & 0.280 & 0.017 & 0.002 &                  \\
TucIV        & 15.4 &  24.7 & 1.000 & 0.000 & 0.000 &          Near-S  \\
TucV         & 20.6 &  74.3 & 0.000 & 0.015 & 0.000 &                  \\
UMaI         & 36.0 & -50.3 & 0.018 & 0.537 & 0.012 &          Near-S  \\
UMaII        & 55.4 & -76.0 & 0.000 & 1.000 & 0.000 &          Near-S  \\
UMiI         & 21.7 &  25.7 & 1.000 & 0.000 & 0.000 &  Near-S, Dist-S  \\
Wil1         & 39.1 &  73.1 & 0.000 & 1.000 & 0.000 &          Near-S  \\
\enddata
\tablecomments{
Column 1: name of the object, Col. 2: angle between VPOS normal and predicted orbital pole (best-possible alignment), Col. 3: angle between VPOS normal and median measured orbital pole, Col. 4: fraction of realizations whose orbital pole falls into the 10\% circles around the VPOS normal vector, Col. 5: probability that an intrinsically perfectly aligned orbital pole is found outside of the 10\% circles given the proper motion measurement uncertainties, Col. 6: probability to find an orbital pole at least as far inclined from the VPOS as the median measured orbital pole if the intrinsic alignment is as close as possible, and Col. 7 membership to the nearby volume-complete sub-sample (quoted $Near-S$) up to the 90.5 kpc slice from \citealt{drlica-wagner20Milky}, or the distant (quoted $Dist-S$) volume complete sub-sample up to the 181 kpc slice.
}
\end{deluxetable*}

Of the 46 dwarf galaxies in our sample, six cannot orbit within the VPOS because their positions place them well outside of the structure ($\theta^\mathrm{predicted}_\mathrm{VPOS-3} > \theta_\mathrm{inVPOS}$). These are Hercules\,I, Segue\,II, Sagittarius\,II, Triangulum\,II, Ursa Major\,II, and Willman\,1. %Of the remaining 40 satellites that could potentially orbit within the VPOS, 11 have a chance of aligning that is smaller than $f_\mathrm{inVPOS} < 0.05$, while the remaining 29 satellites either have well aligned orbital poles, or are sufficiently unconstrained to remain consistent with orbiting along the VPOS.
Of the remaining 40 objects for which an alignment is feasible, 20 have median orbital poles that align to better than $\theta_\mathrm{inVPOS}$\ with the VPOS normal vector\footnote{The dwarf galaxies with median orbital poles aligned with the VPOS are: Canes Venaciti I, Carina I, Carina III, Crater I, Crater II, Draco I, Fornax, Grus I, Grus II, Horologium I, Horologium II, Hydrus, Phoenix II, Pisces II, Reticulum II, Reticulum III, Sculptor, Segue I, Tucana IV, and Ursa Minor.}. Most of these are well constrained, with 13 having almost all their Monte Carlo realizations within this angle from the VPOS plane ($f_\mathrm{inVPOS} \geq 0.99$). Only one object, RetIII with $f_\mathrm{inVPOS} = 0.44$, is relatively poorly constrained while the remaining six have $0.85 > f_\mathrm{inVPOS} > 0.55$. Only three (Grus I, Sculptor, Segue I) of these 20 are counter-orbiting with respect to the bulk orbital sense of the VPOS members. Together with the LMC and SMC, which also orbit along the VPOS, the counter-orbiting fraction is thus $f_\mathrm{counter} = \frac{3}{22} = 0.14$. This is intriguingly close to the counter-orbiting fraction considering only the 11 bright, classical MW satellites \citep{PawlowskiKroupa2020}, of which 8 have orbital poles aligned with the VPOS with only Sculptor orbiting on the opposite sense than the others ($f_\mathrm{counter} = \frac{1}{8} = 0.13$).
%Furthermore, of the 15 M31 satellites associated with its satellite plane, 13 follow a coherent line-of-sight velocity trend which is consistent with a similar counter-orbiting fraction of $f_\mathrm{counter} = \frac{2}{15} = 0.13$\ \citep{Ibata2013}, while for the satellite plane around Centaurus\,A the line-of-sight velocity trend indicates a counter-orbiting fraction of between $f_\mathrm{counter} = \frac{2}{16} = 0.13$\ \citep{Mueller2018} to $f_\mathrm{counter} = \frac{7}{28} = 0.25$\ \citep{Mueller2021}.

The remaining 20 dwarf galaxies have median orbital poles that do not align with the VPOS. Of these, 11 can confidently be ruled out as VPOS members (by our adopted criterion) because their chance of aligning is smaller then $f_\mathrm{inVPOS} < 0.05$. Note, however, that the orbital poles of Crater II and Draco II are well constrained but only marginally outside of our adopted maximum accepted alignment angle of $\theta_\mathrm{inVPOS}$, and a small change in the adopted direction of the VPOS normal vector would place both inside of the 10\% region.

The remaining nine dwarf galaxies have orbital poles that are only poorly constrained, mainly due to their large distance and thus large relative proper motion error\footnote{These dwarf galaxies with orbital poles that are poorly constrained but consistent with VPOS alignment are: Aquarius II, Canes Venaciti I, Eridanus II, Leo I, Leo II, Leo IV, Leo V, Tucana III}. However, these nine are all well consistent with being aligned with the VPOS within their uncertainties, with $0.1 < f_\mathrm{inVPOS} < 0.4$. This is further demonstrated by their high $p_\mathrm{outsideVPOS}$\ values of 19 to 60\% (except Tucana\,III with only $p_\mathrm{outsideVPOS} = 0.02$), indicating that even if they were intrinsically well aligned, the substantial proper motion uncertainties would likely place their derived median orbital pole outside of the region around the VPOS normal vector. Six of the nine objects most-likely co-orbit.

The similarity of our study with that of \citet{fritz18Gaia} in sample and data analysis also allows us to judge how the improved data quality of Gaia eDR3 over DR2 affects the VPOS signal, without the danger of being affected by biases due to differing methodologies. The two studies have 37 objects in common, of which 32 can possibly orbit along the VPOS. Of these 32, we find that 16 have orbital poles that are most-likely aligned with the VPOS normal ($f_\mathrm{inVPOS} > 0.5$), and 10 of these are definitively aligned ($f_\mathrm{inVPOS} > 0.98$). In \citet{fritz18Gaia}, these numbers were 14 and 7, respectively. Thus, the improved data has resulted in a substantial increase in the number of orbitally aligned dwarf galaxies, as to be expected if an underlying correlation is obscured by initially larger measurement errors. The same trend has previously been found for the classical MW satellites, whose orbital poles clustered subsequently tighter around the VPOS normal direction as their proper motion errors have decreased \citep{PawlowskiKroupa2020}. Furthermore, it is noteworthy that $f_\mathrm{counter}$\ dropped considerably compared to \citet{fritz18Gaia}, who found $f_\mathrm{counter} = 0.32$\ among their 17 likely VPOS members and the Magallanic Clouds. This change is also consistent with the VPOS being an intrinsic, correlated structure exhibiting a preferred orbital direction that is becoming more apparent as proper motion measurements improve.

In summary, of the 40 Milky Way dwarf galaxies that have spatial positions consistent with being members of the VPOS, at least 20 to 29 objects have orbital poles aligned with the VPOS normal vector. Thus, from 50 to 73\% of potentially aligned dwarf galaxies (or 43 to 63 \% of all) indeed lie and orbit in the VPOS, which confirms the prominence of the VPOS for the MW dwarf galaxy spatial and orbital distributions\footnote{This fraction can be considered a lower limit: For Leo\,II, one of the most distant dwarf galaxies in our sample that has a poorly constrained orbital pole, more accurate Hubble Space Telescope proper motion \citep{Piatek2016} place its orbital pole firmly in the VPOS \citep{PawlowskiKroupa2020}. Furthermore, discussed above Crater II and Draco II are just marginally outside of the adopted VPOS alignment criterion, and both the LMC and SMC are known to co-orbit along the VPOS (though Sagittarius does not).}. Such a large fraction is indicative of a real structure of dwarf galaxies, which could not be associated to the expectations for a cosmological infall of primordial dwarfs, including if they had been accreted along cold streams \citep[and references therein]{pawlowski14Vast}.

\subsection{Locations of dwarf galaxies are excessively near their orbital pericenter}
\label{sec:pericenters}
Based on Gaia DR2, \citet{fritz18Gaia} and \citet{simon18Gaia} noticed that dwarf galaxy locations are excessively concentrated near their pericenters, which is at odd for satellites that are expected to lie mostly near their apocenters. \citet{fritz18Gaia} argued that this effect may be caused by the existence of non-detected ultra faint dwarfs, mostly those lying beyond 100 kpc. The problem is also less pronounced when considering high MW mass \citep[see their Figure 5 and Figure 1, respectively]{fritz18Gaia,Hammer2020}. If persistent, the dwarf galaxy excess near pericenter may challenge their commonly assumed nature as long-lived MW satellites since such a realization would be associated to a very small probabilities especially for moderate MW mass \citep[P$\sim 2 \times 10^{-7}$ for MW mass $\le$ $10^{12}$ $M_{\odot}$]{Hammer2020}. \\

\begin{figure*}[htb!]
	\epsscale{0.5}
	\plotone{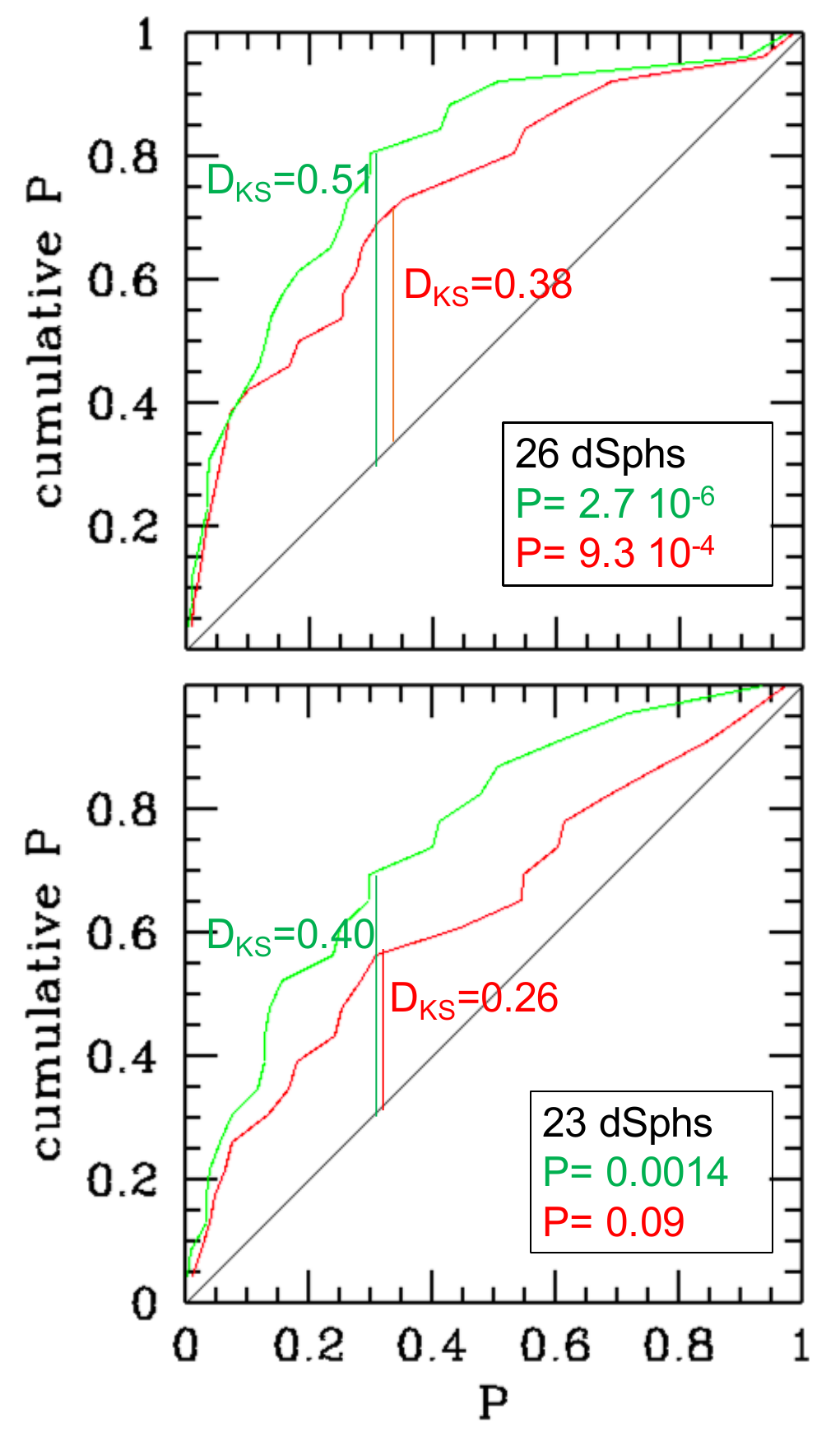}
	\caption{Cumulative distribution of time to reach the pericenter divided by the time taken from pericenter to min(apocenter, 300 kpc). The black solid line shows the null hypothesis, i.e., dwarf galaxies have location randomly distributed from pericenter to min(apocenter, 300 kpc).  Green and red lines represent the $PNFW$ model ($M_{tot}=8.1 \times 10^{11} M_{\odot}$, see also \citealt{eilers19Circular}, and PE$_\mathrm{HM}$ the most massive mass able to fit the MW rotation curve ($M_{tot}= 15 \times 10^{11}M_{\odot}$, see \citealt{jiao21Which}), respectively. $D_\mathrm{max}$ values and associated probabilities of the Kolmogorov Smirnov tests are given in the Figures, top and bottom panels representing the 90.5 (dubbed as small volume) and the 181 kpc (dubbed as large volume) distance slices of \citealt{drlica-wagner20Milky}, respectively.} 
	\label{fig:prob_peri}
\end{figure*}

Here we try to reevaluate these statistics using Gaia EDR3 data and by accounting for possible biases linked to the observability of dwarf galaxies at larger distances. Indeed, one may consider that we are only able to see the closest dwarf galaxies, which then could be those close to their pericenters. Following \citet{Hammer2020} we first define the orbital phase probability:
\begin{equation}
\label{eq:orbitphase}
 P = \frac{t_\mathrm{peri}}{t_\mathrm{peri-apo}},
\end{equation} 
to estimate the chance for a dwarf galaxy to be close to its pericenter. In Equation~\ref{eq:orbitphase}, $t_\mathrm{peri}$ is the orbital time for reaching or leaving the pericenter from the dwarf galaxy's current position (whichever is shorter), and  $t_\mathrm{peri-apo}$ is the time to complete half an orbit. We have used the calculated dwarf orbits from the publicly available code galpy \citep{bovy15galpy} for deriving the time spent to reach the pericenter, as well as that from apocenter to pericenter. We further limit the volume and distance to 300 kpc because beyond this, most dwarf galaxies would have been hard to be detected \citep{Simon2019}. It leads to change $t_\mathrm{peri-apo}$ in Equation~\ref{eq:orbitphase} by the time taken from pericenter to min(apocenter, 300 kpc):
\begin{equation}
\label{eq:orbitphase2}
 P = \frac{t_\mathrm{peri}}{t_\mathrm{peri-min(apo,300kpc)}},
\end{equation} 
In the following, we examine the properties of complete subsamples using the probabilities based on Equation~\ref{eq:orbitphase2}. Non-parametric Kolmogorov-Smirnov tests presented in Figure~\ref{fig:prob_peri} are made by comparing the cumulative probabilities calculated from Equation~\ref{eq:orbitphase2} to those expected in the null hypothesis, which represents a random distribution of time between $t_\mathrm{peri}$ and $t_\mathrm{peri-min(apo,300kpc)}$.  \\%It corresponds also to the limit we have used for defining eccentricities for hyperbolic and very eccentric elliptical orbits(see Section~\ref{sec:orbit_int}).\\

The proximity of dwarf galaxies to their pericenter could be caused by biases, i.e., that we preferentially detect nearby dwarf galaxies more likely to be close to their pericenters, and this at the cost of missing a population of faint dwarfs that would lie at larger distance, e.g., beyond 100 kpc \citep{fritz18Gaia}. To test the possibility of missing non-detected ultra faint dwarfs, we have built 'volume-complete samples' of dwarf galaxies, i.e., samples including only the dwarfs that could be detected within a given volume, independently or their actual distances. We have made use of the detectability study of \citet{drlica-wagner20Milky} and have applied their Equation 2 for estimating the ($M_{V}$, $r_\mathrm{half}$) range to which dwarf galaxies can be observed. It leads to volume-complete samples of 26 ( 23) dwarf galaxies that can be observed up to the whole volume defined by the distance slices of 90.5 (181) kpc, respectively (see \citealt{drlica-wagner20Milky}'s Table 5 and Fig. 6). Membership to the two sub-samples are tabulated in the last column of Table~\ref{tab:VPOS2}, and V-absolute magnitude and $r_{half}$ values used to define the dwarf galaxy detectability have been taken from \citet{Simon2019}.\\

Figure~\ref{fig:prob_peri} presents the the cumulative probability of having an excess of dwarf galaxy locations near the pericenter, assuming they are satellites of the MW modeled with $PNFW$ (green curve, see Table~\ref{tab:halo}) and with the $PE_{HM}$ (red curve, highest mass). As expected (see also \citealt{fritz18Gaia}), the higher the MW mass, the more likely is the satellite hypothesis for dwarf galaxies as it is illustrated in Figure~\ref{fig:r_v}. Gaia EDR3 provides sufficiently accurate orbits and pericenter to gather samples of 26 and  23 dwarf galaxies that are essentially complete in the sense that all included dwarf galaxies can be seen in the whole considered small ($\sim$ 100 kpc) and large ($\sim$ 200 kpc) volume, respectively. Furthermore, this allows us to test the hypothesis of missing ultra-faint dwarfs that would lie at apocenter and then that have supposedly not been detected yet \citep{fritz18Gaia}. \\

Let us consider the small volume-complete sample (top panel of  Figure~\ref{fig:prob_peri}) of 26 dwarf galaxies, which leads to very small probabilities. Examining the apocenters of these dwarf galaxies and considering the $PNFW$ MW mass model, we find that only 34\% of them are within the small volume, i.e., letting  open the possibility that there are undetected missing dwarf galaxies beyond $\sim$ 100 kpc. However, for the most massive MW mass model (PE$_\mathrm{HM}$), most (58\%) apocenters lie within the small volume, and it becomes non plausible that the small probabilities (P$= 9.3 \times 10^{-4}$) can be due to missing ultra faint dwarfs. Now let consider the large volume-complete sample of  23 dwarf galaxies, which is undoubtedly more affected by the difficulty in detecting distant faint dwarfs. Assuming the $PNFW$ MW mass model, 66\% of apocenters are within that volume, which means that there could be only few missing dwarfs lying further away near their apocenter to explain the small probability given in the bottom panel of Figure~\ref{fig:prob_peri}. At first glance, the probability for the  most massive MW mass model (PE$_\mathrm{HM}$) could be interpreted as a significant alleviation of the proximity-to-pericenter problem. However, in such a case, 92\% of the apocenters are within the large volume, letting no possibility for missing dwarfs at larger distances. 

In summary, we have examined the the proximity-to-pericenter problem that may affect the satellite nature of most dwarf galaxies. For this, we have only considered the two most massive MW mass  models of this paper, because they are the ones for which most dwarf galaxies are bound. Having defined volume-complete samples following the detectability procedure of \citet{drlica-wagner20Milky}, we find that  by increasing the MW mass, and by assuming a putative population of distant faint dwarfs may alleviate part of the problem, but certainly not completely. For example, assuming the largest MW mass available to fit the MW rotation curve, the probability that  MW dwarf galaxies behave as satellites is well below 1 in the two considered complete volumes of the above analysis.

\section{Conclusion}
We have determined the proper motions of 46 dwarf galaxies using Gaia EDR3, with a robust evaluation of errors accounting both for effects of their statistics and of Gaia systematics. The gain compared to former DR2 analysis by \citet{fritz18Gaia} is twofold. First, the accuracy for objects in common between the two studies has improved by a factor $\sim$ 2.5, which corresponds to the actual reduction of total errors for the analysis of their tangential and 3D velocities. Second, it allows to increase to 40 the number of dwarf galaxies for which analysis of their motions can be robustly made, i.e., more than twice what was done with Gaia DR2.

We have then derived the 3D phase-space diagram of dwarf galaxies that illustrates the sequence delineated by most dwarf galaxies. Gaia EDR3 errors on 3D velocities are systematically smaller than differences between expectations from various possible MW mass models. It implies that almost all dwarf galaxies are gravitationally bound if the MW total mass is larger than $8\times10^{11}$ $M_{\odot}$, while they would be increasingly unbound for MW mass values going down to $5.1\times10^{11}$ $M_{\odot}$ and $2.8\times10^{11}$ $M_{\odot}$.

In this paper, we have incorporated calculations of integrated orbital parameters after considering the whole range of four MW mass models that can be consistent with the MW rotation curves \citep{eilers19Circular,mroz19Rotation}. It provides a useful library of orbits with different schemes for the MW mass, some of them could be either rosette or hyperbolic. While apocenters and eccentricities are very dependent on the adopted MW potential, we show that pericenter values are very robustly determined, whatever the MW mass model, and down to an accuracy of 10\% for a majority of dwarf galaxies.

We also confirm that many of the dwarf galaxies lie near their pericenters, which cannot simply be due to either selection effects or to an underestimate of the MW mass. This later result appears to be problematic if most dwarf galaxies are MW satellites.  
Finally we identify the strong prominence of the VPOS in the distribution of orbital poles, that includes a majority of dwarf galaxy locations and orbits, confirming it as an important feature of the outer Galactic halo. 
Such a large fraction of VPOS members and their strong kinematic correlation is indicative of a real structure of dwarf galaxies, which could be in conflict with expectations for a cosmological infall of primordial dwarfs.

\begin{acknowledgments}
\par This work has made use of data from the European Space Agency (ESA) mission  {\it Gaia} (\url{https://www.cosmos.esa.int/gaia}), processed by the {\it Gaia} Data Processing and Analysis Consortium (DPAC,
\url{https://www.cosmos.esa.int/web/gaia/dpac/consortium}).
\par Funding for the DPAC has been provided by national institutions, in particular the institutions participating in the {\it Gaia} Multilateral Agreement. This work has been supported by the National Natural Foundation of China (NSFC No. 11973042 and No. 11973052). It was also supported by the Fundamental Research Funds for the Central Universities and  the China Scholarship Council (CSC). We also thank for its support the International Research Program Tianguan, which is an agreement between the CNRS, NAOC and the Yunnan University.  MSP thanks the Klaus Tschira Stiftung gGmbH and German Scholars Organization e.V. for support via a Klaus Tschira Boost Fund. 
\end{acknowledgments}

\appendix

\section{Comparison of PMs with other papers.}
\label{compPMs}

\par
\begin{figure*}
\figurenum{A1}
\epsscale{1}
\plotone{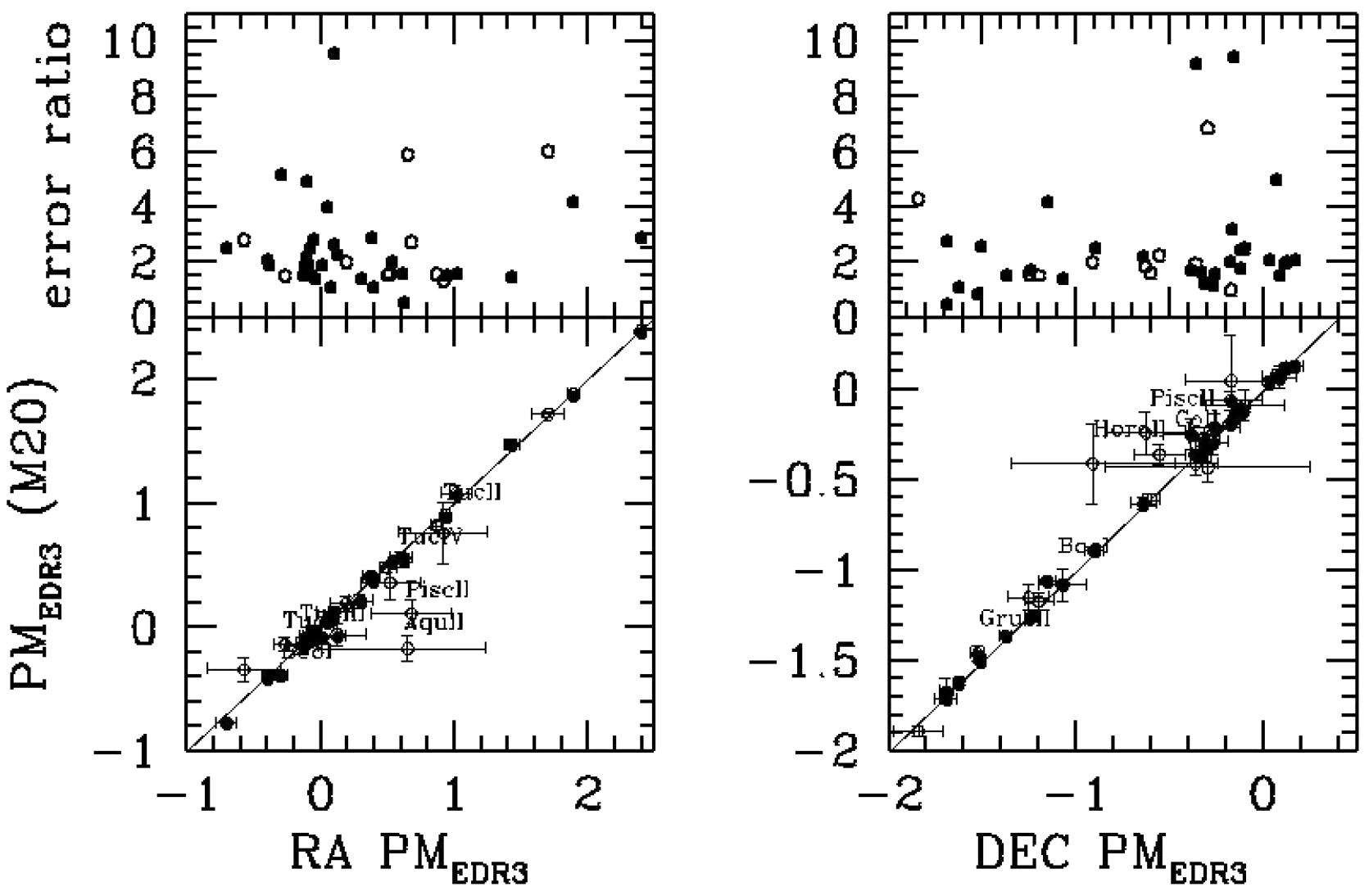}
\caption{Comparison of proper motions (left: RA, right: DEC) of Table~\ref{tab:pm} with those calculated by \citet{mcconnachie20Updated}. Few dwarf galaxies names are indicated, pointing out to dwarf galaxies for which the two measurements are discrepant at $>$ 1 $\sigma$ level. The top panels indicate the ratio of error from this paper to that tabulated by \citet{mcconnachie20Updated} both extracted from Gaia EDR3. Notice that the error on PM RA of Fornax is so small (1$\mu$as yr$^{-1}$) in \citet{mcconnachie20Updated} that it leads to a ratio in excess of 20 (not shown in the Figure). 
\label{fig:compPM}}
\end{figure*}

 \citet{mcconnachie20Updated} published PM based on Gaia EDR3 for a large number of dwarf galaxies. Figure~\ref{fig:compPM} compares values of Table~\ref{tab:pm} to theirs. Although values are often consistent within error bars found in this paper, those tabulated by \citet{mcconnachie20Updated} appear to be extremely small. This is because they do not account for Gaia systematics as we have made in using \citet[see also Section~\ref{sec:pm} of this paper]{vasiliev19Systematic}, though mentioning their impact in the text. We also notice a good agreement within our uncertainties between our values for 9 dwarfs that have been studied by \citet{Vitral2021}, though our uncertainties are larger and there is a small, $<$ 2$\sigma$ discrepancy for Bootes I.

\begin{figure*}
\figurenum{A2}
\epsscale{1}
\plotone{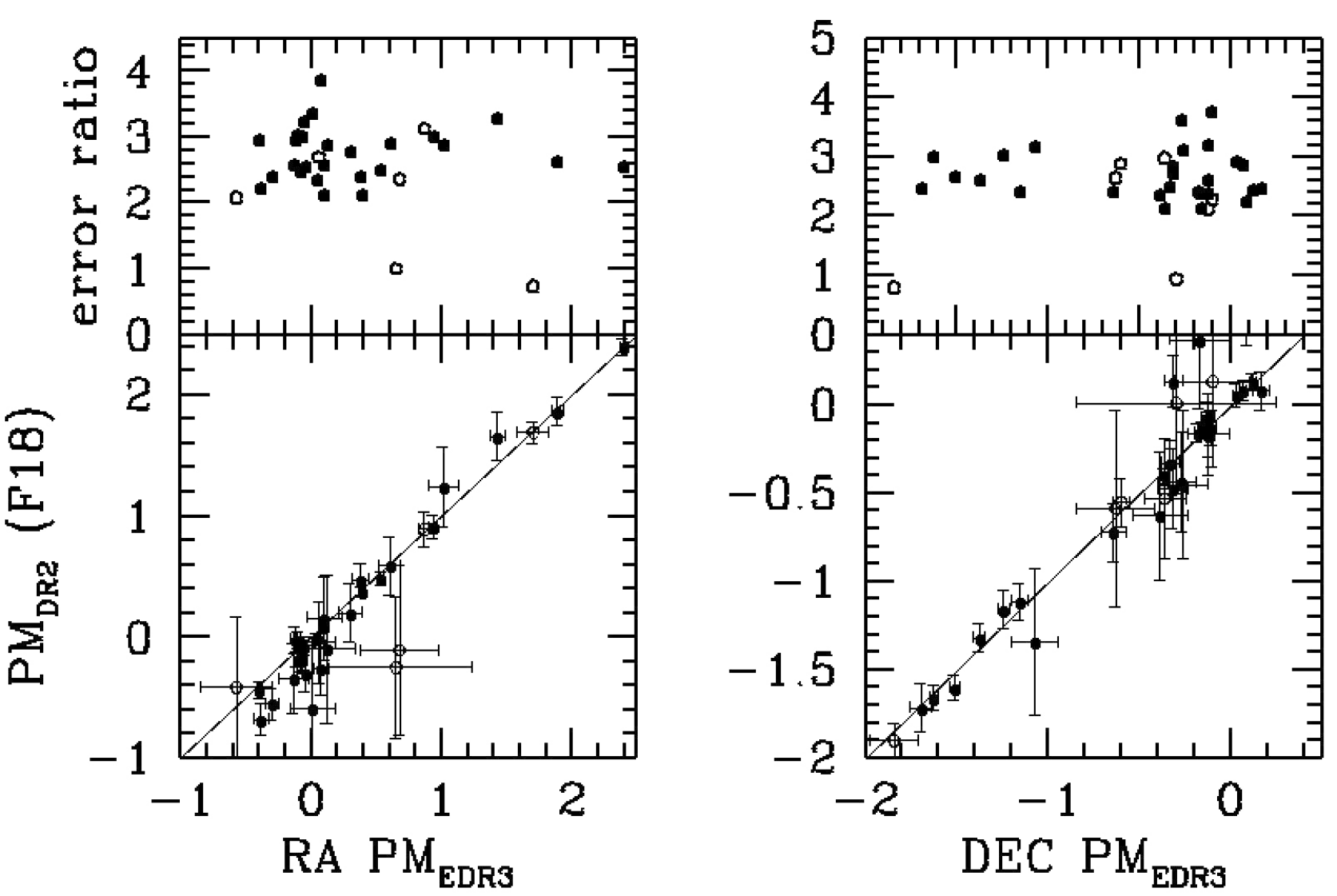}
\caption{Comparison of proper motions (left: RA, right: DEC) of Table~\ref{tab:pm} with those calculated by \citet{fritz18Gaia} with Gaia DR2. The top panels indicate the ratio of error from \citet{fritz18Gaia} Gaia DR2 to that of this paper, Gaia EDR3. \label{fig:compPMDR2}}
\end{figure*}

Figure~\ref{fig:compPMDR2} compares PM values and their errors from EDR3 to those from Gaia DR2 from \citet{fritz18Gaia}.

%

\bibliography{hefan,wp944,francois,add}

\begin{thebibliography}{}
\expandafter\ifx\csname natexlab\endcsname\relax\def\natexlab#1{#1}\fi
\providecommand{\url}[1]{\href{#1}{#1}}
\providecommand{\dodoi}[1]{doi:~\href{http://doi.org/#1}{\nolinkurl{#1}}}
\providecommand{\doeprint}[1]{\href{http://ascl.net/#1}{\nolinkurl{http://ascl.net/#1}}}
\providecommand{\doarXiv}[1]{\href{https://arxiv.org/abs/#1}{\nolinkurl{https://arxiv.org/abs/#1}}}

\bibitem[{{Ad{\'e}n} {et~al.}(2009){Ad{\'e}n}, {Feltzing}, {Koch}, {Wilkinson},
  {Grebel}, {Lundstr{\"o}m}, {Gilmore}, {Zucker}, {Belokurov}, {Evans}, \&
  {Faria}}]{2009A&A...506.1147A}
{Ad{\'e}n}, D., {Feltzing}, S., {Koch}, A., {et~al.} 2009, \aap, 506, 1147,
  \dodoi{10.1051/0004-6361/200912718}

\bibitem[{Aparicio {et~al.}(2001)Aparicio, Carrera, \&
  {Mart{\'i}nez-Delgado}}]{aparicio01Star}
Aparicio, A., Carrera, R., \& {Mart{\'i}nez-Delgado}, D. 2001, AJ, 122, 2524,
  \dodoi{10.1086/323535}

\bibitem[{{Armandroff} {et~al.}(1995){Armandroff}, {Olszewski}, \&
  {Pryor}}]{1995AJ....110.2131A}
{Armandroff}, T.~E., {Olszewski}, E.~W., \& {Pryor}, C. 1995, \aj, 110, 2131,
  \dodoi{10.1086/117675}

\bibitem[{{Battaglia} {et~al.}(2011){Battaglia}, {Tolstoy}, {Helmi}, {Irwin},
  {Parisi}, {Hill}, \& {Jablonka}}]{2011MNRAS.411.1013B}
{Battaglia}, G., {Tolstoy}, E., {Helmi}, A., {et~al.} 2011, \mnras, 411, 1013,
  \dodoi{10.1111/j.1365-2966.2010.17745.x}

\bibitem[{{Battaglia} {et~al.}(2006){Battaglia}, {Tolstoy}, {Helmi}, {Irwin},
  {Letarte}, {Jablonka}, {Hill}, {Venn}, {Shetrone}, {Arimoto}, {Primas},
  {Kaufer}, {Francois}, {Szeifert}, {Abel}, \&
  {Sadakane}}]{2006A&A...459..423B}
---. 2006, \aap, 459, 423, \dodoi{10.1051/0004-6361:20065720}

\bibitem[{Besla {et~al.}(2012)Besla, Kallivayalil, Hernquist, {van der Marel},
  Cox, \& Kere{\v s}}]{besla12role}
Besla, G., Kallivayalil, N., Hernquist, L., {et~al.} 2012, MNRAS, 421, 2109,
  \dodoi{10.1111/j.1365-2966.2012.20466.x}

\bibitem[{Binney \& Tremaine(1987)}]{binney87Galactic}
Binney, J., \& Tremaine, S. 1987, Princeton, NJ, Princeton University Press,
  1987, 747 p.

\bibitem[{{Bland-Hawthorn} \& Gerhard(2016)}]{bland-hawthorn16Galaxy}
{Bland-Hawthorn}, J., \& Gerhard, O. 2016, ARA\&A, 54, 529,
  \dodoi{10.1146/annurev-astro-081915-023441}

\bibitem[{Bovy(2015)}]{bovy15galpy}
Bovy, J. 2015, ApJS, 216, 29, \dodoi{10.1088/0067-0049/216/2/29}

\bibitem[{{Caldwell} {et~al.}(2017){Caldwell}, {Walker}, {Mateo}, {Olszewski},
  {Koposov}, {Belokurov}, {Torrealba}, {Geringer-Sameth}, \&
  {Johnson}}]{2017ApJ...839...20C}
{Caldwell}, N., {Walker}, M.~G., {Mateo}, M., {et~al.} 2017, \apj, 839, 20,
  \dodoi{10.3847/1538-4357/aa688e}

\bibitem[{Cautun {et~al.}(2020)Cautun, {Ben{\'i}tez-Llambay}, Deason, Frenk,
  Fattahi, G{\'o}mez, Grand, Oman, Navarro, \& Simpson}]{cautun20milky}
Cautun, M., {Ben{\'i}tez-Llambay}, A., Deason, A.~J., {et~al.} 2020, MNRAS,
  494, 4291, \dodoi{10.1093/mnras/staa1017}

\bibitem[{{de Boer} {et~al.}(2012){de Boer}, Tolstoy, Hill, Saha, Olszewski,
  Mateo, Starkenburg, Battaglia, \& Walker}]{deboer12star}
{de Boer}, T. J.~L., Tolstoy, E., Hill, V., {et~al.} 2012, A\&A, 544, A73,
  \dodoi{10.1051/0004-6361/201219547}

\bibitem[{{de Salas} {et~al.}(2019){de Salas}, Malhan, Freese, Hattori, \&
  Valluri}]{desalas19estimation}
{de Salas}, P.~F., Malhan, K., Freese, K., Hattori, K., \& Valluri, M. 2019,
  JCAP, 10, 037, \dodoi{10.1088/1475-7516/2019/10/037}

\bibitem[{{Drlica-Wagner} {et~al.}(2020){Drlica-Wagner}, Bechtol, Mau, McNanna,
  Nadler, Pace, Li, Pieres, Rozo, Simon, Walker, Wechsler, Abbott, Allam,
  Annis, Bertin, Brooks, Burke, Rosell, Carrasco~Kind, Carretero, Costanzi, {da
  Costa}, De~Vicente, Desai, Diehl, Doel, Eifler, Everett, Flaugher, Frieman,
  {Garc{\'i}a-Bellido}, Gaztanaga, Gruen, Gruendl, Gschwend, Gutierrez,
  Honscheid, James, Krause, Kuehn, Kuropatkin, Lahav, Maia, Marshall, Melchior,
  Menanteau, Miquel, Palmese, Plazas, Sanchez, Scarpine, Schubnell, Serrano,
  {Sevilla-Noarbe}, Smith, Suchyta, Tarle, \& {DES
  Collaboration}}]{drlica-wagner20Milky}
{Drlica-Wagner}, A., Bechtol, K., Mau, S., {et~al.} 2020, ApJ, 893, 47,
  \dodoi{10.3847/1538-4357/ab7eb9}

\bibitem[{Eilers {et~al.}(2019)Eilers, Hogg, Rix, \& Ness}]{eilers19Circular}
Eilers, A.-C., Hogg, D.~W., Rix, H.-W., \& Ness, M.~K. 2019, ApJ, 871, 120,
  \dodoi{10.3847/1538-4357/aaf648}

\bibitem[{Erkal {et~al.}(2019)Erkal, Belokurov, Laporte, Koposov, Li,
  Grillmair, Kallivayalil, {Price-Whelan}, Evans, Hawkins, Hendel, Mateu,
  Navarro, {del Pino}, Slater, Sohn, \& {Orphan Aspen Treasury
  Collaboration}}]{erkal19total}
Erkal, D., Belokurov, V., Laporte, C. F.~P., {et~al.} 2019, MNRAS, 487, 2685,
  \dodoi{10.1093/mnras/stz1371}

\bibitem[{{Fox} {et~al.}(2014){Fox}, {Wakker}, {Barger}, {Hernandez},
  {Richter}, {Lehner}, {Bland-Hawthorn}, {Charlton}, {Westmeier}, {Thom},
  {Tumlinson}, {Misawa}, {Howk}, {Haffner}, {Ely}, {Rodriguez-Hidalgo}, \&
  {Kumari}}]{Fox2014}
{Fox}, A.~J., {Wakker}, B.~P., {Barger}, K.~A., {et~al.} 2014, \apj, 787, 147,
  \dodoi{10.1088/0004-637X/787/2/147}

\bibitem[{Fritz {et~al.}(2018)Fritz, Battaglia, Pawlowski, Kallivayalil,
  van~der Marel, Sohn, Brook, \& Besla}]{fritz18Gaia}
Fritz, T.~K., Battaglia, G., Pawlowski, M.~S., {et~al.} 2018, A\&A, 619, A103,
  \dodoi{10.1051/0004-6361/201833343}

\bibitem[{{Fritz} {et~al.}(2019){Fritz}, {Carrera}, {Battaglia}, \&
  {Taibi}}]{2019A&A...623A.129F}
{Fritz}, T.~K., {Carrera}, R., {Battaglia}, G., \& {Taibi}, S. 2019, \aap, 623,
  A129, \dodoi{10.1051/0004-6361/201833458}

\bibitem[{{Gaia Collaboration} {et~al.}(2020){Gaia Collaboration}, {Brown},
  {Vallenari}, {Prusti}, {de Bruijne}, {Babusiaux}, \&
  {Biermann}}]{2020arXiv201201533G}
{Gaia Collaboration}, {Brown}, A.~G.~A., {Vallenari}, A., {et~al.} 2020, arXiv
  e-prints, arXiv:2012.01533.
\newblock \doarXiv{2012.01533}

\bibitem[{{Gaia Collaboration} {et~al.}(2016){Gaia Collaboration}, {Prusti},
  {de Bruijne}, {Brown}, {Vallenari}, {Babusiaux}, {Bailer-Jones}, {Bastian},
  {Biermann}, {Evans}, {Eyer}, {Jansen}, {Jordi}, {Klioner}, {Lammers},
  {Lindegren}, {Luri}, {Mignard}, {Milligan}, {Panem}, {Poinsignon},
  {Pourbaix}, {Randich}, {Sarri}, {Sartoretti}, {Siddiqui}, {Soubiran},
  {Valette}, {van Leeuwen}, {Walton}, {Aerts}, {Arenou}, {Cropper}, {Drimmel},
  {H{\o}g}, {Katz}, {Lattanzi}, {O'Mullane}, {Grebel}, {Holland}, {Huc},
  {Passot}, {Bramante}, {Cacciari}, {Casta{\~n}eda}, {Chaoul}, {Cheek}, {De
  Angeli}, {Fabricius}, {Guerra}, {Hern{\'a}ndez}, {Jean-Antoine-Piccolo},
  {Masana}, {Messineo}, {Mowlavi}, {Nienartowicz}, {Ord{\'o}{\~n}ez-Blanco},
  {Panuzzo}, {Portell}, {Richards}, {Riello}, {Seabroke}, {Tanga},
  {Th{\'e}venin}, {Torra}, {Els}, {Gracia-Abril}, {Comoretto},
  {Garcia-Reinaldos}, {Lock}, {Mercier}, {Altmann}, {Andrae}, {Astraatmadja},
  {Bellas-Velidis}, {Benson}, {Berthier}, {Blomme}, {Busso}, {Carry},
  {Cellino}, {Clementini}, {Cowell}, {Creevey}, {Cuypers}, {Davidson}, {De
  Ridder}, {de Torres}, {Delchambre}, {Dell'Oro}, {Ducourant}, {Fr{\'e}mat},
  {Garc{\'\i}a-Torres}, {Gosset}, {Halbwachs}, {Hambly}, {Harrison}, {Hauser},
  {Hestroffer}, {Hodgkin}, {Huckle}, {Hutton}, {Jasniewicz}, {Jordan},
  {Kontizas}, {Korn}, {Lanzafame}, {Manteiga}, {Moitinho}, {Muinonen},
  {Osinde}, {Pancino}, {Pauwels}, {Petit}, {Recio-Blanco}, {Robin}, {Sarro},
  {Siopis}, {Smith}, {Smith}, {Sozzetti}, {Thuillot}, {van Reeven}, {Viala},
  {Abbas}, {Abreu Aramburu}, {Accart}, {Aguado}, {Allan}, {Allasia},
  {Altavilla}, {{\'A}lvarez}, {Alves}, {Anderson}, {Andrei}, {Anglada Varela},
  {Antiche}, {Antoja}, {Ant{\'o}n}, {Arcay}, {Atzei}, {Ayache}, {Bach},
  {Baker}, {Balaguer-N{\'u}{\~n}ez}, {Barache}, {Barata}, {Barbier}, {Barblan},
  {Baroni}, {Barrado y Navascu{\'e}s}, {Barros}, {Barstow}, {Becciani},
  {Bellazzini}, {Bellei}, {Bello Garc{\'\i}a}, {Belokurov}, {Bendjoya},
  {Berihuete}, {Bianchi}, {Bienaym{\'e}}, {Billebaud}, {Blagorodnova},
  {Blanco-Cuaresma}, {Boch}, {Bombrun}, {Borrachero}, {Bouquillon}, {Bourda},
  {Bouy}, {Bragaglia}, {Breddels}, {Brouillet}, {Br{\"u}semeister},
  {Bucciarelli}, {Budnik}, {Burgess}, {Burgon}, {Burlacu}, {Busonero}, {Buzzi},
  {Caffau}, {Cambras}, {Campbell}, {Cancelliere}, {Cantat-Gaudin}, {Carlucci},
  {Carrasco}, {Castellani}, {Charlot}, {Charnas}, {Charvet}, {Chassat},
  {Chiavassa}, {Clotet}, {Cocozza}, {Collins}, {Collins}, {Costigan}, {Crifo},
  {Cross}, {Crosta}, {Crowley}, {Dafonte}, {Damerdji}, {Dapergolas}, {David},
  {David}, {De Cat}, {de Felice}, {de Laverny}, {De Luise}, {De March}, {de
  Martino}, {de Souza}, {Debosscher}, {del Pozo}, {Delbo}, {Delgado},
  {Delgado}, {di Marco}, {Di Matteo}, {Diakite}, {Distefano}, {Dolding}, {Dos
  Anjos}, {Drazinos}, {Dur{\'a}n}, {Dzigan}, {Ecale}, {Edvardsson}, {Enke},
  {Erdmann}, {Escolar}, {Espina}, {Evans}, {Eynard Bontemps}, {Fabre},
  {Fabrizio}, {Faigler}, {Falc{\~a}o}, {Farr{\`a}s Casas}, {Faye}, {Federici},
  {Fedorets}, {Fern{\'a}ndez-Hern{\'a}ndez}, {Fernique}, {Fienga}, {Figueras},
  {Filippi}, {Findeisen}, {Fonti}, {Fouesneau}, {Fraile}, {Fraser}, {Fuchs},
  {Furnell}, {Gai}, {Galleti}, {Galluccio}, {Garabato}, {Garc{\'\i}a-Sedano},
  {Gar{\'e}}, {Garofalo}, {Garralda}, {Gavras}, {Gerssen}, {Geyer}, {Gilmore},
  {Girona}, {Giuffrida}, {Gomes}, {Gonz{\'a}lez-Marcos},
  {Gonz{\'a}lez-N{\'u}{\~n}ez}, {Gonz{\'a}lez-Vidal}, {Granvik}, {Guerrier},
  {Guillout}, {Guiraud}, {G{\'u}rpide}, {Guti{\'e}rrez-S{\'a}nchez}, {Guy},
  {Haigron}, {Hatzidimitriou}, {Haywood}, {Heiter}, {Helmi}, {Hobbs},
  {Hofmann}, {Holl}, {Holland}, {Hunt}, {Hypki}, {Icardi}, {Irwin}, {Jevardat
  de Fombelle}, {Jofr{\'e}}, {Jonker}, {Jorissen}, {Julbe}, {Karampelas},
  {Kochoska}, {Kohley}, {Kolenberg}, {Kontizas}, {Koposov}, {Kordopatis},
  {Koubsky}, {Kowalczyk}, {Krone-Martins}, {Kudryashova}, {Kull}, {Bachchan},
  {Lacoste-Seris}, {Lanza}, {Lavigne}, {Le Poncin-Lafitte}, {Lebreton},
  {Lebzelter}, {Leccia}, {Leclerc}, {Lecoeur-Taibi}, {Lemaitre}, {Lenhardt},
  {Leroux}, {Liao}, {Licata}, {Lindstr{\o}m}, {Lister}, {Livanou}, {Lobel},
  {L{\"o}ffler}, {L{\'o}pez}, {Lopez-Lozano}, {Lorenz}, {Loureiro},
  {MacDonald}, {Magalh{\~a}es Fernandes}, {Managau}, {Mann}, {Mantelet},
  {Marchal}, {Marchant}, {Marconi}, {Marie}, {Marinoni}, {Marrese},
  {Marschalk{\'o}}, {Marshall}, {Mart{\'\i}n-Fleitas}, {Martino}, {Mary},
  {Matijevi{\v{c}}}, {Mazeh}, {McMillan}, {Messina}, {Mestre}, {Michalik},
  {Millar}, {Miranda}, {Molina}, {Molinaro}, {Molinaro}, {Moln{\'a}r},
  {Moniez}, {Montegriffo}, {Monteiro}, {Mor}, {Mora}, {Morbidelli}, {Morel},
  {Morgenthaler}, {Morley}, {Morris}, {Mulone}, {Muraveva}, {Musella},
  {Narbonne}, {Nelemans}, {Nicastro}, {Noval}, {Ord{\'e}novic},
  {Ordieres-Mer{\'e}}, {Osborne}, {Pagani}, {Pagano}, {Pailler}, {Palacin},
  {Palaversa}, {Parsons}, {Paulsen}, {Pecoraro}, {Pedrosa}, {Pentik{\"a}inen},
  {Pereira}, {Pichon}, {Piersimoni}, {Pineau}, {Plachy}, {Plum}, {Poujoulet},
  {Pr{\v{s}}a}, {Pulone}, {Ragaini}, {Rago}, {Rambaux}, {Ramos-Lerate},
  {Ranalli}, {Rauw}, {Read}, {Regibo}, {Renk}, {Reyl{\'e}}, {Ribeiro},
  {Rimoldini}, {Ripepi}, {Riva}, {Rixon}, {Roelens}, {Romero-G{\'o}mez},
  {Rowell}, {Royer}, {Rudolph}, {Ruiz-Dern}, {Sadowski}, {Sagrist{\`a}
  Sell{\'e}s}, {Sahlmann}, {Salgado}, {Salguero}, {Sarasso}, {Savietto},
  {Schnorhk}, {Schultheis}, {Sciacca}, {Segol}, {Segovia}, {Segransan},
  {Serpell}, {Shih}, {Smareglia}, {Smart}, {Smith}, {Solano}, {Solitro},
  {Sordo}, {Soria Nieto}, {Souchay}, {Spagna}, {Spoto}, {Stampa}, {Steele},
  {Steidelm{\"u}ller}, {Stephenson}, {Stoev}, {Suess}, {S{\"u}veges}, {Surdej},
  {Szabados}, {Szegedi-Elek}, {Tapiador}, {Taris}, {Tauran}, {Taylor},
  {Teixeira}, {Terrett}, {Tingley}, {Trager}, {Turon}, {Ulla}, {Utrilla},
  {Valentini}, {van Elteren}, {Van Hemelryck}, {van Leeuwen}, {Varadi},
  {Vecchiato}, {Veljanoski}, {Via}, {Vicente}, {Vogt}, {Voss}, {Votruba},
  {Voutsinas}, {Walmsley}, {Weiler}, {Weingrill}, {Werner}, {Wevers},
  {Whitehead}, {Wyrzykowski}, {Yoldas}, {{\v{Z}}erjal}, {Zucker}, {Zurbach},
  {Zwitter}, {Alecu}, {Allen}, {Allende Prieto}, {Amorim},
  {Anglada-Escud{\'e}}, {Arsenijevic}, {Azaz}, {Balm}, {Beck}, {Bernstein},
  {Bigot}, {Bijaoui}, {Blasco}, {Bonfigli}, {Bono}, {Boudreault}, {Bressan},
  {Brown}, {Brunet}, {Bunclark}, {Buonanno}, {Butkevich}, {Carret}, {Carrion},
  {Chemin}, {Ch{\'e}reau}, {Corcione}, {Darmigny}, {de Boer}, {de Teodoro}, {de
  Zeeuw}, {Delle Luche}, {Domingues}, {Dubath}, {Fodor}, {Fr{\'e}zouls},
  {Fries}, {Fustes}, {Fyfe}, {Gallardo}, {Gallegos}, {Gardiol}, {Gebran},
  {Gomboc}, {G{\'o}mez}, {Grux}, {Gueguen}, {Heyrovsky}, {Hoar}, {Iannicola},
  {Isasi Parache}, {Janotto}, {Joliet}, {Jonckheere}, {Keil}, {Kim},
  {Klagyivik}, {Klar}, {Knude}, {Kochukhov}, {Kolka}, {Kos}, {Kutka}, {Lainey},
  {LeBouquin}, {Liu}, {Loreggia}, {Makarov}, {Marseille}, {Martayan},
  {Martinez-Rubi}, {Massart}, {Meynadier}, {Mignot}, {Munari}, {Nguyen},
  {Nordlander}, {Ocvirk}, {O'Flaherty}, {Olias Sanz}, {Ortiz}, {Osorio},
  {Oszkiewicz}, {Ouzounis}, {Palmer}, {Park}, {Pasquato}, {Peltzer}, {Peralta},
  {P{\'e}turaud}, {Pieniluoma}, {Pigozzi}, {Poels}, {Prat}, {Prod'homme},
  {Raison}, {Rebordao}, {Risquez}, {Rocca-Volmerange}, {Rosen}, {Ruiz-Fuertes},
  {Russo}, {Sembay}, {Serraller Vizcaino}, {Short}, {Siebert}, {Silva},
  {Sinachopoulos}, {Slezak}, {Soffel}, {Sosnowska}, {Strai{\v{z}}ys}, {ter
  Linden}, {Terrell}, {Theil}, {Tiede}, {Troisi}, {Tsalmantza}, {Tur},
  {Vaccari}, {Vachier}, {Valles}, {Van Hamme}, {Veltz}, {Virtanen}, {Wallut},
  {Wichmann}, {Wilkinson}, {Ziaeepour}, \& {Zschocke}}]{2016A&A...595A...1G}
{Gaia Collaboration}, {Prusti}, T., {de Bruijne}, J.~H.~J., {et~al.} 2016,
  \aap, 595, A1, \dodoi{10.1051/0004-6361/201629272}

\bibitem[{{Gravity Collaboration} {et~al.}(2018){Gravity Collaboration},
  Abuter, Amorim, Anugu, Baub{\"o}ck, Benisty, Berger, Blind, Bonnet, Brandner,
  Buron, Collin, Chapron, Cl{\'e}net, Coud{\'e} Du~Foresto, {de Zeeuw}, Deen,
  {Delplancke-Str{\"o}bele}, Dembet, Dexter, Duvert, Eckart, Eisenhauer,
  Finger, F{\"o}rster~Schreiber, F{\'e}dou, Garcia, Garcia~Lopez, Gao, Gendron,
  Genzel, Gillessen, Gordo, Habibi, Haubois, Haug, Hau{\ss}mann, Henning,
  Hippler, Horrobin, Hubert, Hubin, Jimenez~Rosales, Jochum, Jocou, Kaufer,
  Kellner, Kendrew, Kervella, Kok, Kulas, Lacour, Lapeyr{\`e}re, Lazareff,
  Le~Bouquin, L{\'e}na, Lippa, Lenzen, M{\'e}rand, M{\"u}ler, Neumann, Ott,
  Palanca, Paumard, Pasquini, Perraut, Perrin, Pfuhl, Plewa, Rabien,
  Ram{\'i}rez, Ramos, Rau, {Rodr{\'i}guez-Coira}, Rohloff, Rousset,
  {Sanchez-Bermudez}, Scheithauer, Sch{\"o}ller, Schuler, Spyromilio, Straub,
  Straubmeier, Sturm, Tacconi, Tristram, Vincent, {von Fellenberg}, Wank,
  Waisberg, Widmann, Wieprecht, Wiest, Wiezorrek, Woillez, Yazici, Ziegler, \&
  Zins}]{gravitycollaboration18Detection}
{Gravity Collaboration}, Abuter, R., Amorim, A., {et~al.} 2018, A\&A, 615, L15,
  \dodoi{10.1051/0004-6361/201833718}

\bibitem[{{Hammer} {et~al.}(2021){Hammer}, {Wang}, {Pawlowski}, {Yang},
  {Bonifacio}, {Li}, {Babusiaux}, \& {Arenou}}]{hammer21}
{Hammer}, F., {Wang}, J., {Pawlowski}, M.~S., {et~al.} 2021, submitted to ApJ

\bibitem[{{Hammer} {et~al.}(2020){Hammer}, {Yang}, {Arenou}, {Wang}, {Li},
  {Bonifacio}, \& {Babusiaux}}]{Hammer2020}
{Hammer}, F., {Yang}, Y., {Arenou}, F., {et~al.} 2020, The Astrophysical
  Journal, 892, 3, \dodoi{10.3847/1538-4357/ab77be}

\bibitem[{Hammer {et~al.}(2019)Hammer, Yang, Wang, Arenou, Puech, Flores, \&
  Babusiaux}]{hammer19Absence}
Hammer, F., Yang, Y., Wang, J., {et~al.} 2019, ApJ, 883, 171,
  \dodoi{10.3847/1538-4357/ab36b6}

\bibitem[{Hammer {et~al.}(2015)Hammer, Yang, Flores, Puech, \&
  Fouquet}]{hammer15Magellanic}
Hammer, F., Yang, Y.~B., Flores, H., Puech, M., \& Fouquet, S. 2015, ApJ, 813,
  110, \dodoi{10.1088/0004-637X/813/2/110}

\bibitem[{Helmi {et~al.}(2018)Helmi, van Leeuwen, McMillan, Massari, Antoja,
  Robin, Lindegren, Bastian, Arenou, Babusiaux, Biermann, Breddels, Hobbs,
  Jordi, Pancino, Reyl{\'e}, Veljanoski, Brown, Vallenari, Prusti, de~Bruijne,
  {Bailer-Jones}, Evans, Eyer, Jansen, Klioner, Lammers, Luri, Mignard, Panem,
  Pourbaix, Randich, Sartoretti, Siddiqui, Soubiran, Walton, Cropper, Drimmel,
  Katz, Lattanzi, Bakker, Cacciari, Casta{\~n}eda, Chaoul, Cheek, Angeli,
  Fabricius, Guerra, Holl, Masana, Messineo, Mowlavi, Nienartowicz, Panuzzo,
  Portell, Riello, Seabroke, Tanga, Th{\'e}venin, {Gracia-Abril}, Comoretto,
  {Garcia-Reinaldos}, Teyssier, Altmann, Andrae, Audard, {Bellas-Velidis},
  Benson, Berthier, Blomme, Burgess, Busso, Carry, Cellino, Clementini, Clotet,
  Creevey, Davidson, Ridder, Delchambre, Dell'Oro, Ducourant,
  {Fern{\'a}ndez-Hern{\'a}ndez}, Fouesneau, Fr{\'e}mat, Galluccio,
  {Garc{\'i}a-Torres}, {Gonz{\'a}lez-N{\'u}{\~n}ez}, {Gonz{\'a}lez-Vidal},
  Gosset, Guy, Halbwachs, Hambly, Harrison, Hern{\'a}ndez, Hestroffer, Hodgkin,
  Hutton, Jasniewicz, {Jean-Antoine-Piccolo}, Jordan, Korn, {Krone-Martins},
  Lanzafame, Lebzelter, L{\"o}ffler, Manteiga, Marrese, {Mart{\'i}n-Fleitas},
  Moitinho, Mora, Muinonen, Osinde, Pauwels, Petit, {Recio-Blanco}, Richards,
  Rimoldini, Sarro, Siopis, Smith, Sozzetti, S{\"u}veges, Torra, van Reeven,
  Abbas, Aramburu, Accart, Aerts, Altavilla, {\'A}lvarez, Alvarez, Alves,
  Anderson, Andrei, Varela, Antiche, Arcay, Astraatmadja, Bach, Baker,
  {Balaguer-N{\'u}{\~n}ez}, Balm, Barache, Barata, Barbato, Barblan, Barklem,
  Barrado, Barros, Barstow, Mu{\~n}oz, Bassilana, Becciani, Bellazzini,
  Berihuete, Bertone, Bianchi, Bienaym{\'e}, {Blanco-Cuaresma}, Boch, Boeche,
  Bombrun, Borrachero, Bossini, Bouquillon, Bourda, Bragaglia, Bramante,
  Bressan, Brouillet, Br{\"u}semeister, Brugaletta, Bucciarelli, Burlacu,
  Busonero, Butkevich, Buzzi, Caffau, Cancelliere, Cannizzaro, {Cantat-Gaudin},
  Carballo, Carlucci, Carrasco, Casamiquela, Castellani, {Castro-Ginard},
  Charlot, Chemin, Chiavassa, Cocozza, Costigan, Cowell, Crifo, Crosta,
  Crowley, Cuypers, Dafonte, Damerdji, Dapergolas, David, David, de~Laverny,
  Luise, March, de~Martino, de~Souza, de~Torres, Debosscher, del Pozo, Delbo,
  Delgado, Delgado, Matteo, Diakite, Diener, Distefano, Dolding, Drazinos,
  Dur{\'a}n, Edvardsson, Enke, Eriksson, Esquej, Bontemps, Fabre, Fabrizio,
  Faigler, Falc{\~a}o, Casas, Federici, Fedorets, Fernique, Figueras, Filippi,
  Findeisen, Fonti, Fraile, Fraser, Fr{\'e}zouls, Gai, Galleti, Garabato,
  {Garc{\'i}a-Sedano}, Garofalo, Garralda, Gavel, Gavras, Gerssen, Geyer,
  Giacobbe, Gilmore, Girona, Giuffrida, Glass, Gomes, Granvik, Gueguen,
  Guerrier, Guiraud, {Guti{\'e}rrez-S{\'a}nchez}, Haigron, Hatzidimitriou,
  Hauser, Haywood, Heiter, Heu, Hilger, Hofmann, Holland, Huckle, Hypki,
  Icardi, Jan{\ss}en, de~Fombelle, Jonker, Juh{\'a}sz, Julbe, Karampelas,
  Kewley, Klar, Kochoska, Kohley, Kolenberg, Kontizas, Kontizas, Koposov,
  Kordopatis, {Kostrzewa-Rutkowska}, Koubsky, Lambert, Lanza, Lasne, Lavigne,
  Fustec, {Poncin-Lafitte}, Lebreton, Leccia, Leclerc, {Lecoeur-Taibi},
  Lenhardt, Leroux, Liao, Licata, Lindstr{\o}m, Lister, Livanou, Lobel,
  L{\'o}pez, Managau, Mann, Mantelet, Marchal, Marchant, Marconi, Marinoni,
  Marschalk{\'o}, Marshall, Martino, Marton, Mary, Matijevi{\v c}, Mazeh,
  Messina, Michalik, Millar, Molina, Molinaro, Moln{\'a}r, Montegriffo, Mor,
  Morbidelli, Morel, Morris, Mulone, Muraveva, Musella, Nelemans, Nicastro,
  Noval, O'Mullane, Ord{\'e}novic, {Ord{\'o}{\~n}ez-Blanco}, Osborne, Pagani,
  Pagano, Pailler, Palacin, Palaversa, Panahi, Pawlak, Piersimoni, Pineau,
  Plachy, Plum, Poggio, Poujoulet, Pr{\v s}a, Pulone, Racero, Ragaini, Rambaux,
  {Ramos-Lerate}, Regibo, Riclet, Ripepi, Riva, Rivard, Rixon, Roegiers,
  Roelens, {Romero-G{\'o}mez}, Rowell, Royer, {Ruiz-Dern}, Sadowski,
  Sell{\'e}s, Sahlmann, Salgado, Salguero, Sanna, {Santana-Ros}, Sarasso,
  Savietto, Schultheis, Sciacca, Segol, Segovia, S{\'e}gransan, Shih, Siltala,
  Silva, Smart, Smith, Solano, Solitro, Sordo, Nieto, Souchay, Spagna, Spoto,
  Stampa, Steele, Steidelm{\"u}ller, Stephenson, Stoev, Suess, Surdej,
  Szabados, {Szegedi-Elek}, Tapiador, Taris, Tauran, Taylor, Teixeira, Terrett,
  Teyssandier, Thuillot, Titarenko, Clotet, Turon, Ulla, Utrilla, Uzzi,
  Vaillant, Valentini, Valette, van Elteren, Hemelryck, van Leeuwen, Vaschetto,
  Vecchiato, Viala, Vicente, Vogt, von Essen, Voss, Votruba, Voutsinas,
  Walmsley, Weiler, Wertz, Wevems, Wyrzykowski, Yoldas, {\v Z}erjal, Ziaeepour,
  Zorec, Zschocke, Zucker, Zurbach, \& Zwitter}]{helmi18Gaia}
Helmi, A., van Leeuwen, F., McMillan, P.~J., {et~al.} 2018, A\&A, 616, A12,
  \dodoi{10.1051/0004-6361/201832698}

\bibitem[{{Hill} {et~al.}(2019){Hill}, {Sk{\'u}lad{\'o}ttir}, {Tolstoy},
  {Venn}, {Shetrone}, {Jablonka}, {Primas}, {Battaglia}, {de Boer},
  {Fran{\c{c}}ois}, {Helmi}, {Kaufer}, {Letarte}, {Starkenburg}, \&
  {Spite}}]{2019A&A...626A..15H}
{Hill}, V., {Sk{\'u}lad{\'o}ttir}, {\'A}., {Tolstoy}, E., {et~al.} 2019, \aap,
  626, A15, \dodoi{10.1051/0004-6361/201833950}

\bibitem[{{Jenkins} {et~al.}(2020){Jenkins}, {Li}, {Pace}, {Ji}, {Koposov}, \&
  {Mutlu-Pakdil}}]{2021arXiv210100013J}
{Jenkins}, S., {Li}, T.~S., {Pace}, A.~B., {et~al.} 2020, arXiv e-prints,
  arXiv:2101.00013.
\newblock \doarXiv{2101.00013}

\bibitem[{Jiao {et~al.}(2021)Jiao, Hammer, Wang, \& Yang}]{jiao21Which}
Jiao, Y., Hammer, F., Wang, J., \& Yang, Y. 2021, arXiv e-prints, accepted in
  Astronomy \& Astrophysics, arXiv:2107.00014

\bibitem[{Johnson \& Soderblom(1987)}]{johnson87Calculating}
Johnson, D. R.~H., \& Soderblom, D.~R. 1987, AJ, 93, 864,
  \dodoi{10.1086/114370}

\bibitem[{Juri{\'c} {et~al.}(2008)Juri{\'c}, Ivezi{\'c}, Brooks, Lupton,
  Schlegel, Finkbeiner, Padmanabhan, Bond, Sesar, Rockosi, Knapp, Gunn, Sumi,
  Schneider, Barentine, Brewington, Brinkmann, Fukugita, Harvanek, Kleinman,
  Krzesinski, Long, Neilsen, Nitta, Snedden, \& York}]{juric08Milky}
Juri{\'c}, M., Ivezi{\'c}, {\v Z}., Brooks, A., {et~al.} 2008, ApJ, 673, 864,
  \dodoi{10.1086/523619}

\bibitem[{{Kacharov} {et~al.}(2017){Kacharov}, {Battaglia}, {Rejkuba}, {Cole},
  {Carrera}, {Fraternali}, {Wilkinson}, {Gallart}, {Irwin}, \&
  {Tolstoy}}]{2017MNRAS.466.2006K}
{Kacharov}, N., {Battaglia}, G., {Rejkuba}, M., {et~al.} 2017, \mnras, 466,
  2006, \dodoi{10.1093/mnras/stw3188}

\bibitem[{Kallivayalil {et~al.}(2013)Kallivayalil, {van der Marel}, Besla,
  Anderson, \& Alcock}]{kallivayalil13Thirdepoch}
Kallivayalil, N., {van der Marel}, R.~P., Besla, G., Anderson, J., \& Alcock,
  C. 2013, ApJ, 764, 161, \dodoi{10.1088/0004-637X/764/2/161}

\bibitem[{Karukes {et~al.}(2020)Karukes, Benito, Iocco, Trotta, \&
  {Geringer-Sameth}}]{karukes20robust}
Karukes, E.~V., Benito, M., Iocco, F., Trotta, R., \& {Geringer-Sameth}, A.
  2020, JCAP, 05, 033, \dodoi{10.1088/1475-7516/2020/05/033}

\bibitem[{{Kirby} {et~al.}(2013){Kirby}, {Boylan-Kolchin}, {Cohen}, {Geha},
  {Bullock}, \& {Kaplinghat}}]{2013ApJ...770...16K}
{Kirby}, E.~N., {Boylan-Kolchin}, M., {Cohen}, J.~G., {et~al.} 2013, \apj, 770,
  16, \dodoi{10.1088/0004-637X/770/1/16}

\bibitem[{{Kirby} {et~al.}(2017){Kirby}, {Cohen}, {Simon}, {Guhathakurta},
  {Thygesen}, \& {Duggan}}]{2017ApJ...838...83K}
{Kirby}, E.~N., {Cohen}, J.~G., {Simon}, J.~D., {et~al.} 2017, \apj, 838, 83,
  \dodoi{10.3847/1538-4357/aa6570}

\bibitem[{{Kirby} {et~al.}(2015){Kirby}, {Simon}, \&
  {Cohen}}]{2015ApJ...810...56K}
{Kirby}, E.~N., {Simon}, J.~D., \& {Cohen}, J.~G. 2015, \apj, 810, 56,
  \dodoi{10.1088/0004-637X/810/1/56}

\bibitem[{{Kleyna} {et~al.}(2002){Kleyna}, {Wilkinson}, {Evans}, {Gilmore}, \&
  {Frayn}}]{2002MNRAS.330..792K}
{Kleyna}, J., {Wilkinson}, M.~I., {Evans}, N.~W., {Gilmore}, G., \& {Frayn}, C.
  2002, \mnras, 330, 792, \dodoi{10.1046/j.1365-8711.2002.05155.x}

\bibitem[{{Koch} {et~al.}(2007){Koch}, {Kleyna}, {Wilkinson}, {Grebel},
  {Gilmore}, {Evans}, {Wyse}, \& {Harbeck}}]{2007AJ....134..566K}
{Koch}, A., {Kleyna}, J.~T., {Wilkinson}, M.~I., {et~al.} 2007, \aj, 134, 566,
  \dodoi{10.1086/519380}

\bibitem[{{Koch} {et~al.}(2009){Koch}, {Wilkinson}, {Kleyna}, {Irwin},
  {Zucker}, {Belokurov}, {Gilmore}, {Fellhauer}, \&
  {Evans}}]{2009ApJ...690..453K}
{Koch}, A., {Wilkinson}, M.~I., {Kleyna}, J.~T., {et~al.} 2009, \apj, 690, 453,
  \dodoi{10.1088/0004-637X/690/1/453}

\bibitem[{{Koposov} {et~al.}(2011){Koposov}, {Gilmore}, {Walker}, {Belokurov},
  {Evans}, {Fellhauer}, {Gieren}, {Geisler}, {Monaco}, {Norris}, {Okamoto},
  {Pe{\~n}arrubia}, {Wilkinson}, {Wyse}, \& {Zucker}}]{2011ApJ...736..146K}
{Koposov}, S.~E., {Gilmore}, G., {Walker}, M.~G., {et~al.} 2011, \apj, 736,
  146, \dodoi{10.1088/0004-637X/736/2/146}

\bibitem[{{Koposov} {et~al.}(2015){Koposov}, {Casey}, {Belokurov}, {Lewis},
  {Gilmore}, {Worley}, {Hourihane}, {Randich}, {Bensby}, {Bragaglia},
  {Bergemann}, {Carraro}, {Costado}, {Flaccomio}, {Francois}, {Heiter}, {Hill},
  {Jofre}, {Lando}, {Lanzafame}, {de Laverny}, {Monaco}, {Morbidelli},
  {Sbordone}, {Mikolaitis}, \& {Ryde}}]{2015ApJ...811...62K}
{Koposov}, S.~E., {Casey}, A.~R., {Belokurov}, V., {et~al.} 2015, \apj, 811,
  62, \dodoi{10.1088/0004-637X/811/1/62}

\bibitem[{{Koposov} {et~al.}(2018){Koposov}, {Walker}, {Belokurov}, {Casey},
  {Geringer-Sameth}, {Mackey}, {Da Costa}, {Erkal}, {Jethwa}, {Mateo},
  {Olszewski}, \& {Bailey}}]{2018MNRAS.479.5343K}
{Koposov}, S.~E., {Walker}, M.~G., {Belokurov}, V., {et~al.} 2018, \mnras, 479,
  5343, \dodoi{10.1093/mnras/sty1772}

\bibitem[{{Li} {et~al.}(2017){Li}, {Simon}, {Drlica-Wagner}, {Bechtol}, {Wang},
  {Garc{\'\i}a-Bellido}, {Frieman}, {Marshall}, {James}, {Strigari}, {Pace},
  {Balbinot}, {Zhang}, {Abbott}, {Allam}, {Benoit-L{\'e}vy}, {Bernstein},
  {Bertin}, {Brooks}, {Burke}, {Carnero Rosell}, {Carrasco Kind}, {Carretero},
  {Cunha}, {D'Andrea}, {da Costa}, {DePoy}, {Desai}, {Diehl}, {Eifler},
  {Flaugher}, {Goldstein}, {Gruen}, {Gruendl}, {Gschwend}, {Gutierrez},
  {Krause}, {Kuehn}, {Lin}, {Maia}, {March}, {Menanteau}, {Miquel}, {Plazas},
  {Romer}, {Sanchez}, {Santiago}, {Schubnell}, {Sevilla-Noarbe}, {Smith},
  {Sobreira}, {Suchyta}, {Tarle}, {Thomas}, {Tucker}, {Walker}, {Wechsler},
  {Wester}, {Yanny}, \& {DES Collaboration}}]{2017ApJ...838....8L}
{Li}, T.~S., {Simon}, J.~D., {Drlica-Wagner}, A., {et~al.} 2017, \apj, 838, 8,
  \dodoi{10.3847/1538-4357/aa6113}

\bibitem[{{Li} {et~al.}(2018){Li}, {Simon}, {Kuehn}, {Pace}, {Erkal},
  {Bechtol}, {Yanny}, {Drlica-Wagner}, {Marshall}, {Lidman}, {Balbinot},
  {Carollo}, {Jenkins}, {Mart{\'\i}nez-V{\'a}zquez}, {Shipp}, {Stringer},
  {Vivas}, {Walker}, {Wechsler}, {Abdalla}, {Allam}, {Annis}, {Avila},
  {Bertin}, {Brooks}, {Buckley-Geer}, {Burke}, {Carnero Rosell}, {Carrasco
  Kind}, {Carretero}, {Cunha}, {D'Andrea}, {da Costa}, {Davis}, {De Vicente},
  {Doel}, {Eifler}, {Evrard}, {Flaugher}, {Frieman}, {Garc{\'\i}a-Bellido},
  {Gaztanaga}, {Gerdes}, {Gruen}, {Gruendl}, {Gschwend}, {Gutierrez},
  {Hartley}, {Hollowood}, {Honscheid}, {James}, {Krause}, {Maia}, {March},
  {Menanteau}, {Miquel}, {Plazas}, {Sanchez}, {Santiago}, {Scarpine},
  {Schindler}, {Schubnell}, {Sevilla-Noarbe}, {Smith}, {Smith},
  {Soares-Santos}, {Sobreira}, {Suchyta}, {Swanson}, {Tarle}, {Tucker}, \& {DES
  Collaboration}}]{Li2018}
{Li}, T.~S., {Simon}, J.~D., {Kuehn}, K., {et~al.} 2018, \apj, 866, 22,
  \dodoi{10.3847/1538-4357/aadf91}

\bibitem[{Lindegren {et~al.}(2020)Lindegren, Klioner, Hern{\'a}ndez, Bombrun,
  {Ramos-Lerate}, Steidelm{\"u}ller, Bastian, Biermann, {de Torres}, Gerlach,
  Geyer, Hilger, Hobbs, Lammers, McMillan, Stephenson, Casta{\~n}eda, Davidson,
  Fabricius, {Gracia-Abril}, Portell, Rowell, Teyssier, Torra, Bartolom{\'e},
  Clotet, Garralda, {Gonz{\'a}lez-Vidal}, Torra, Abbas, Altmann,
  Anglada~Varela, {Balaguer-N{\'u}{\~n}ez}, Balog, Barache, Becciani, Bernet,
  Bertone, Bianchi, Bouquillon, Brown, Bucciarelli, Busonero, Butkevich, Buzzi,
  Cancelliere, Carlucci, Charlot, Cioni, Crosta, Crowley, {del Peloso}, {del
  Pozo}, Drimmel, Esquej, Fienga, Fraile, Gai, {Garcia-Reinaldos}, Guerra,
  Hambly, Hauser, Jan{\ss}en, Jordan, {Kostrzewa-Rutkowska}, Lattanzi, Liao,
  Licata, Lister, L{\"o}ffler, Marchant, Masip, Mignard, Mints, Molina, Mora,
  Morbidelli, Murphy, Pagani, Panuzzo, Pe{\~n}alosa~Esteller, Poggio,
  Re~Fiorentin, Riva, Sagrist{\`a}~Sell{\'e}s, Sanchez~Gimenez, Sarasso,
  Sciacca, Siddiqui, Smart, Souami, Spagna, Steele, Taris, Utrilla, {van
  Reeven}, \& Vecchiato}]{lindegren20Gaiaa}
Lindegren, L., Klioner, S.~A., Hern{\'a}ndez, J., {et~al.} 2020, arXiv
  e-prints, arXiv:2012.03380

\bibitem[{{Longeard} {et~al.}(2020){Longeard}, {Martin}, {Starkenburg},
  {Ibata}, {Collins}, {Laevens}, {Mackey}, {Rich}, {Aguado}, {Arentsen},
  {Jablonka}, {Gonz{\'a}lez Hern{\'a}ndez}, {Navarro}, \&
  {S{\'a}nchez-Janssen}}]{2020MNRAS.491..356L}
{Longeard}, N., {Martin}, N., {Starkenburg}, E., {et~al.} 2020, \mnras, 491,
  356, \dodoi{10.1093/mnras/stz2854}

\bibitem[{{Martin} {et~al.}(2007){Martin}, {Ibata}, {Chapman}, {Irwin}, \&
  {Lewis}}]{2007MNRAS.380..281M}
{Martin}, N.~F., {Ibata}, R.~A., {Chapman}, S.~C., {Irwin}, M., \& {Lewis},
  G.~F. 2007, \mnras, 380, 281, \dodoi{10.1111/j.1365-2966.2007.12055.x}

\bibitem[{{Martin} {et~al.}(2016){Martin}, {Geha}, {Ibata}, {Collins},
  {Laevens}, {Bell}, {Rix}, {Ferguson}, {Chambers}, {Wainscoat}, \&
  {Waters}}]{2016MNRAS.458L..59M}
{Martin}, N.~F., {Geha}, M., {Ibata}, R.~A., {et~al.} 2016, \mnras, 458, L59,
  \dodoi{10.1093/mnrasl/slw013}

\bibitem[{{Mateo} {et~al.}(2008){Mateo}, {Olszewski}, \&
  {Walker}}]{2008ApJ...675..201M}
{Mateo}, M., {Olszewski}, E.~W., \& {Walker}, M.~G. 2008, \apj, 675, 201,
  \dodoi{10.1086/522326}

\bibitem[{McConnachie \& Venn(2020)}]{mcconnachie20Updated}
McConnachie, A.~W., \& Venn, K.~A. 2020, arXiv e-prints, arXiv:2012.03904

\bibitem[{Miyamoto \& Nagai(1975)}]{miyamoto75Threedimensional}
Miyamoto, M., \& Nagai, R. 1975, Publications of the Astronomical Society of
  Japan, 27, 533

\bibitem[{Mr{\'o}z {et~al.}(2019)Mr{\'o}z, Udalski, Skowron, Skowron,
  Soszy{\'n}ski, Pietrukowicz, Szyma{\'n}ski, Poleski, Koz{\l}owski, \&
  Ulaczyk}]{mroz19Rotation}
Mr{\'o}z, P., Udalski, A., Skowron, D.~M., {et~al.} 2019, ApJL, 870, L10,
  \dodoi{10.3847/2041-8213/aaf73f}

\bibitem[{{Mu{\~n}oz} {et~al.}(2006){Mu{\~n}oz}, {Majewski}, {Zaggia},
  {Kunkel}, {Frinchaboy}, {Nidever}, {Crnojevic}, {Patterson}, {Crane},
  {Johnston}, {Sohn}, {Bernstein}, \& {Shectman}}]{2006ApJ...649..201M}
{Mu{\~n}oz}, R.~R., {Majewski}, S.~R., {Zaggia}, S., {et~al.} 2006, \apj, 649,
  201, \dodoi{10.1086/505620}

\bibitem[{Navarro {et~al.}(1997)Navarro, Frenk, \& White}]{navarro97Universal}
Navarro, J.~F., Frenk, C.~S., \& White, S. D.~M. 1997, ApJ, 490, 493,
  \dodoi{10.1086/304888}

\bibitem[{{Pace} {et~al.}(2020){Pace}, {Kaplinghat}, {Kirby}, {Simon},
  {Tollerud}, {Mu{\~n}oz}, {C{\^o}t{\'e}}, {Djorgovski}, \& {Geha}}]{pace2020}
{Pace}, A.~B., {Kaplinghat}, M., {Kirby}, E., {et~al.} 2020, \mnras, 495, 3022,
  \dodoi{10.1093/mnras/staa1419}

\bibitem[{Patel {et~al.}(2020)Patel, Kallivayalil, {Garavito-Camargo}, Besla,
  Weisz, {van der Marel}, {Boylan-Kolchin}, Pawlowski, \&
  G{\'o}mez}]{patel20Orbital}
Patel, E., Kallivayalil, N., {Garavito-Camargo}, N., {et~al.} 2020, ApJ, 893,
  121, \dodoi{10.3847/1538-4357/ab7b75}

\bibitem[{{Pawlowski} \& {Kroupa}(2013)}]{Pawlowski2013}
{Pawlowski}, M.~S., \& {Kroupa}, P. 2013, \mnras, 435, 2116,
  \dodoi{10.1093/mnras/stt1429}

\bibitem[{Pawlowski \& Kroupa(2014)}]{pawlowski14Vast}
Pawlowski, M.~S., \& Kroupa, P. 2014, ApJ, 790, 74,
  \dodoi{10.1088/0004-637X/790/1/74}

\bibitem[{{Pawlowski} \& {Kroupa}(2020)}]{PawlowskiKroupa2020}
{Pawlowski}, M.~S., \& {Kroupa}, P. 2020, \mnras, 491, 3042,
  \dodoi{10.1093/mnras/stz3163}

\bibitem[{Pawlowski {et~al.}(2015)Pawlowski, McGaugh, \&
  Jerjen}]{pawlowski15new}
Pawlowski, M.~S., McGaugh, S.~S., \& Jerjen, H. 2015, MNRAS, 453, 1047,
  \dodoi{10.1093/mnras/stv1588}

\bibitem[{{Piatek} {et~al.}(2016){Piatek}, {Pryor}, \&
  {Olszewski}}]{Piatek2016}
{Piatek}, S., {Pryor}, C., \& {Olszewski}, E.~W. 2016, \aj, 152, 166,
  \dodoi{10.3847/0004-6256/152/6/166}

\bibitem[{Pouliasis {et~al.}(2017)Pouliasis, Matteo, \&
  Haywood}]{pouliasis17Milky}
Pouliasis, E., Matteo, P.~D., \& Haywood, M. 2017, A\&A, 598, A66,
  \dodoi{10.1051/0004-6361/201527346}

\bibitem[{{Retana-Montenegro} {et~al.}(2012){Retana-Montenegro}, {van Hese},
  Gentile, Baes, \& {Frutos-Alfaro}}]{retana-montenegro12Analytical}
{Retana-Montenegro}, E., {van Hese}, E., Gentile, G., Baes, M., \&
  {Frutos-Alfaro}, F. 2012, A\&A, 540, A70, \dodoi{10.1051/0004-6361/201118543}

\bibitem[{Ripepi {et~al.}(2017)Ripepi, Cioni, Moretti, Marconi, Bekki,
  Clementini, {de Grijs}, Emerson, Groenewegen, Ivanov, Molinaro, Muraveva,
  Oliveira, Piatti, Subramanian, \& {van Loon}}]{ripepi17VMC}
Ripepi, V., Cioni, M.-R.~L., Moretti, M.~I., {et~al.} 2017, MNRAS, 472, 808,
  \dodoi{10.1093/mnras/stx2096}

\bibitem[{Savino {et~al.}(2015)Savino, Salaris, \& Tolstoy}]{savino15Inclusion}
Savino, A., Salaris, M., \& Tolstoy, E. 2015, A\&A, 583, A126,
  \dodoi{10.1051/0004-6361/201527072}

\bibitem[{Simon(2018)}]{simon18Gaia}
Simon, J.~D. 2018, ApJ, 863, 89, \dodoi{10.3847/1538-4357/aacdfb}

\bibitem[{{Simon}(2019)}]{Simon2019}
{Simon}, J.~D. 2019, Annual Review of Astronomy and Astrophysics, 57, 375,
  \dodoi{10.1146/annurev-astro-091918-104453}

\bibitem[{{Simon} \& {Geha}(2007)}]{2007ApJ...670..313S}
{Simon}, J.~D., \& {Geha}, M. 2007, \apj, 670, 313, \dodoi{10.1086/521816}

\bibitem[{{Simon} {et~al.}(2011){Simon}, {Geha}, {Minor}, {Martinez}, {Kirby},
  {Bullock}, {Kaplinghat}, {Strigari}, {Willman}, {Choi}, {Tollerud}, \&
  {Wolf}}]{2011ApJ...733...46S}
{Simon}, J.~D., {Geha}, M., {Minor}, Q.~E., {et~al.} 2011, \apj, 733, 46,
  \dodoi{10.1088/0004-637X/733/1/46}

\bibitem[{{Simon} {et~al.}(2015){Simon}, {Drlica-Wagner}, {Li}, {Nord}, {Geha},
  {Bechtol}, {Balbinot}, {Buckley-Geer}, {Lin}, {Marshall}, {Santiago},
  {Strigari}, {Wang}, {Wechsler}, {Yanny}, {Abbott}, {Bauer}, {Bernstein},
  {Bertin}, {Brooks}, {Burke}, {Capozzi}, {Carnero Rosell}, {Carrasco Kind},
  {D'Andrea}, {da Costa}, {DePoy}, {Desai}, {Diehl}, {Dodelson}, {Cunha},
  {Estrada}, {Evrard}, {Fausti Neto}, {Fernandez}, {Finley}, {Flaugher},
  {Frieman}, {Gaztanaga}, {Gerdes}, {Gruen}, {Gruendl}, {Honscheid}, {James},
  {Kent}, {Kuehn}, {Kuropatkin}, {Lahav}, {Maia}, {March}, {Martini}, {Miller},
  {Miquel}, {Ogando}, {Romer}, {Roodman}, {Rykoff}, {Sako}, {Sanchez},
  {Schubnell}, {Sevilla}, {Smith}, {Soares-Santos}, {Sobreira}, {Suchyta},
  {Swanson}, {Tarle}, {Thaler}, {Tucker}, {Vikram}, {Walker}, {Wester}, \& {DES
  Collaboration}}]{2015ApJ...808...95S}
{Simon}, J.~D., {Drlica-Wagner}, A., {Li}, T.~S., {et~al.} 2015, \apj, 808, 95,
  \dodoi{10.1088/0004-637X/808/1/95}

\bibitem[{{Simon} {et~al.}(2017){Simon}, {Li}, {Drlica-Wagner}, {Bechtol},
  {Marshall}, {James}, {Wang}, {Strigari}, {Balbinot}, {Kuehn}, {Walker},
  {Abbott}, {Allam}, {Annis}, {Benoit-L{\'e}vy}, {Brooks}, {Buckley-Geer},
  {Burke}, {Carnero Rosell}, {Carrasco Kind}, {Carretero}, {Cunha}, {D'Andrea},
  {da Costa}, {DePoy}, {Desai}, {Doel}, {Fernandez}, {Flaugher}, {Frieman},
  {Garc{\'\i}a-Bellido}, {Gaztanaga}, {Goldstein}, {Gruen}, {Gutierrez},
  {Kuropatkin}, {Maia}, {Martini}, {Menanteau}, {Miller}, {Miquel}, {Neilsen},
  {Nord}, {Ogando}, {Plazas}, {Romer}, {Rykoff}, {Sanchez}, {Santiago},
  {Scarpine}, {Schubnell}, {Sevilla-Noarbe}, {Smith}, {Sobreira}, {Suchyta},
  {Swanson}, {Tarle}, {Whiteway}, {Yanny}, \& {DES
  Collaboration}}]{2017ApJ...838...11S}
{Simon}, J.~D., {Li}, T.~S., {Drlica-Wagner}, A., {et~al.} 2017, \apj, 838, 11,
  \dodoi{10.3847/1538-4357/aa5be7}

\bibitem[{{Simon} {et~al.}(2020){Simon}, {Li}, {Erkal}, {Pace},
  {Drlica-Wagner}, {James}, {Marshall}, {Bechtol}, {Hansen}, {Kuehn}, {Lidman},
  {Allam}, {Annis}, {Avila}, {Bertin}, {Brooks}, {Burke}, {Rosell}, {Carrasco
  Kind}, {Carretero}, {da Costa}, {De Vicente}, {Desai}, {Doel}, {Eifler},
  {Everett}, {Fosalba}, {Frieman}, {Garc{\'\i}a-Bellido}, {Gaztanaga},
  {Gerdes}, {Gruen}, {Gruendl}, {Gschwend}, {Gutierrez}, {Hollowood},
  {Honscheid}, {Krause}, {Kuropatkin}, {MacCrann}, {Maia}, {March}, {Miquel},
  {Palmese}, {Paz-Chinch{\'o}n}, {Plazas}, {Reil}, {Roodman}, {Sanchez},
  {Santiago}, {Scarpine}, {Schubnell}, {Serrano}, {Smith}, {Suchyta}, {Tarle},
  {Walker}, \& {DES Collaboration}}]{2020ApJ...892..137S}
{Simon}, J.~D., {Li}, T.~S., {Erkal}, D., {et~al.} 2020, \apj, 892, 137,
  \dodoi{10.3847/1538-4357/ab7ccb}

\bibitem[{{Sohn} {et~al.}(2007){Sohn}, {Majewski}, {Mu{\~n}oz}, {Kunkel},
  {Johnston}, {Ostheimer}, {Guhathakurta}, {Patterson}, {Siegel}, \&
  {Cooper}}]{2007ApJ...663..960S}
{Sohn}, S.~T., {Majewski}, S.~R., {Mu{\~n}oz}, R.~R., {et~al.} 2007, \apj, 663,
  960, \dodoi{10.1086/518302}

\bibitem[{{Spencer} {et~al.}(2017){Spencer}, {Mateo}, {Walker}, \&
  {Olszewski}}]{2017ApJ...836..202S}
{Spencer}, M.~E., {Mateo}, M., {Walker}, M.~G., \& {Olszewski}, E.~W. 2017,
  \apj, 836, 202, \dodoi{10.3847/1538-4357/836/2/202}

\bibitem[{{Tolstoy} {et~al.}(2009){Tolstoy}, {Hill}, \& {Tosi}}]{Tolstoy2009}
{Tolstoy}, E., {Hill}, V., \& {Tosi}, M. 2009, \araa, 47, 371,
  \dodoi{10.1146/annurev-astro-082708-101650}

\bibitem[{Tolstoy {et~al.}(2004)Tolstoy, Irwin, Helmi, Battaglia, Jablonka,
  Hill, Venn, Shetrone, Letarte, Cole, Primas, Francois, Arimoto, Sadakane,
  Kaufer, Szeifert, \& Abel}]{tolstoy04Two}
Tolstoy, E., Irwin, M.~J., Helmi, A., {et~al.} 2004, ApJL, 617, L119,
  \dodoi{10.1086/427388}

\bibitem[{{Torrealba} {et~al.}(2016){Torrealba}, {Koposov}, {Belokurov},
  {Irwin}, {Collins}, {Spencer}, {Ibata}, {Mateo}, {Bonaca}, \&
  {Jethwa}}]{2016MNRAS.463..712T}
{Torrealba}, G., {Koposov}, S.~E., {Belokurov}, V., {et~al.} 2016, \mnras, 463,
  712, \dodoi{10.1093/mnras/stw2051}

\bibitem[{{Torrealba} {et~al.}(2019){Torrealba}, {Belokurov}, {Koposov}, {Li},
  {Walker}, {Sanders}, {Geringer-Sameth}, {Zucker}, {Kuehn}, {Evans}, \&
  {Dehnen}}]{2019MNRAS.488.2743T}
{Torrealba}, G., {Belokurov}, V., {Koposov}, S.~E., {et~al.} 2019, \mnras, 488,
  2743, \dodoi{10.1093/mnras/stz1624}

\bibitem[{Torrealba {et~al.}(2019)Torrealba, Belokurov, Koposov, Li, Walker,
  Sanders, {Geringer-Sameth}, Zucker, Kuehn, Evans, \&
  Dehnen}]{torrealba19hidden}
Torrealba, G., Belokurov, V., Koposov, S.~E., {et~al.} 2019, MNRAS, 488, 2743,
  \dodoi{10.1093/mnras/stz1624}

\bibitem[{{Ural} {et~al.}(2010){Ural}, {Wilkinson}, {Koch}, {Gilmore}, {Beers},
  {Belokurov}, {Evans}, {Grebel}, {Vidrih}, \& {Zucker}}]{2010MNRAS.402.1357U}
{Ural}, U., {Wilkinson}, M.~I., {Koch}, A., {et~al.} 2010, \mnras, 402, 1357,
  \dodoi{10.1111/j.1365-2966.2009.15975.x}

\bibitem[{{van der Marel} \& Guhathakurta(2008)}]{vandermarel08M31}
{van der Marel}, R.~P., \& Guhathakurta, P. 2008, ApJ, 678, 187,
  \dodoi{10.1086/533430}

\bibitem[{Vasiliev(2019)}]{vasiliev19Systematic}
Vasiliev, E. 2019, MNRAS, 489, 623, \dodoi{10.1093/mnras/stz2100}

\bibitem[{{Vitral}(2021)}]{Vitral2021}
{Vitral}, E. 2021, arXiv e-prints, arXiv:2102.04841.
\newblock \doarXiv{2102.04841}

\bibitem[{{Voggel} {et~al.}(2016){Voggel}, {Hilker}, {Baumgardt}, {Collins},
  {Grebel}, {Husemann}, {Richtler}, \& {Frank}}]{2016MNRAS.460.3384V}
{Voggel}, K., {Hilker}, M., {Baumgardt}, H., {et~al.} 2016, \mnras, 460, 3384,
  \dodoi{10.1093/mnras/stw1132}

\bibitem[{{Walker} {et~al.}(2009{\natexlab{a}}){Walker}, {Belokurov}, {Evans},
  {Irwin}, {Mateo}, {Olszewski}, \& {Gilmore}}]{2009ApJ...694L.144W}
{Walker}, M.~G., {Belokurov}, V., {Evans}, N.~W., {et~al.} 2009{\natexlab{a}},
  \apjl, 694, L144, \dodoi{10.1088/0004-637X/694/2/L144}

\bibitem[{{Walker} {et~al.}(2009{\natexlab{b}}){Walker}, {Mateo}, \&
  {Olszewski}}]{2009AJ....137.3100W}
{Walker}, M.~G., {Mateo}, M., \& {Olszewski}, E.~W. 2009{\natexlab{b}}, \aj,
  137, 3100, \dodoi{10.1088/0004-6256/137/2/3100}

\bibitem[{{Walker} {et~al.}(2015){Walker}, {Olszewski}, \&
  {Mateo}}]{2015MNRAS.448.2717W}
{Walker}, M.~G., {Olszewski}, E.~W., \& {Mateo}, M. 2015, \mnras, 448, 2717,
  \dodoi{10.1093/mnras/stv099}

\bibitem[{{Walker} {et~al.}(2016){Walker}, {Mateo}, {Olszewski}, {Koposov},
  {Belokurov}, {Jethwa}, {Nidever}, {Bonnivard}, {Bailey}, {Bell}, \&
  {Loebman}}]{2016ApJ...819...53W}
{Walker}, M.~G., {Mateo}, M., {Olszewski}, E.~W., {et~al.} 2016, \apj, 819, 53,
  \dodoi{10.3847/0004-637X/819/1/53}

\bibitem[{Wang \& Hammer(2021)}]{wang21}
Wang, J., \& Hammer, F. 2021, to be submitted

\bibitem[{Wang {et~al.}(2019)Wang, Hammer, Yang, Ripepi, Cioni, Puech, \&
  Flores}]{wang19complete}
Wang, J., Hammer, F., Yang, Y., {et~al.} 2019, MNRAS, 486, 5907,
  \dodoi{10.1093/mnras/stz1274}

\end{thebibliography}
\bibliographystyle{aasjournal}

\listofchanges

\end{document}